\documentclass{elsart}
\usepackage[english]{babel}
\usepackage{graphicx,subfigure}
\usepackage{amsfonts,amsmath}
\usepackage{comment}
\graphicspath{{./Figsnew2/}}
\newcommand{\im}{\mathrm{i}}

\newcommand{\co}{\cos(\psi_0)}
\newcommand{\so}{\sin(\psi_0)}
\begin{document}
\begin{frontmatter}
\title{Birhythmicity, Synchronization, and Turbulence in an Oscillatory
System with Nonlocal Inertial Coupling}
\author{Vanessa Casagrande\thanksref{kuramoto}}
and
\author{Alexander S. Mikhailov\corauthref{cor}\thanksref{kuramoto}}
\address{Abteilung Physikalische Chemie,\\
Fritz-Haber-Institut der Max-Planck-Gesellschaft,\\
Faradayweg 4-6, 14195 Berlin, Germany}
\corauth[cor]{Corresponding author}
\thanks[kuramoto]{Dedicated to Prof. Y. Kuramoto on the occasion of 
his retirement}
\ead{mikhailov@fhi-berlin.mpg.de}
\begin{abstract}
We consider a model where a population of diffusively coupled 
limit-cycle oscillators, described by the complex Ginzburg-Landau 
equation, interacts nonlocally via an inertial field. 
For sufficiently high intensity of nonlocal inertial coupling, the 
system exhibits birhythmicity with two oscillation modes at largely 
different frequencies. Stability of uniform oscillations in the 
birhythmic region is analyzed by means of the phase
dynamics approximation. Numerical simulations show that, depending on 
its parameters, the system has irregular intermittent regimes with 
local bursts of synchronization or desynchronization. 
\end{abstract}
\end{frontmatter}

\section{Introduction}
Beginning with the pioneering contributions by Kuramoto \cite{kurbook} 
and Winfree \cite{winfree}, studies of synchronization and 
spatiotemporal chaos (turbulence) in populations of coupled active
oscillators have developed into a classical field of research. 
In chemical systems, each oscillator 
represents a reaction element and coupling between such elements is 
usually due to diffusion of reactants in the system. This coupling 
extends only within a short diffusion length and is therefore local. 
The complex Ginzburg-Landau equation is the canonical model for 
oscillatory systems with local coupling
near a supercritical Hopf bifurcation. In surface chemical reactions
\cite{ertl}, reaction elements are however additionally interacting
through the gas phase where instantaneous complete mixing occurs. 
As a result, global coupling between chemical oscillators arises 
\cite{global}.
Moreover, global delayed feedbacks through the gas phase could be
artificially introduced to control turbulence in surface reactions 
\cite{batt96,batt97,kawamura,science,bertram1,bertram2}.
A special class of systems are arrays of active oscillators that do not
directly interact one with another, but are all coupled to a single
diffusing field (in biochemistry, an example of such a system would be
provided by arrays of allosteric enzymes interacting through a chemical
messenger \cite{enzymes}). 
Adiabatically eliminating this field, models with nonlocal
coupling between the oscillators, described by a finite-range integral,
were derived \cite{kur95,kur98}. 
Investigations of nonlocally coupled oscillator
arrays have revealed that such form of coupling does not always 
synchronize neighbouring oscillators and discontinuous distributions 
with scaleless fractal structure can develop \cite{kur97,kur98}. 
Spatiotemporal chaos can develop in such systems even in the 
Benjamin-Feir stable regime \cite{batt00}
and spiral waves with phase-randomized cores can exist here 
\cite{shima,chapter9}.

Furthermore, arrays with both local and nonlocal coupling between
oscillators are possible. This situation is characteristic, for example,
for surface chemical reactions. In such systems, diffusion provides local
coupling between neighbouring surface oscillators, whereas much faster
heat conduction is responsible for nonlocal coupling between them. 

Recently, Tanaka and Kuramoto \cite{tanaka} have shown how, in the 
vicinity of a supercritical Hopf bifurcation, the description of 
arrays of nonlocally coupled oscillators can be reduced to the complex 
Ginzburg-Landau equation with nonlocal coupling. Because of the critical
slowing down near the bifurcation point, the coupling is effectively 
instantaneous in the reduced description.

In realistic systems, which are not too close to the Hopf bifurcation, one
can however also encounter the opposite situation, with a very slow
inertial field giving rise to nonlocal coupling. The aim of our study is
to analyze what new effects, primarily due to slow nonlocal coupling, are
possible in this class of systems. For our investigations, we have chosen
an abstract model of the complex Ginzburg-Landau equation interacting with
an additional slow field, so that the coupling is nonlocal both with
respect to space and time. We expect that the behaviour found in this
general model would be typical for a broad class of realistic systems. 

The principal effect of coupling inertiality is that birhythmicity,
leading to chaotic intermittency, can develop in such systems. In
contrast to the birhythmicity in reaction-diffusion systems near a 
pitchfork-Hopf bifurcation (see \cite{michael01,michael02}), the two 
oscillatory states 
are characterized here by very different frequencies and oscillation 
amplitudes. For the rapid oscillatory state, the additional field 
responsible for the inertial long-range coupling is almost absent
and only diffusive local coupling between the oscillators is 
important. In slow oscillations, the elements are however entrained by 
the long-range inertial field.

Depending on the parameters, two new kinds of intermittency are found
here. When rapid oscillations are modulationally unstable and give rise to
turbulence, spatial regions occupied by slow entrained oscillations
spontaneously develop and die out in the medium. They can be interpreted
as bursts of synchronization on a turbulent background. On the other hand,
relatively small, irregularly evolving islands filled with rapid
turbulence can persist on the background of slow almost uniform
oscillations. They can be therefore described as bursts of
desynchronization on the background of regular slow oscillations.  Our
analysis of such phenomena is based on the phase dynamics approximation,
derived for the birhythmic system. It is complemented by 1D and 2D
numerical simulations.

\section{The Model}

The investigated model describes a system of diffusively coupled active
oscillators coupled to an additional inertial diffusive field. 
Introducing the local complex oscillation amplitude $\eta (x,t)$ and 
denoting as $z(x,t)$ the additional complex-valued diffusive field, 
we have therefore a system of two equations 
\begin{subequations}
\begin{align}
&\dot\eta&=&\quad(1+\im\omega)\eta-(1+\im\alpha)|\eta|^2\eta+
(1+\im\beta)\nabla^2\eta+K(z-\eta)\label{model:one}\\
&\tau\dot z&=&\quad\eta-z+l^2\nabla ^2 z\label{model:two}
\end{align}\label{model}
\end{subequations}
The additional last term in the complex Ginzburg-Landau equation
(\ref{model:one}) takes into account coupling of the oscillatory 
subsystem with the field $z$ whose evolution obeys equation 
(\ref{model:two}); $K$ is the respective coupling constant. 
The equations are brought into a dimensionless form by choosing
the characteristic diffusion length in the oscillatory subsystem as the
length unit and taking the characteristic relaxation time scale of the
oscillators as the time unit. The parameters $\tau$ and $l$ determine
characteristic time and length scales of the additional field $z$. 
We assume that this field is inertial ($\tau \gg 1$) and slowly 
varying in space ($l\gg 1$).

The linear equation (\ref{model:two}) can easily be solved as 
\begin{equation}
z(x,t)=\int_{-\infty }^{\infty }\int_{0}^{t}G(x-x^{\prime },t-t^{\prime
})\eta (x^{\prime },t^{\prime })dx^{\prime }dt^{\prime }.
\end{equation}
where the kernel is given by 
\begin{equation}
G(x,t)=\frac{1}{2\tau \sqrt{\pi Dt}}\exp \left( -\frac{x^{2}}{4Dt}-
\frac{t}{\tau }\right) .  \label{kernel}
\end{equation}
Substituting this into equation (\ref{model:one}), we obtain an 
equivalent integro-differential form of the considered model, 
\begin{eqnarray}
\dot\eta &=&(1+\im\omega)\eta-(1+\im\alpha)|\eta|^{2}\eta 
+(1+\im\beta)\nabla^2\eta  \nonumber\\
&&+K\int_{-\infty }^{\infty }\int_{0}^{t}G(x-x^{\prime },t-t^{\prime
})\left[ \eta (x^{\prime },t^{\prime })-\eta (x,t)\right] dx^{\prime
}dt^{\prime } \label{integralform}
\end{eqnarray}
We see that, besides of diffusion, the model also includes an 
additional coupling, nonlocal both with respect to space and time.

Note that very close to a supercritical Hopf bifurcation all processes
become fast as compared to the relaxation time scale of individual
oscillators, because this time is inversely proportional to the 
distance from the bifurcation point (critical slowing down). 
However, it is known
that the complex Ginzburg-Landau equation yields qualitatively correct
description even relatively far from the bifurcation. 
Equations (\ref{model}) and (\ref{integralform}) can be viewed as 
providing a simple model of an oscillating system coupled to an 
inertial diffusive field.

\section{Birhythmicity}

The system described by equations (\ref{model}) is birhythmic,
i.e. it can have two different oscillatory uniform states. 
Assuming that
$\eta (t)=\rho e^{-\im\gamma t}$ and $z(t)=re^{-\im\gamma t}$ and
substituting this into equations (\ref{model:one}) and 
(\ref{model:two}), we obtain a
cubic equation for the oscillation frequency $\gamma $, 
\begin{equation}
\tau ^{2}\gamma ^{3}+\tau ^{2}(\omega -\alpha +\alpha K)
\gamma ^{2}+(1+\tau
K)\gamma +\omega -\alpha =0.  \label{3rd_order}
\end{equation}

When the frequency $\gamma$ is known, the oscillation amplitude 
$\rho$ of the field $\eta$ is given by 
\begin{equation}
\rho ^{2}=1-\frac{K\tau ^{2}\gamma ^{2}}{1+\tau ^{2}\gamma ^{2}}.
\label{ampl}
\end{equation}
The respective oscillation amplitude of the field $z$ is 
\begin{equation}
r=\frac{\rho}{\sqrt{1+\tau ^{2}\gamma ^{2}}}.
\end{equation}
Depending on its parameters, equation (\ref{3rd_order}) can have 
either one or three real roots. The three roots correspond to three 
possible modes of uniform oscillations with frequencies 
$\gamma_{1,2,3}$, such that $\gamma_{1}<\gamma_{2}<\gamma_{3}$. 
It can be checked that oscillations with the
middle frequency $\gamma_{2}$ are always unstable. 
In contrast to this,
uniform oscillations with frequencies $\gamma_{1}$ and $\gamma_{3}$ 
are possible (but may still be unstable with respect to nonuniform
perturbations, see the discussion below). 

\begin{figure}
\begin{center}
\subfigure[]{\scalebox{0.4}[0.4]
{\label{ak_bif}\includegraphics*{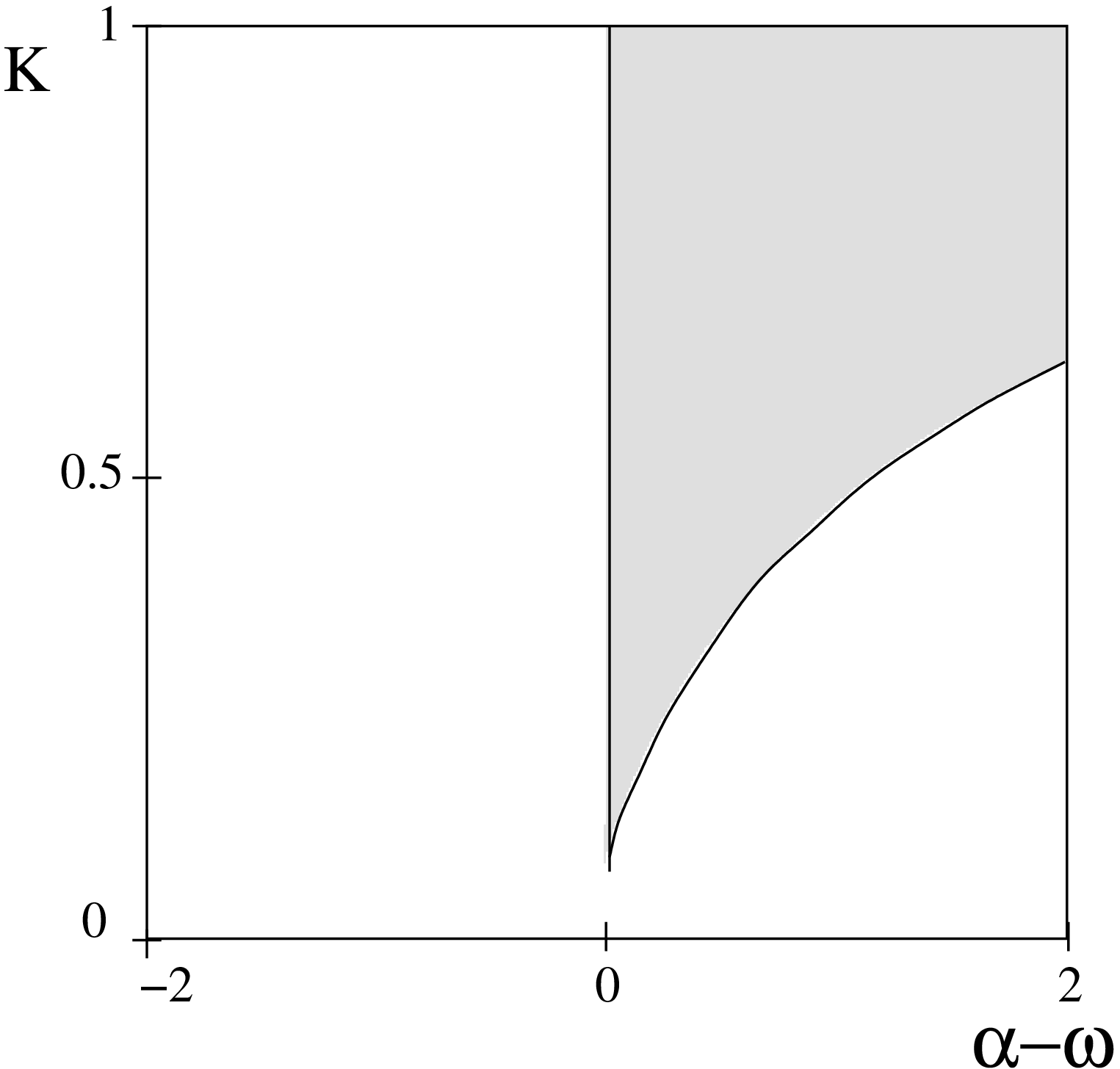}}}
\subfigure[]{\scalebox{0.4}[0.4]
{\label{tk_bif}\includegraphics*{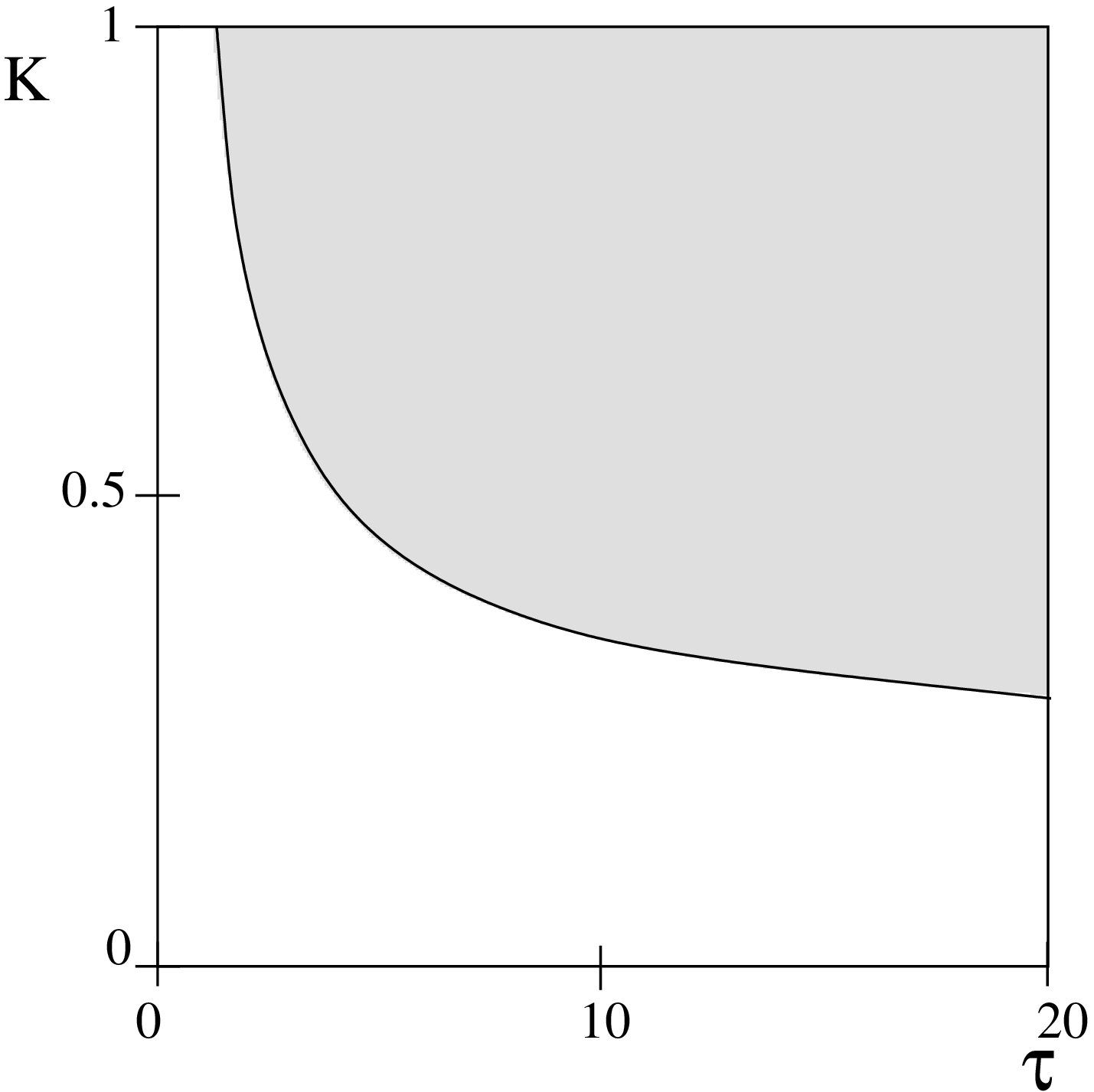}}}
\caption{Birhythimicity regions (gray) for the model (\ref{model}) 
in the parameter
planes $(\alpha-\omega ,K)$ and $(\tau ,K).$ The fixed parameters are 
$\tau=10$ in the first diagram and $\omega =2,\alpha =2.5$ in the 
second diagram.
The boundaries of the displayed regions do not depend on the parameter 
$\beta$ of the model.}
\label{bif}
\end{center}
\end{figure}

Figure \ref{ak_bif} shows the region in the parameter planes 
($\alpha-\omega,K$)  and  ($\tau,K$) where birhythmicity exists. 
The birhythmicity is possible only for sufficiently strong coupling
$K$. Moreover, it develops only if the 
characteristic time $\tau$ of the field $z$ is sufficiently large
(Fig. \ref{tk_bif}).

\begin{figure}
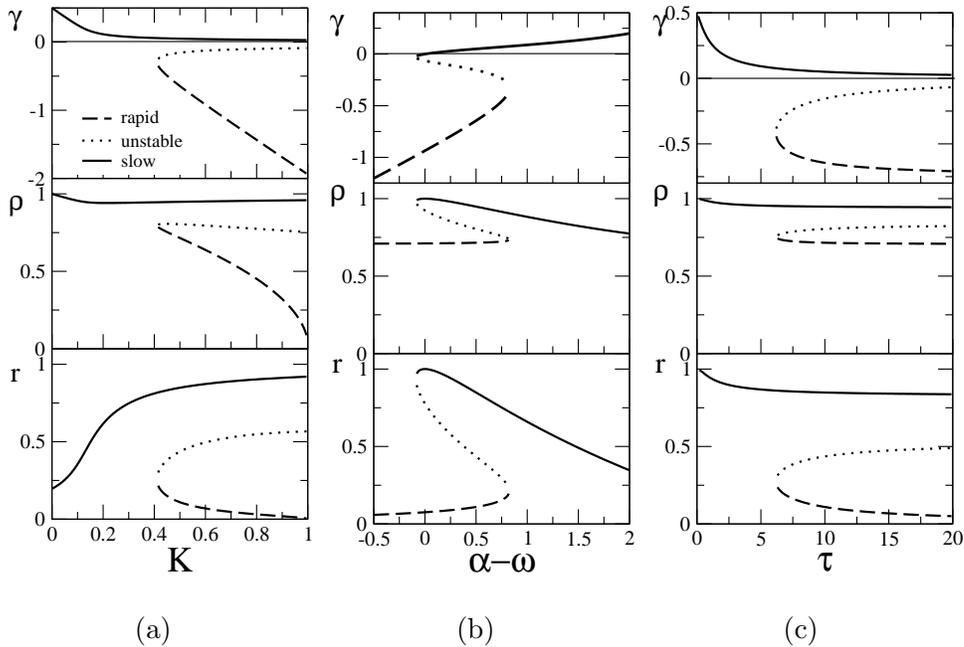

\begin{center}
\subfigure[]{\scalebox{.45}[.45]{\label{bif_k}
\includegraphics*{bifurcation_k.eps}}}
\subfigure[]{\scalebox{.45}[.45]{\label{bif_a}
\includegraphics*{bifurcation_a.eps}}}
\subfigure[]{\scalebox{.45}[.45]{\label{bif_t}
\includegraphics*{bifurcation_tau.eps}}}
\caption{Bifurcation diagrams. Frequencies $\gamma$ of uniform 
oscillations, together with the respective amplitudes $\rho =|\eta|$ 
and $r=|z|$ as functions of $K$, $\alpha -\omega $ and $\tau $ for
slow (bold line), rapid (dashed) and absolutely unstable (dotted) 
uniform oscillation modes. The parameters are (a) $\omega =2,
\alpha =2.5,\tau =10$ (b) $\omega =2,K=0.5,\tau =10,$ (c) 
$\omega =2,\alpha =2.5,K=0.5.$}
\label{bif_diag}
\end{center}
\end{figure}

Numerical solutions of equation (\ref{3rd_order}) are displayed in Fig.
\ref{bif_diag}. If the parameter $\alpha$ is kept constant and the
coupling strength $K$ is varied (Fig. \ref{bif_k}), birhythmicity 
is found inside the interval $0.4<K<1$. 
When $K\ge 1,$ the system has a stable stationary state coexisting
with oscillations. The dependence of the effective oscillation 
frequency $\gamma$ on the difference $\alpha-\omega$ for $K=0.5$ is 
shown in Fig.\ref{bif_a}. 
Two stable limit cycles coexist within the interval $-0.1<\alpha-\omega
<0.8$.

The difference between the two limit cycles becomes clear if we compare
the amplitudes of oscillations of the fields $\eta$ and $z$ which are 
presented below in the same figures. As we have already observed, 
at $K=0$ only one oscillating mode is present, as we are in this limit 
reduced to the standard CGLE. 
Thus, we can expect that the branch starting at $K=0$ (which is the
upper one in the bifurcation diagrams of Fig. \ref{bif_diag})
would have the strongest similarities to the CGLE, even though for
finite $K$ it would be affected by the coupling to the second field. 
This branch is characterized by a large amplitude of the field $\eta$, 
which never gets significantly smaller than 1. 
The corresponding frequency is quite small, starting from $\gamma =0.5$
at $K=0$ and slowly decreasing as $K$ increases.
Below, we refer to this attractor as corresponding to the {\em slow\/} 
limit cycle.

In contrast to this, oscillations in the lower branch have a smaller
amplitude $\rho$, which decreases for higher coupling strengths $K$ and
eventually vanishes at $K=1$. Note that $\gamma$ is negative for this 
mode and therefore such oscillations are counter-rotating with respect 
to the upper branch. The frequency varies significantly for this mode. 
Inside the birhythmicity interval, the frequency of such {\em rapid\/} 
limit cycle is always much larger (in its magnitude) than the frequency
of the slow mode. A characteristic feature of the rapid limit cycle is 
that the amplitude $r$ of the field $z$ is very small here. 
This is because the oscillations of the
field $\eta$ are so rapid here that the inertial field $z$ cannot 
respond to them. Hence, the coupling term $Kz$ in the system 
(\ref{model}) is almost negligible for the rapid limit cycle so that 
the nonlocal coupling is ineffective and oscillators interact only via 
the local diffusional coupling in this mode.

\section{The Phase Dynamics Approximation}

Above, uniform oscillations in the model have been discussed. 
Now, we want to consider the properties of {\em almost\/} uniform 
oscillations, characterized by small phase gradients. 
Because the system is invariant with
respect to the uniform shift of all phases, the evolution of phase
distributions with small phase gradients should be slow, in contrast 
to fast local relaxation of the amplitude perturbations. 
Therefore, a reduced description, known as the phase dynamics 
approximation, can be constructed in such cases \cite{kurbook}. 
When birhythmicity is present, the coefficients
of the phase dynamics equation are different for the two oscillation 
modes.

The derivation of the phase dynamics equation for this system is 
lengthy and has been performed by using a computer program for 
algebraic calculations ({\em Mathematica\/}). 
Below in this section we show the scheme of derivation
and discuss the results. The complete derivation is presented in 
Appendix A.

We define the phases ($\phi$ and $\varphi$) and amplitudes 
($\rho$ and $r$) of complex fields $\eta$ and $z$ as 
$\eta =\rho e^{\im\phi }$, $z=re^{\im\varphi}$. 
It is convenient to use instead of $\phi $ and $\varphi $
the variables $\psi =\phi-\varphi $ and $\Theta =\phi+\varphi$. 
Note that the phase sum $\Theta$ is the slow variable in this system. 
The phases of the two fields are rigidly correlated in the uniform 
state and their difference $\psi $ undergoes rapid relaxation, 
similar to oscillation amplitudes $\rho$ and $z$.

In terms of the new variables, the model (\ref{model}) takes the form 
\begin{subequations}
\begin{eqnarray}
\dot\rho&=&\rho-\rho^3+K r \cos\psi-K\rho+\rho\left
(-\displaystyle\frac{\nabla\Theta^2}{4}+\displaystyle\frac{1}{2}
\beta\nabla^2\Theta\right)\\
\dot r&=&\displaystyle\frac{\rho}{\tau}\cos{\psi}-
\displaystyle\frac{r}{\tau}-\displaystyle\frac{l^2}{4\tau}r
\nabla\Theta^2\\
\dot\psi&=&-\omega+\alpha\rho^2-\left(\displaystyle\frac{k r}{\rho}
+\displaystyle\frac{\rho}{\tau r}\right)\sin\psi
+\displaystyle\frac{\beta}{4}\nabla\Theta^2
+\displaystyle\frac{1}{2}\left(1-
\displaystyle\frac{l^2}{\tau}\right)\nabla^2\Theta\\
\dot\Theta&=&-\omega+\alpha\rho^2-\left(\displaystyle\frac{k r}{\rho}
-\displaystyle\frac{\rho}{\tau r}\right)\sin\psi+\displaystyle
\frac{\beta}{4}\nabla\Theta^2+\displaystyle\frac{1}{2}\left(
1+\displaystyle\frac{l^2}{\tau}\right)\nabla^2\Theta
\end{eqnarray}\label{four}
\end{subequations}
Suppose now that the birhythmic system is close to one of its two
oscillatory states, characterized by some amplitudes $\rho_0$ and 
$r_0$, and the relative phase $\psi_0$, whereas the phase sum 
$\Theta$ remains arbitrary. 
If the phase sum $\Theta$ is slowly varying in space,
the variables $\rho$, $r$ and $\psi $ would only slightly deviate from
their equilibrium values. 
Therefore, we can introduce small perturbations
of such variables, $\rho =\rho_0+\delta \rho$, $r=r_0+\delta r$, 
$\psi=\psi_0+\delta \psi $, and linearize the equations for with 
respect to such perturbations. 
The linearized evolution equation for the phase $\Theta $ is 
\begin{eqnarray}
\dot\Theta&=&-\omega+\alpha\rho_0^2-\left(\frac{Kr_0}{\rho_0}-
\frac{\rho_0}{\tau r_0}\right)\sin\psi_0\nonumber\\
&+&\delta\rho\left[2\alpha\rho_0+\left(\frac{Kr_0}{\rho_0^2}
+\frac{1}{\tau r_0}\right)\sin\psi_0\right]\nonumber\\
&+&\delta r\left(-\frac{K}{\rho_0}-\frac{\rho_0}{\tau r_0^2}\right)
\sin \psi_0+\delta\psi\left(-\frac{Kr_0}{\rho_0}
+\frac{\rho_0}{\tau r_0}\right) \cos\psi_0\nonumber\\
&+&\frac{\beta}{4}\nabla\Theta^2+\frac{1}{2}\left(1
+\frac{l^2}{\tau}\right)\nabla^2\Theta
\label{sum}
\end{eqnarray}
The linearized equations for the variables $\delta \rho ,\delta r,
\delta\psi$ can be solved in the adiabatic approximation, because 
these fast variables are adjusting to slow variation of $\Theta$. 
Substituting the result into (\ref{sum}), a closed evolution equation 
for the phase variable $\Theta $ is derived. 
This equation has the form:
\begin{equation}
\dot\Theta=C_0+C_1(\nabla\Theta)^2+C_2\nabla^2\Theta
\end{equation}
The analytical expressions for the coefficients $C_0,$ $C_1$
are given in Appendix A. Note that inside the
birhythmic region there are two uniform oscillation modes with 
different $\rho_0$, $r_0$, and $\psi _0$, and thus with 
different values of these coefficients.

\begin{figure}
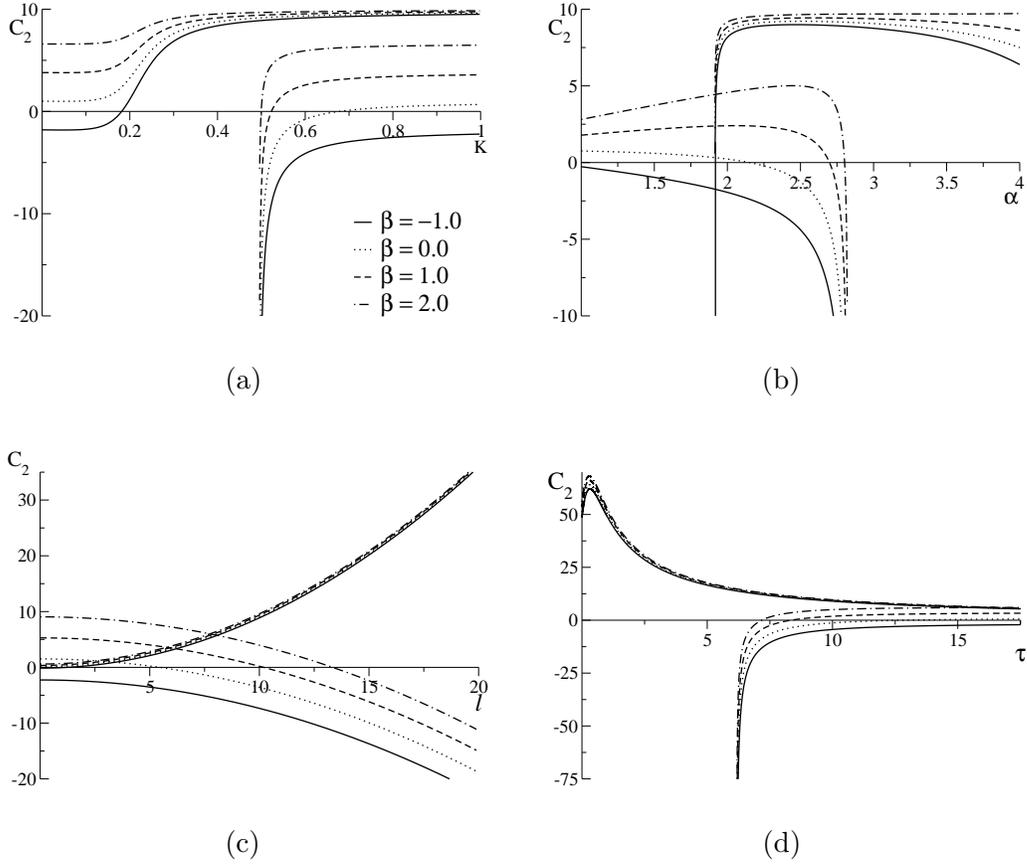

\begin{center}
\subfigure[]{\label{coeff2:k}\scalebox{.27}[.27]
{\includegraphics*{k_stability.eps}}}
\hspace{0.5cm}
\subfigure[]{\label{coeff2:a}\scalebox{.27}[.27]
{\includegraphics*{alpha_stability.eps}}}\\
\subfigure[]{\label{coeff2:l}\scalebox{.27}[.27]
{\includegraphics*{l_stability.eps}}}
\hspace{0.5cm}
\subfigure[]{\label{coeff2:t}\scalebox{.27}[.27]
{\includegraphics*{t_stability.eps}}}
\end{center}
\caption{
The coefficient $C_2$ of the phase dynamics equations as functions
of $K,\alpha ,l$ and $\tau $ for several different values of the 
parameter $\beta $. The upper branches in (a,b,d) and for the larger 
values of $l$ in (c) correspond to slow oscillations, the lower 
branches are for the rapid oscillations. The oscillations are unstable 
if $C_2<0$. The parameters are (a) $\alpha =2.5,$ $l=10$, $\tau =10,$
(b) $K=0.5,l=10,\tau =10,$ (c) $\alpha =2.5,$ $K=0.5,$ $\tau =10,$ 
(d) $\alpha =2.5,$ $K=0.5,$ $l=10;$ for all curves $\omega =2.$
}
\label{coeff2}
\end{figure}

The most important coefficient is $C_2$ because its sign controls
stability of uniform oscillations with respect to phase modulation 
(the Benjamin-Feir instability). 
If $C_2>0$, uniform oscillations are stable,
if $C_2<0$ they are unstable. We have computed this coefficient as
functions of several model parameters, by using the obtained analytical
expressions. The computed dependences are shown in Fig. \ref{coeff2}. 
Two different branches correspond to the two different limit cycles. 
The upper curves in Fig. \ref{coeff2}(a,b,d) are for the slow mode, 
whereas the lower curves are for the
rapid oscillations. In Fig. \ref{coeff2}(c), the upper curves 
correspond to the slow mode for larger values of the parameter $l$.

Examining these plots, several observations can be made: Slow mode
oscillations are always stable inside the birhythmic region, even when 
the system without coupling is Benjamin-Feir unstable 
(i.e., $1+\alpha \beta <0$ and therefore $C_2<0$ at $K=0$). 
In contrast to this, rapid oscillations are always unstable when the 
condition $1+\alpha \beta <0$ is satisfied.
Moreover, they are unstable near the boundary of the birhythmic region,
at which they first appear, even when this condition is violated. 
Increasing the diffusion length $l$ of the field $z$ favors instability
of rapid uniform oscillations.

\section{Numerical Simulations}

To investigate nonlinear spatiotemporal dynamics in the considered 
model, simulations of equations (\ref{model:one}) and (\ref{model:two})
have been performed. For numerical integration of these equations, 
the fourth-order Runge-Kutta algorithm has been used. 
The mesh size for space discretization and the time
step have been chosen to optimize the computational time for each 
parameter choice. Since the diffusion length and the diffusion constant
of the oscillatory field have both been chosen to be equal to unity, 
$\Delta x$ varied between 0.3 and 0.5, while $\Delta t$ could vary 
between $(\Delta x)^{2}/5$ and $(\Delta x)^{2}/2$. 
Both one- and two-dimensional systems were
investigated. No-flux boundary conditions were employed. 
Initial conditions varied depending on a particular simulation.

To display the results of one-dimensional simulations, space-time 
diagrams have been constructed. 
In such diagrams, time always runs along the
horizontal axis and the vertical axis is corresponding to the spatial
coordinate. In two dimensions, snapshots at some time moments and 
space-time diagrams showing evolution of the pattern along some linear 
cross-section are presented. The patterns are shown by using a gray 
scale where the white color encodes the smallest and the black color 
the largest values of the displayed variable.

\begin{figure}
\begin{center}
\subfigure[Re$(\eta)$]{\label{front:re25}\includegraphics*
[width=3cm,height=5cm, angle=-90]{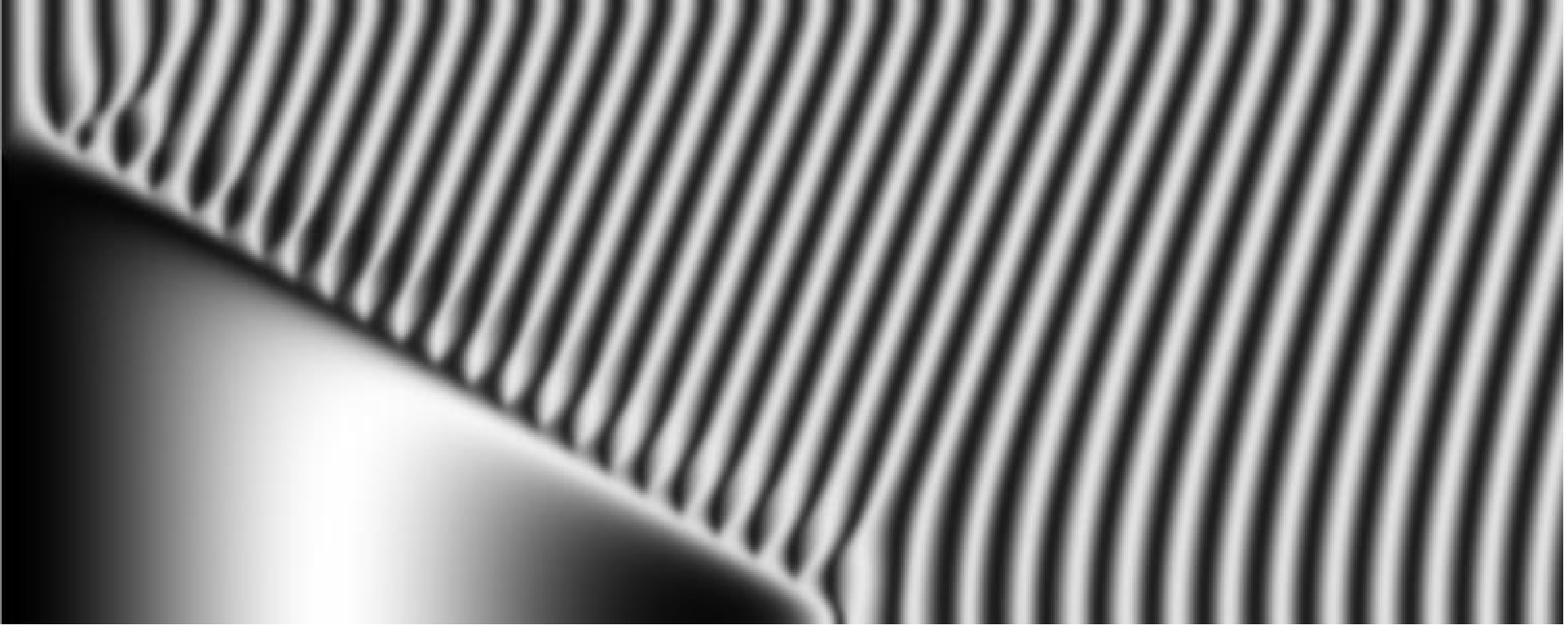}}
\hspace{0.5cm}
\subfigure[$|\eta|$]{\label{front:a25}\includegraphics*
[width=3cm,height=5cm,angle=-90]{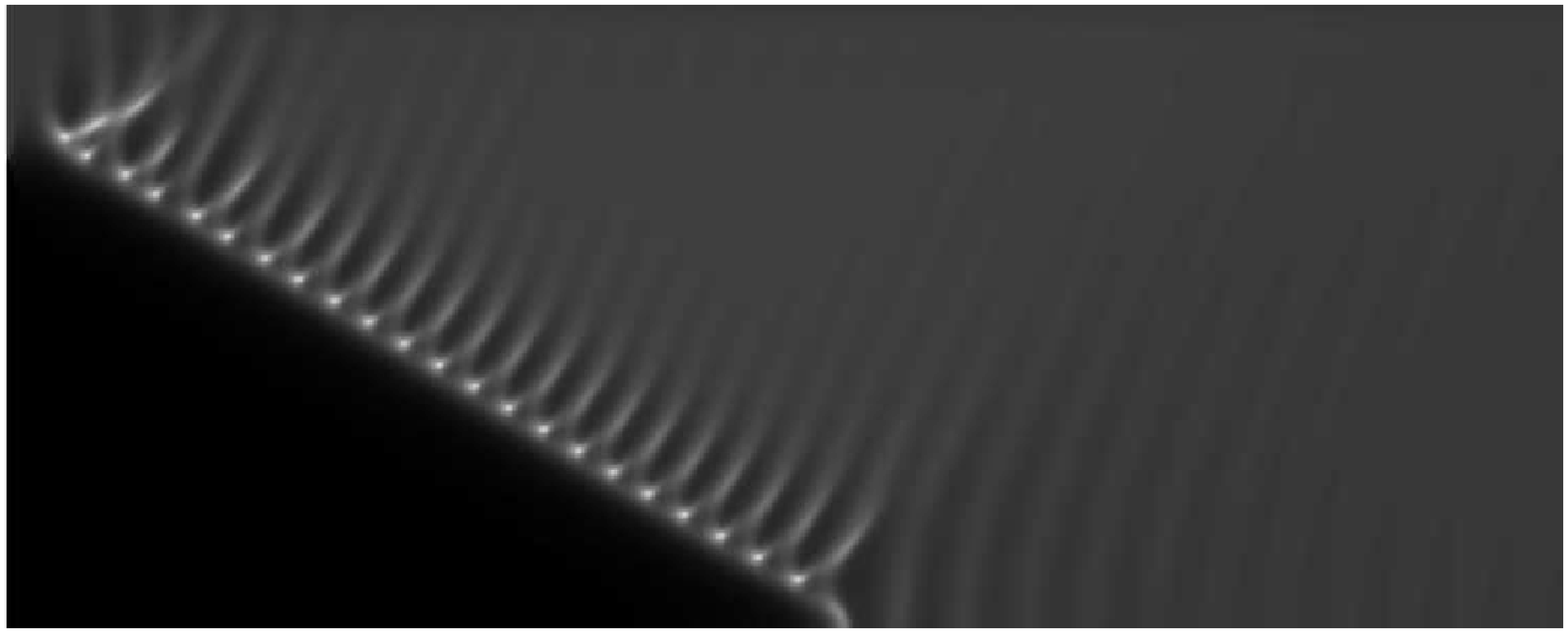}}
\caption{
Spatiotemporal diagrams showing front propagation in the birhythmic
system with $\omega =2,\alpha =2.5,\beta =1,K=0.5,l=10,$ and 
$\tau =10.$ 
The total system length is $L=120$ and the total displayed time 
interval is $T=300$. 
Bright regions correspond to small values of displayed variables.
Time runs from left to right along the horizontal axis, the spatial
coordinate is varied in vertical direction.}
\label{front}
\end{center}
\end{figure}

\begin{figure}
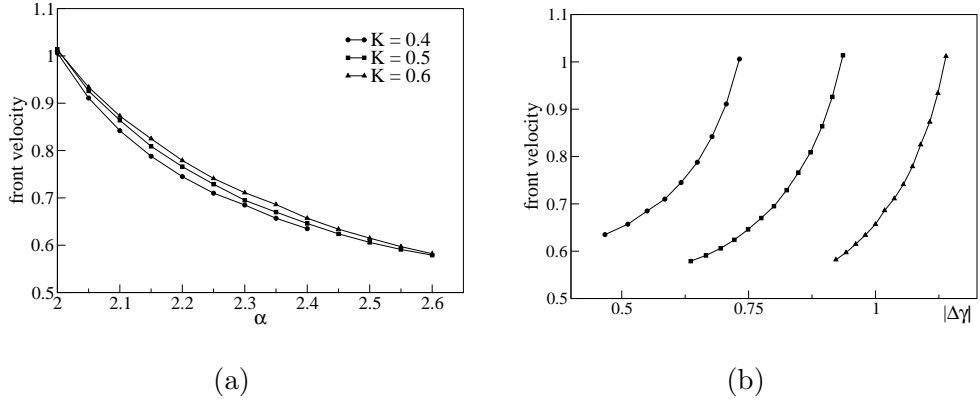

\begin{center}
\subfigure[]{\label{front_vel_a}{\scalebox{0.25}[0.25]
{\includegraphics*{frontvelocity_alpha.eps}}}}
\hspace{0.5cm}
\subfigure[]{\label{front_vel_dom}{\scalebox{0.25}[0.25]
{\includegraphics*{frontvelocity_dom.eps}}}}
\caption{Front velocity as functions of the parameter $\alpha $ and
difference $\Delta \gamma $ of oscillation frequencies in the two modes
in the birhythmic regime. The fixed parameters are $\omega =2,\beta =1,
l=10,$ and $\tau =10.$}
\label{front_vel}
\end{center}
\end{figure}

First, the behavior of fronts separating spatial regions with two 
different oscillatory states in the birhythmic system has been studied 
(Fig. \ref{front}). 
The front travels towards the slow oscillating region, in such a way 
that the system ends up with uniform oscillations of the rapid mode. 
Its propagation is characterized by periodic appearance of amplitude 
defects which occur when the phase difference between the two 
oscillating regions equals $2\pi $ (see Fig \ref{front:a25}). 
The front moves at a constant velocity which depends on the
parameters of the system. Fig. \ref{front_vel_a} 
shows the dependence of the front velocity
on the parameter $\alpha $ for several values of the coupling strength 
$K$. In Fig. \ref{front_vel_dom}, the dependence of the front 
propagation velocity on the frequency difference 
$\Delta \gamma =\gamma _{3}-\gamma _{1}$ of two uniform
oscillatory modes is presented. 
The front always moves faster when this difference is increased.

As follows from the stability analysis of uniform oscillatory states 
(see, e.g., Fig. \ref{coeff2:a}), 
rapid oscillations should be unstable with respect to phase
modulation near the birhythmicity boundary and complex spatiotemporal
regimes can be expected there. We have performed a series of numerical
simulations, exploring pattern formation in the system within the 
intervals $2.55<\alpha <2.8$ and $-1<\beta <2.5$, while keeping fixed 
the other parameters ($\omega =2,K=0.5,l=10,$ and $\tau =10$). 
Each simulation started from independently chosen, completely random 
initial conditions.

\begin{figure}
\begin{center}
\scalebox{.4}[.4]{\includegraphics*{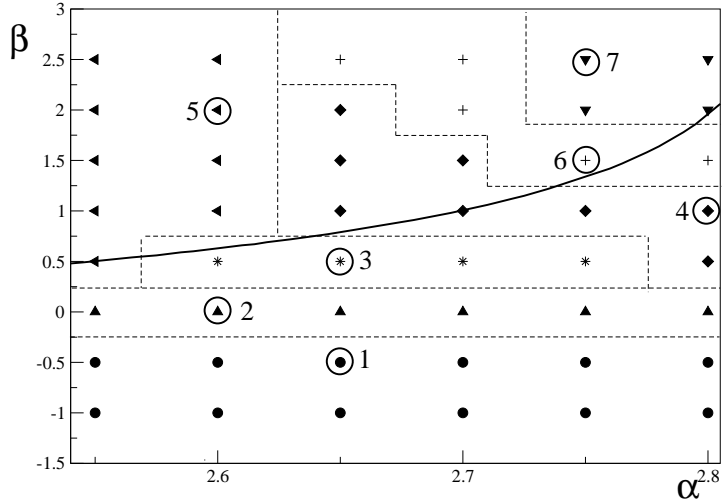}}
\caption{
Schematic phase diagram. Numerical simulations of the
one-dimensional model were performed at the values of $\alpha $ and 
$\beta $ indicated by symbols in this diagram. 
Depending on the observed properties of patterns, the diagram is 
divided into regions 1 to 6. The circles indicate the values of these 
parameters used to produce a typical patterns
for the corresponding region, displayed in Fig. \ref{patterns}. 
The solid line shows the stability boundary
of rapid uniform oscillations, given by the condition $C_2=0$. Other
parameters are $\omega =2,K=0.5,l=10,$ and $\tau =10$.}
\label{phasediagram}
\end{center}
\end{figure}

Figure \ref{phasediagram} is a schematic summary of these 
numerical investigations. The solid
line is the stability boundary of rapid oscillations ($C_2=0$); rapid
oscillations are unstable below this boundary. Slow oscillations are 
always stable in the considered region. Simulations have been performed
for a set of parameter values indicated by symbols in this diagram.

\begin{figure}
\begin{tabular}{cccc}
& Re$(\eta)$ & $|\eta|$ & $|z|$\\
\begin{minipage}[b]{1ex}
\textbf{1}\\\mbox{ }\\\mbox{ }
\end{minipage}
&{\includegraphics*[width=2.5cm,height=4cm,angle=-90,origin=rB]
{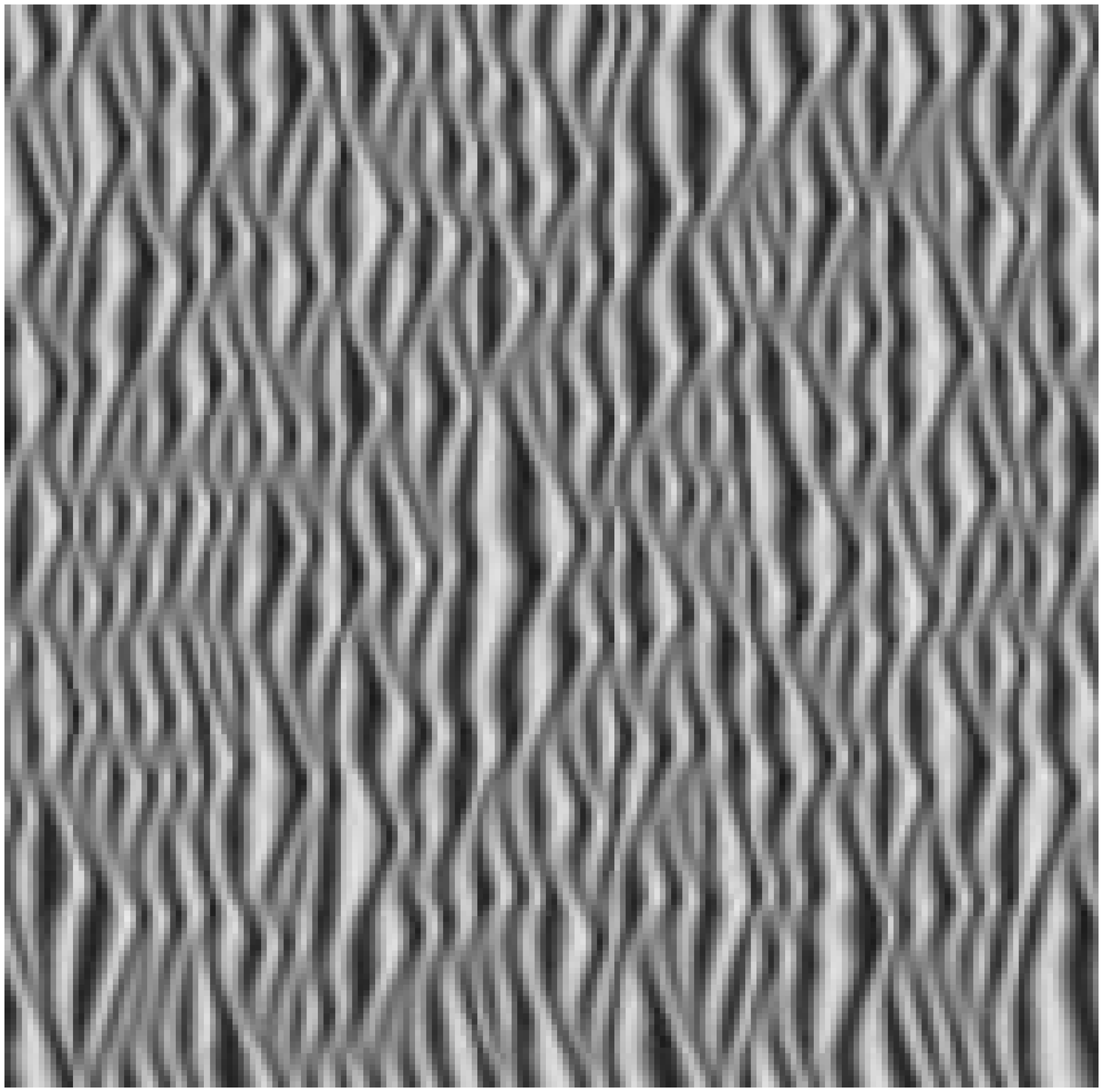}} 
& {\includegraphics*[width=2.5cm,height=4cm,angle=-90,origin=rB]
{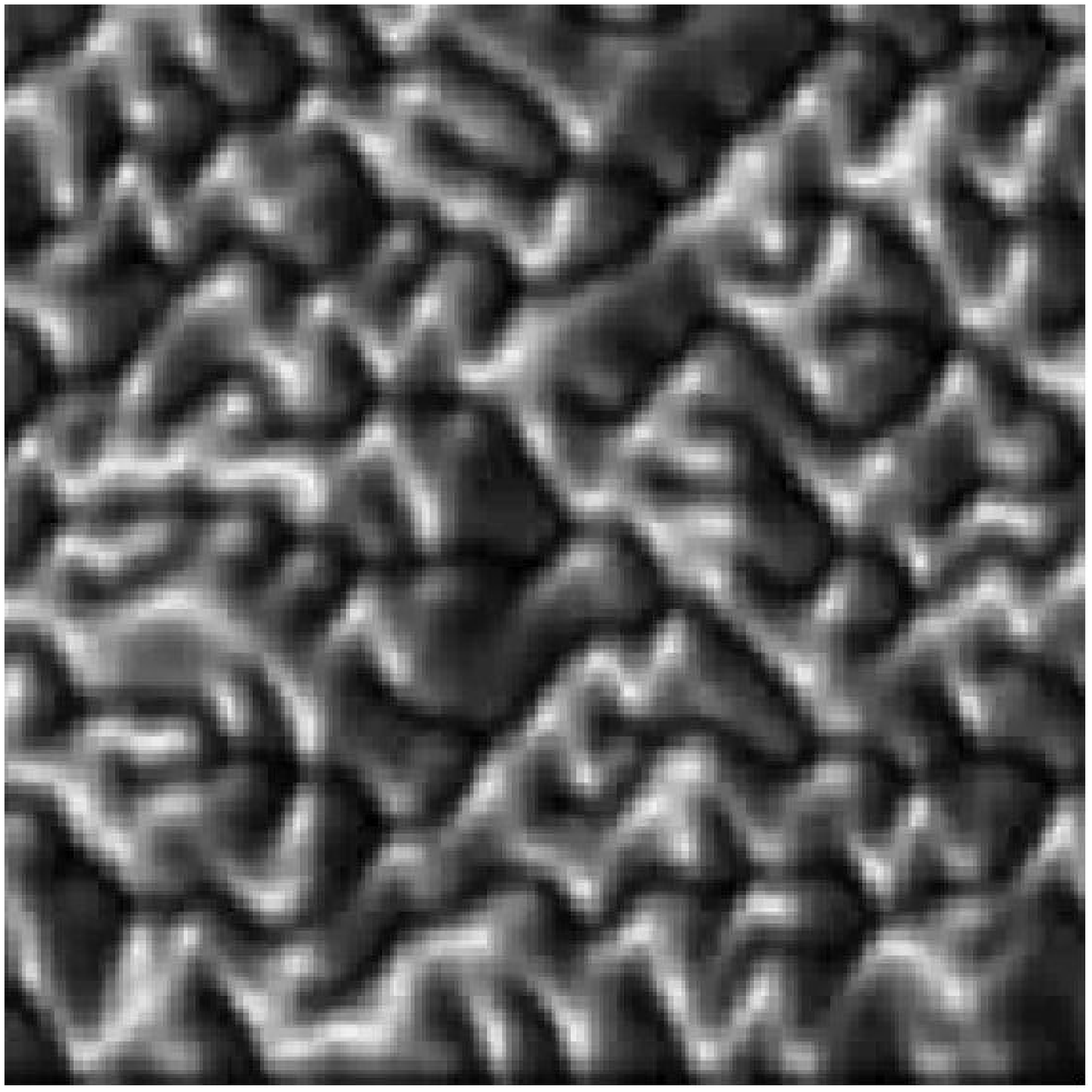}} 
& {\includegraphics*[width=2.5cm,height=4cm,angle=-90,origin=rB]
{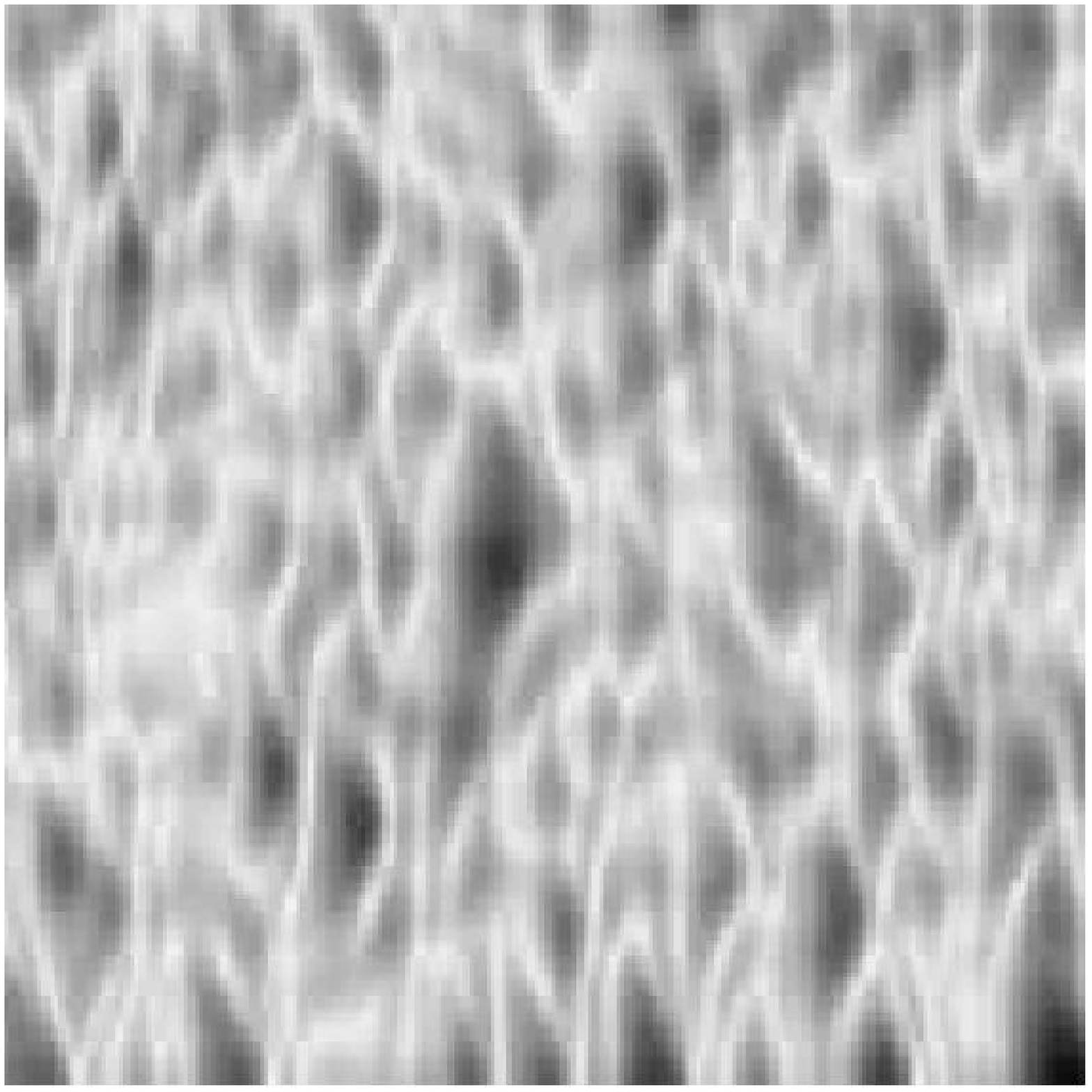}}\\
\begin{minipage}[b]{1ex}
\textbf{2}\\\mbox{ }\\\mbox{ }
\end{minipage}
& {\includegraphics*[width=2.5cm,height=4cm,angle=-90,origin=rB]
{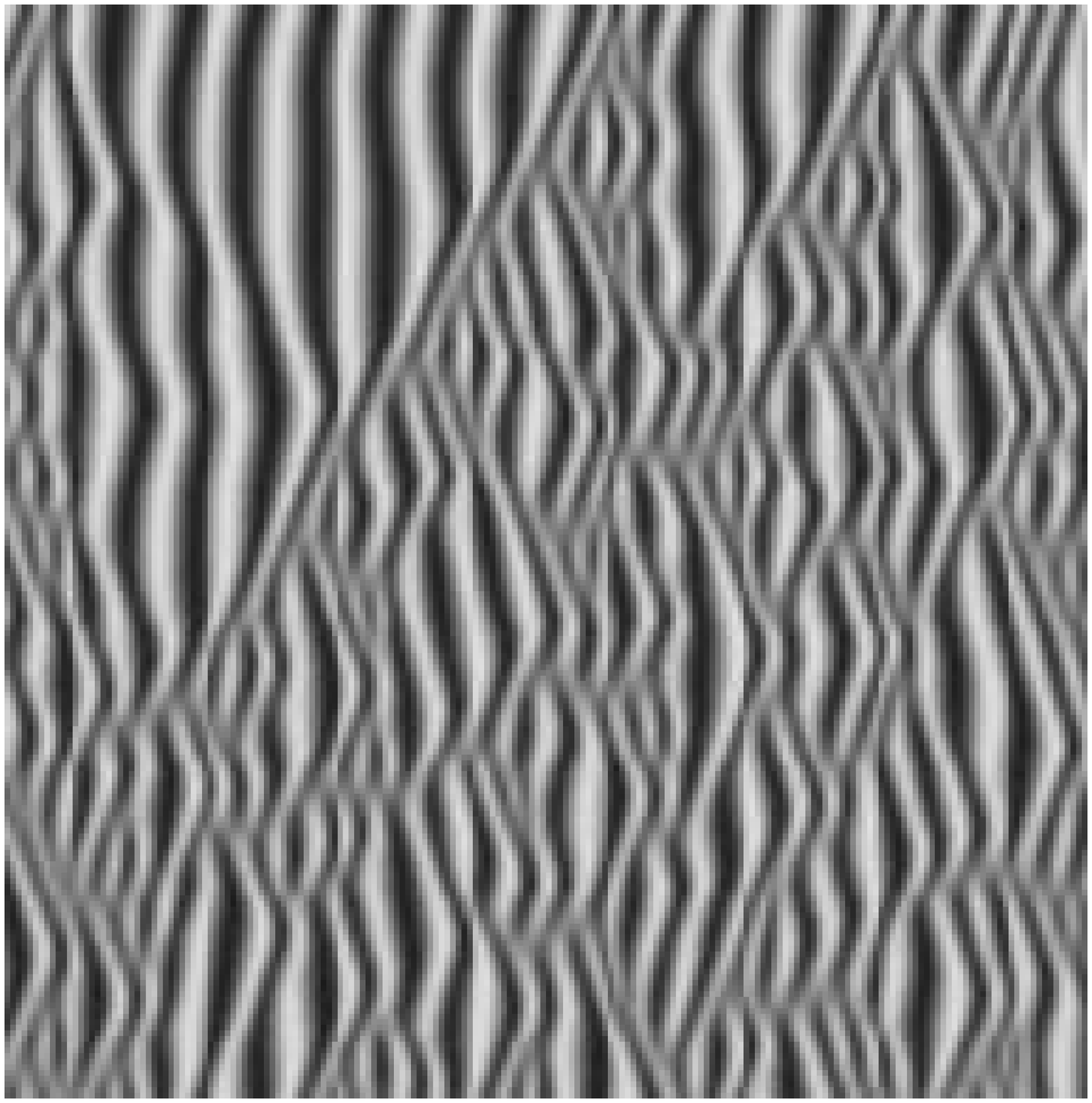}} 
& {\includegraphics*[width=2.5cm,height=4cm,angle=-90,origin=rB]
{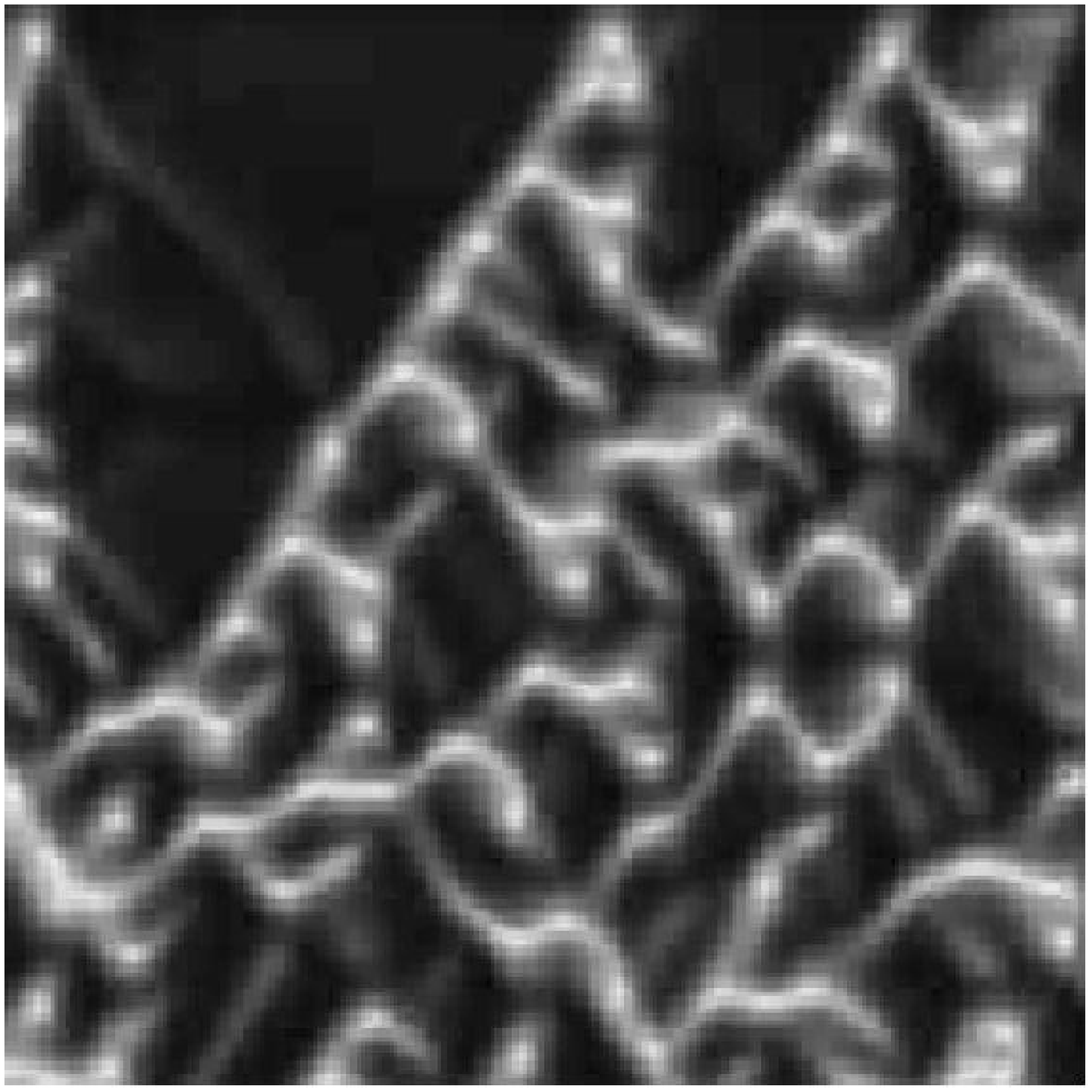}} 
& {\includegraphics*[width=2.5cm,height=4cm,angle=-90,origin=rB]
{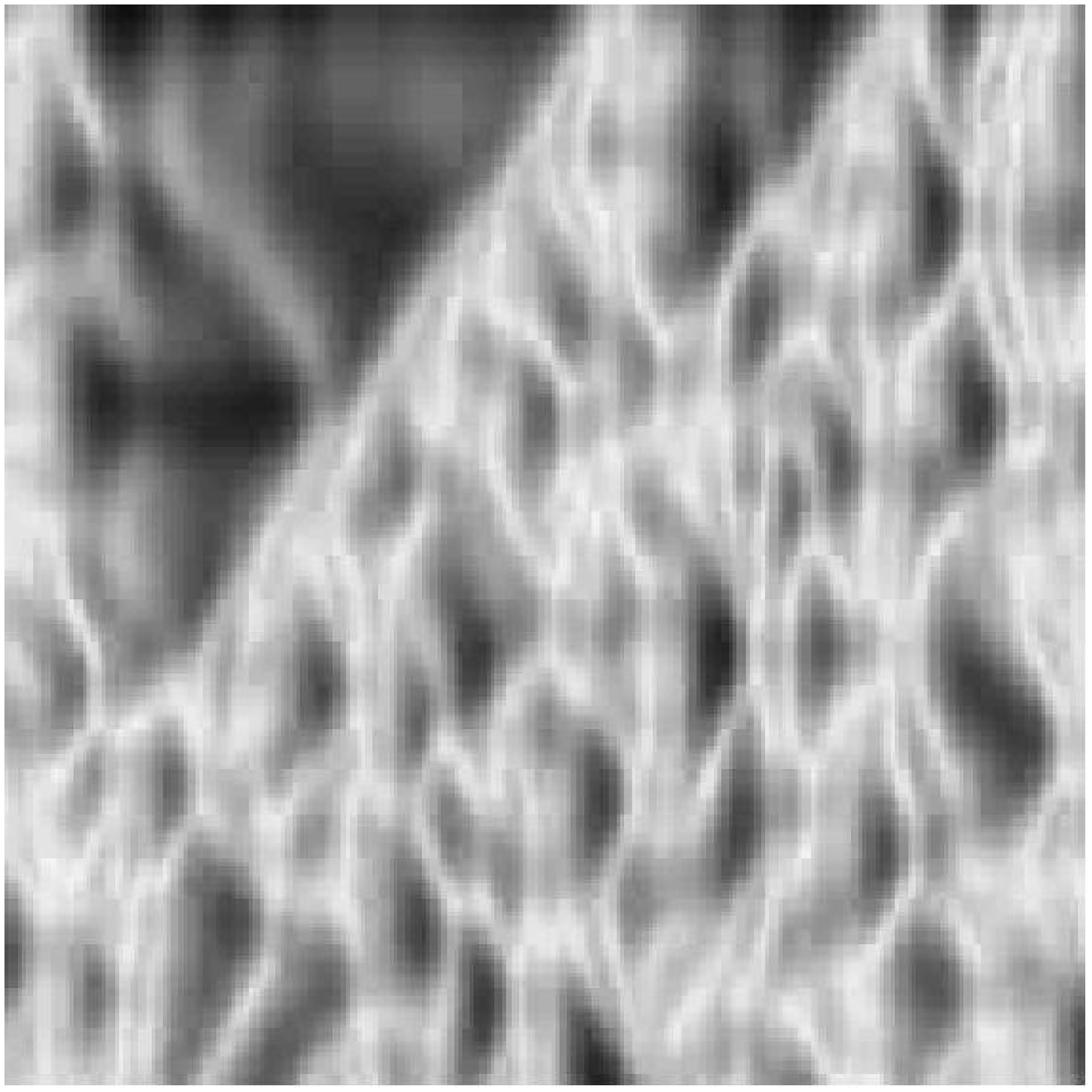}}\\
\begin{minipage}[b]{1ex}
\textbf{3}\\\mbox{ }\\\mbox{ }
\end{minipage}
& {\includegraphics*[width=2.5cm,height=4cm,angle=-90,origin=rB]
{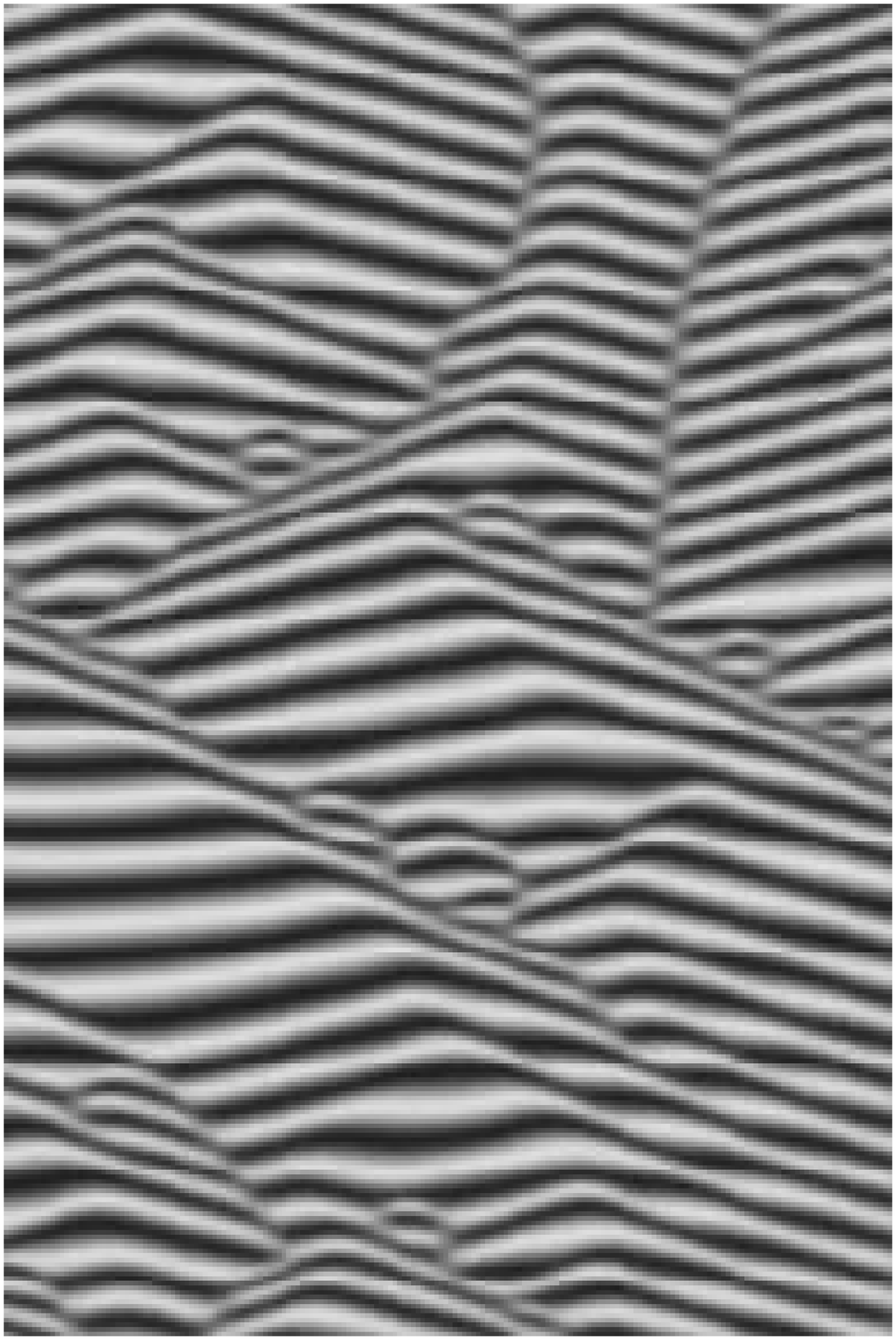}} 
& {\includegraphics*[width=2.5cm,height=4cm,angle=-90,origin=rB]
{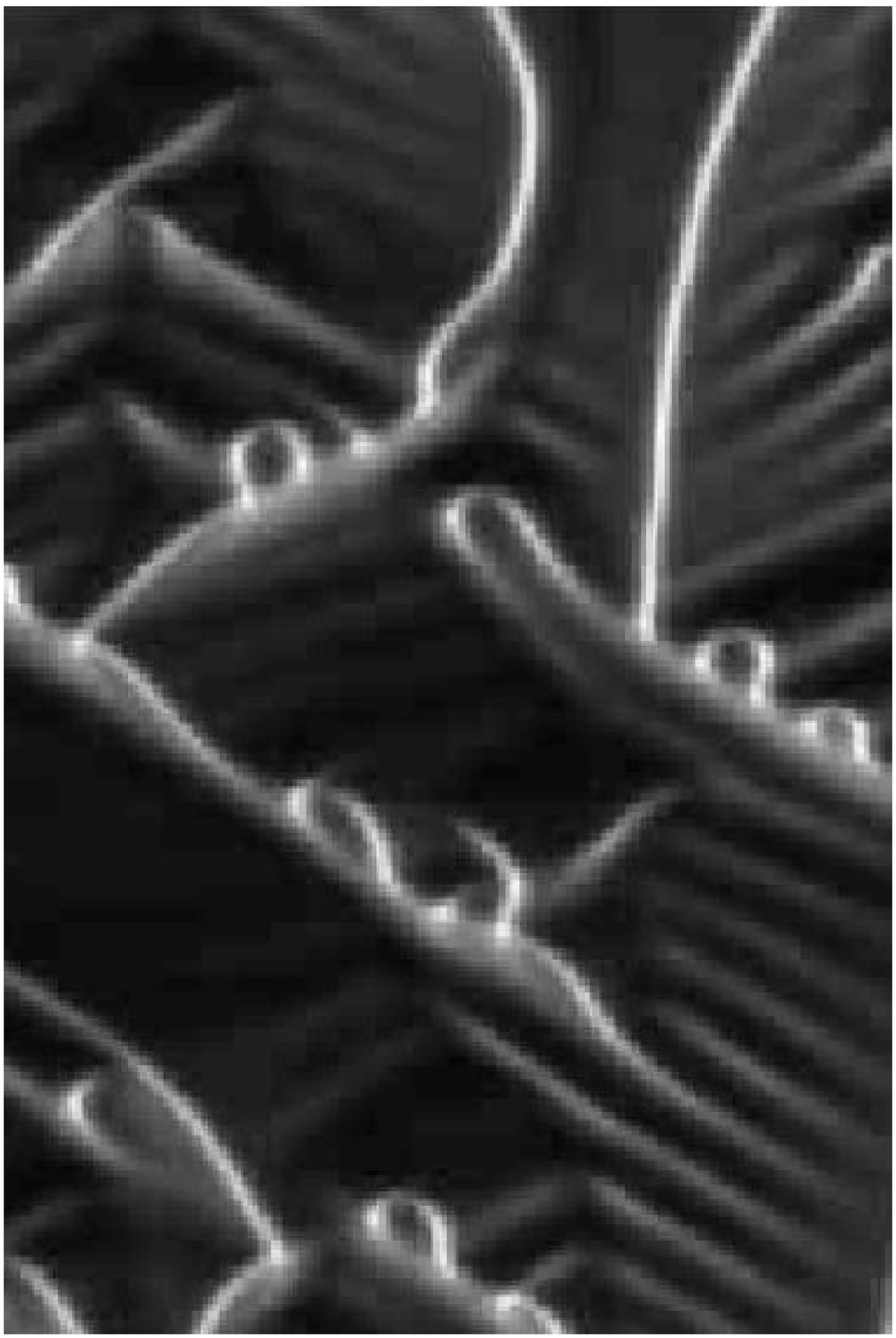}} 
& {\includegraphics*[width=2.5cm,height=4cm,angle=-90,origin=rB]
{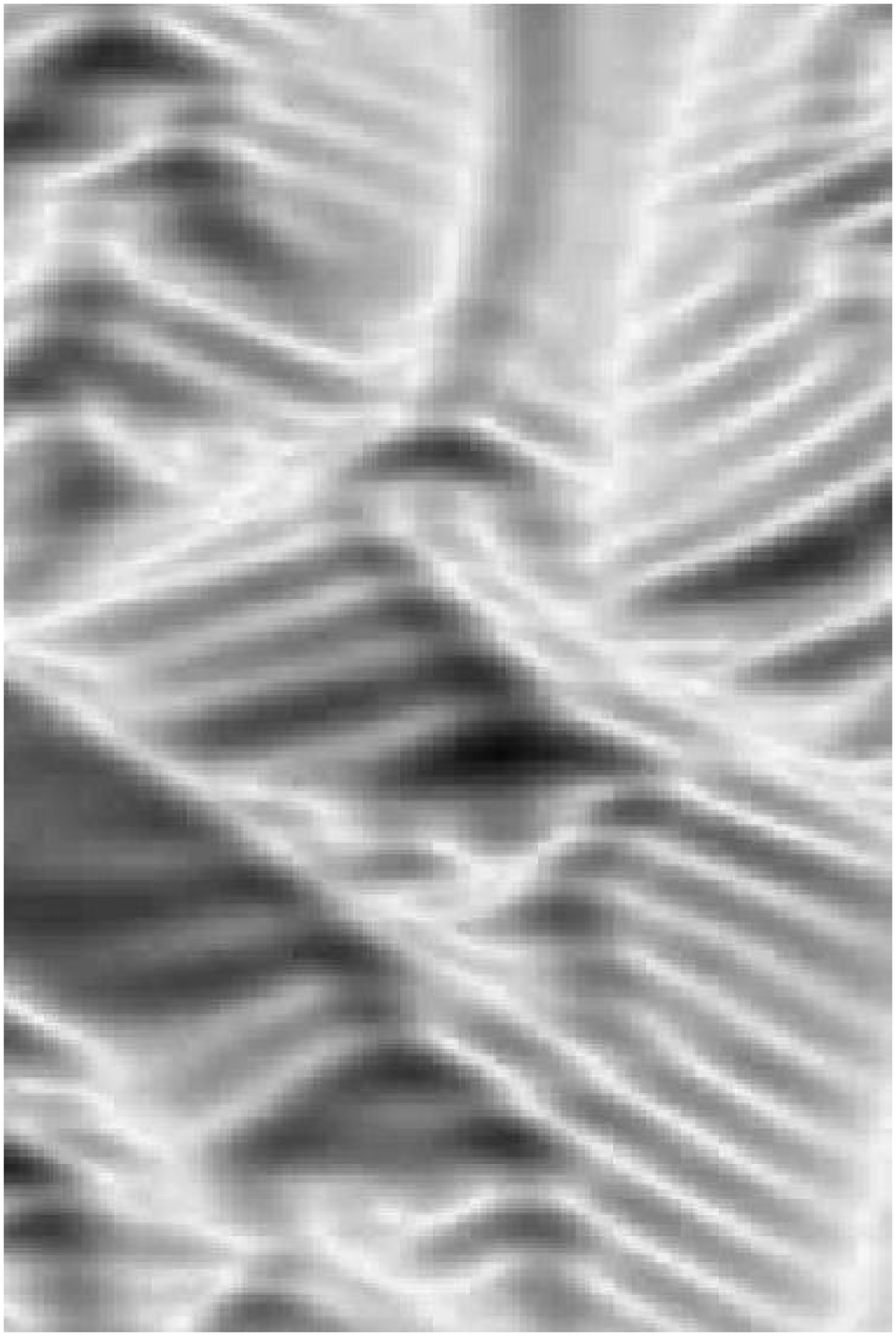}}\\
\begin{minipage}[b]{1ex}
\textbf{4}\\\mbox{ }\\\mbox{ }
\end{minipage}
& {\includegraphics*[width=2.5cm,height=4cm,angle=-90,origin=rB]
{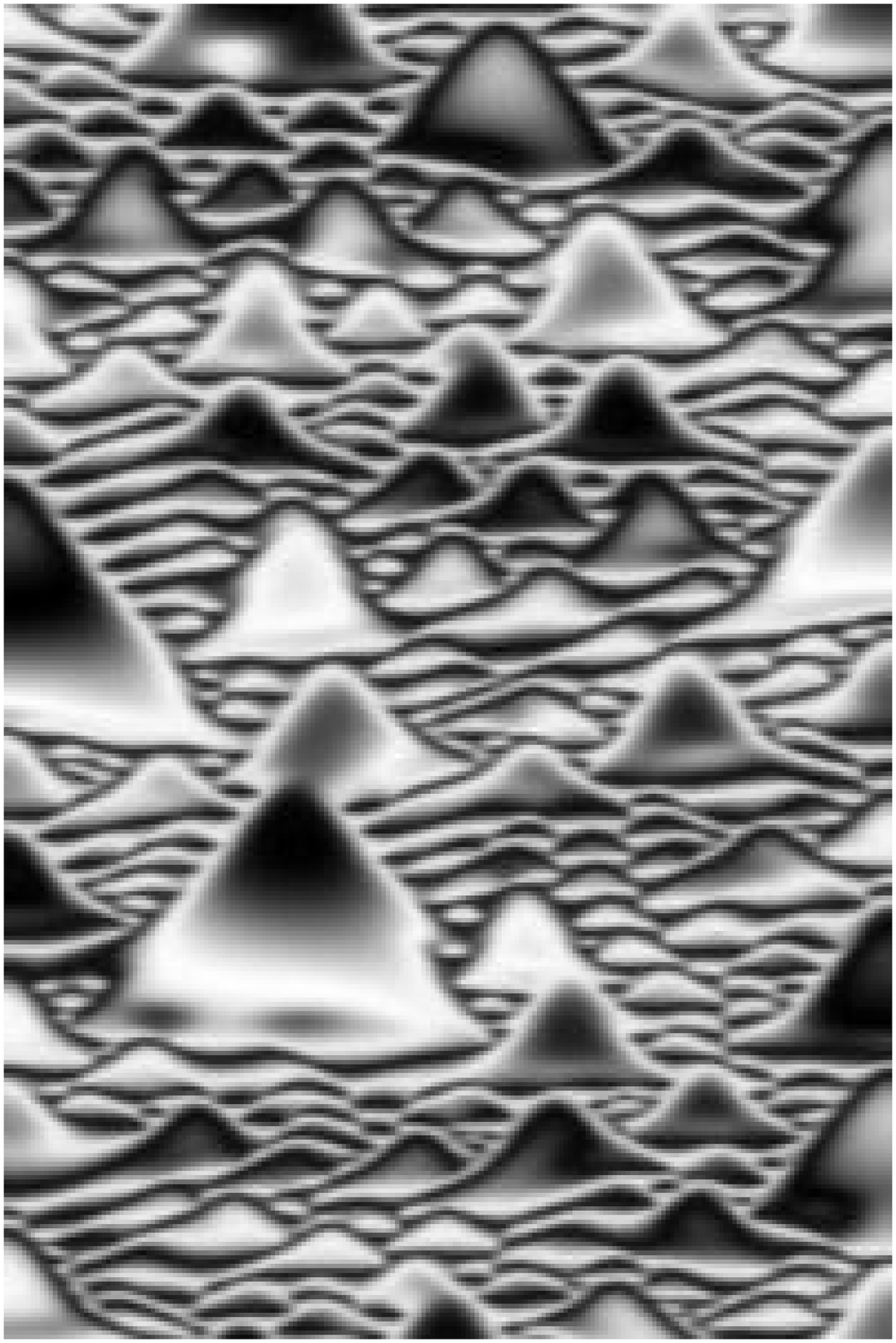}} 
& {\includegraphics*[width=2.5cm,height=4cm,angle=-90,origin=rB]
{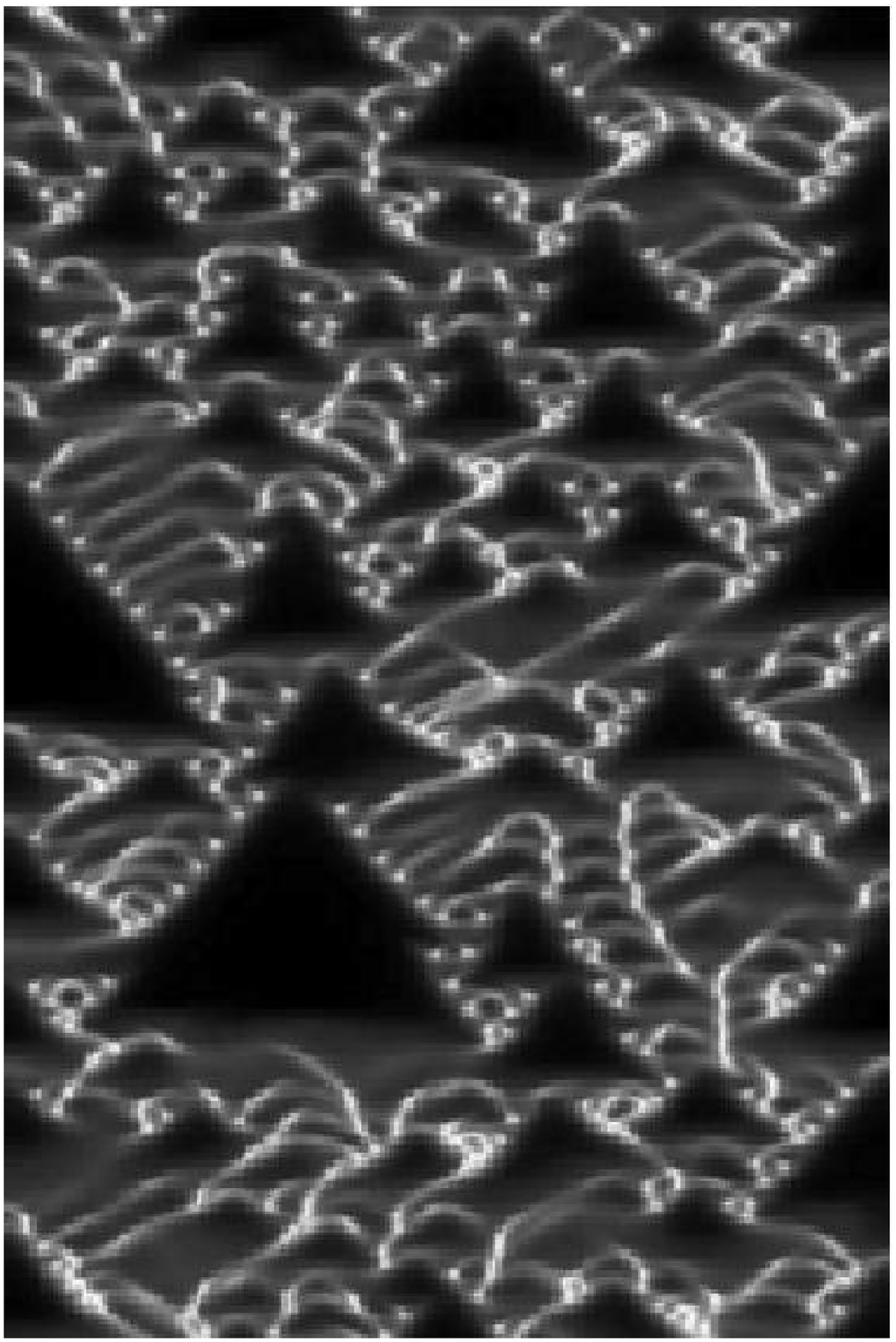}} 
& {\includegraphics*[width=2.5cm,height=4cm,angle=-90,origin=rB]
{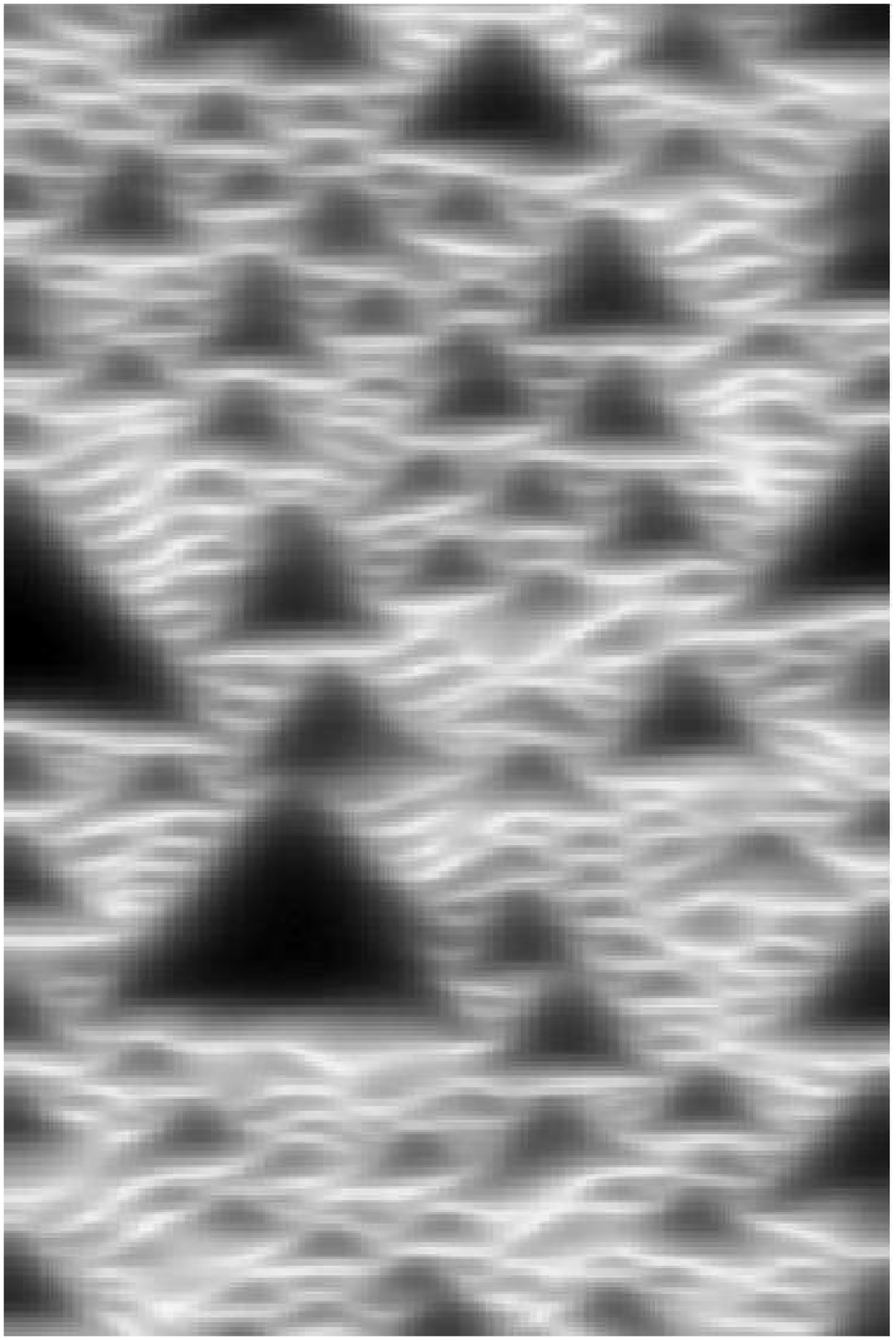}}\\
\begin{minipage}[b]{1ex}
\textbf{5}\\\mbox{ }\\\mbox{ }
\end{minipage}
& {\includegraphics*[width=2.5cm,height=4cm,angle=-90,origin=rB]
{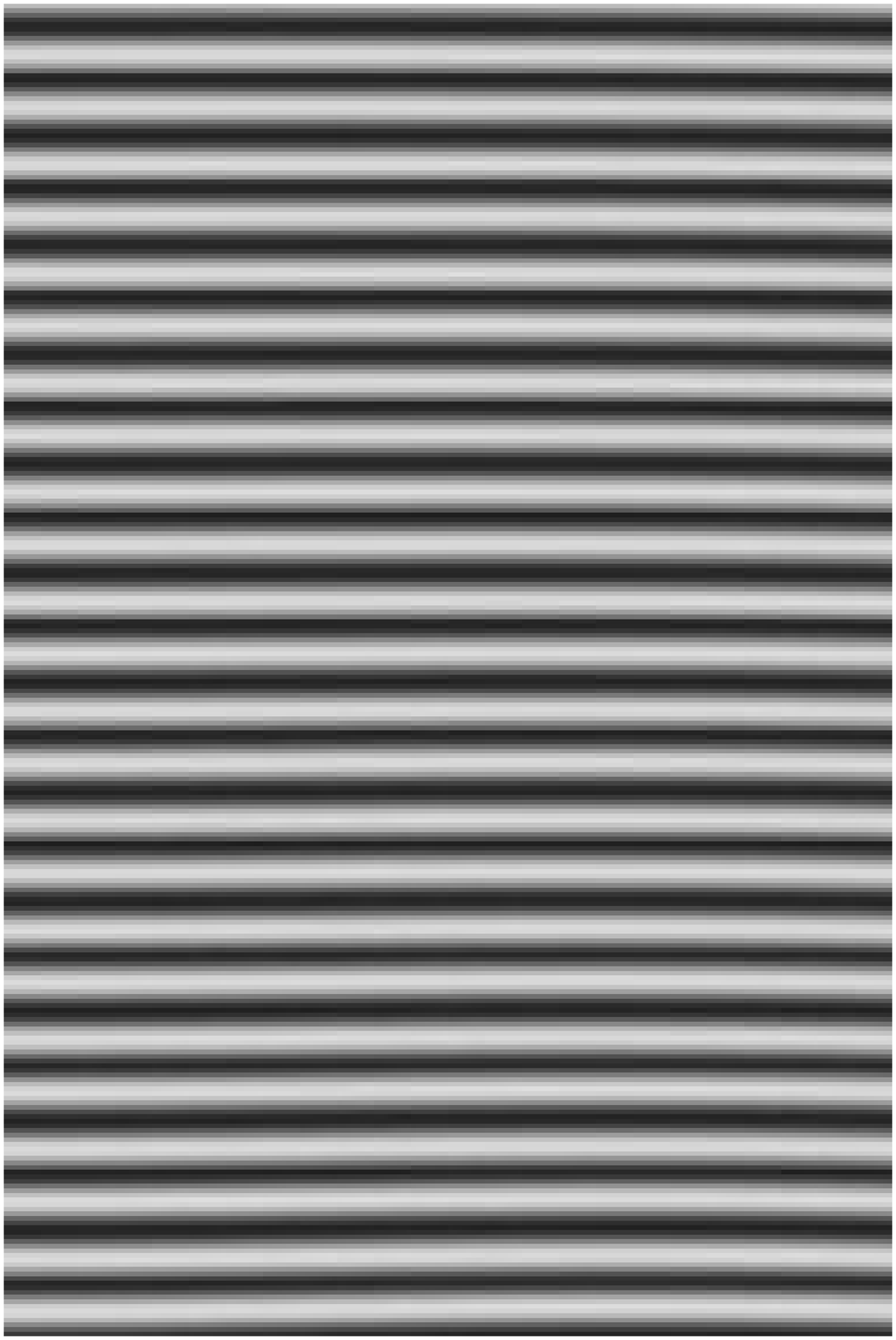}} 
& {\includegraphics*[width=2.5cm,height=4cm,angle=-90,origin=rB]
{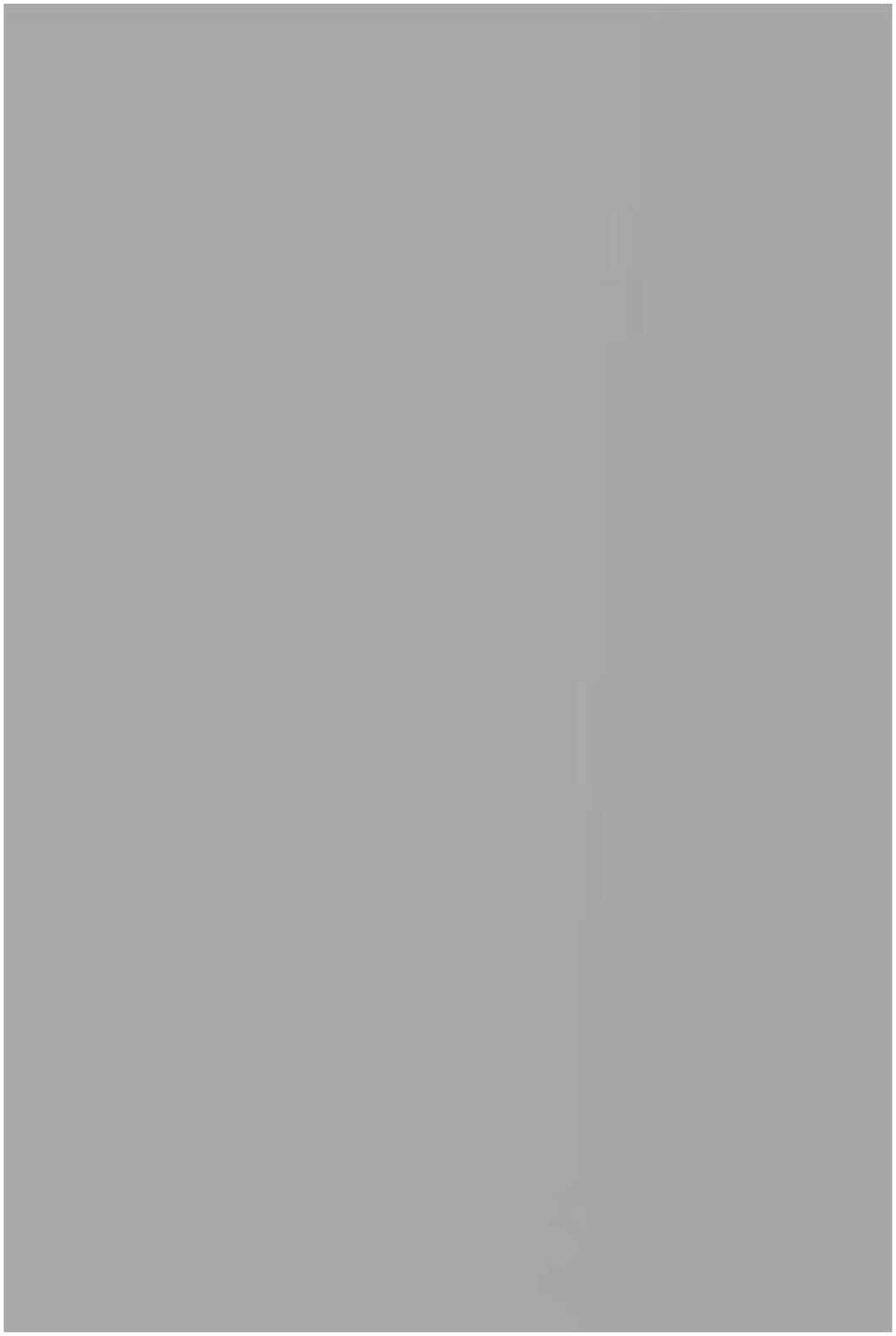}} 
& {\includegraphics*[width=2.5cm,height=4cm,angle=-90,origin=rB]
{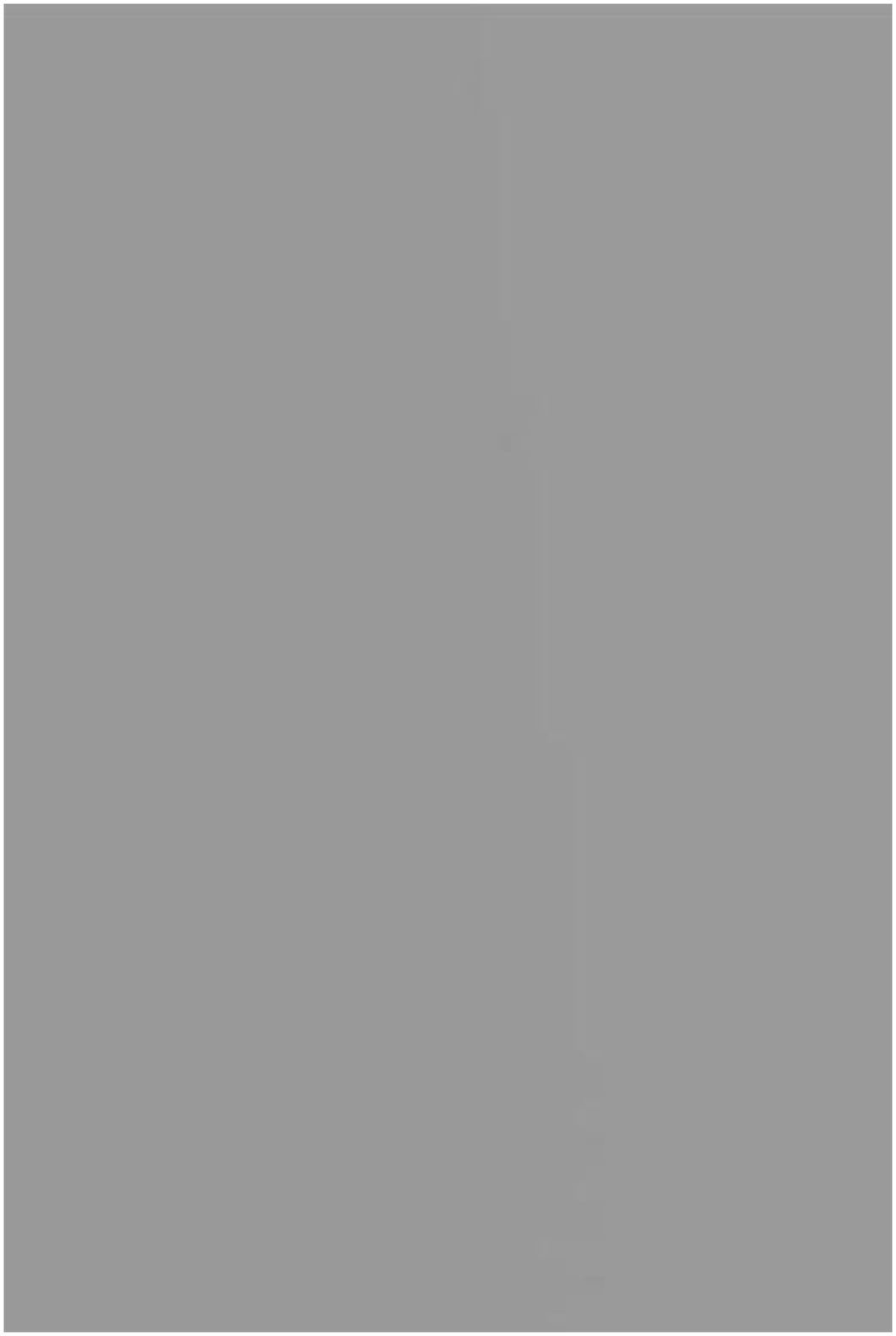}} \\
\begin{minipage}[b]{1ex}
\textbf{6}\\\mbox{ }\\\mbox{ }
\end{minipage}
& {\includegraphics*[width=2.5cm,height=4cm,angle=-90,origin=rB]
{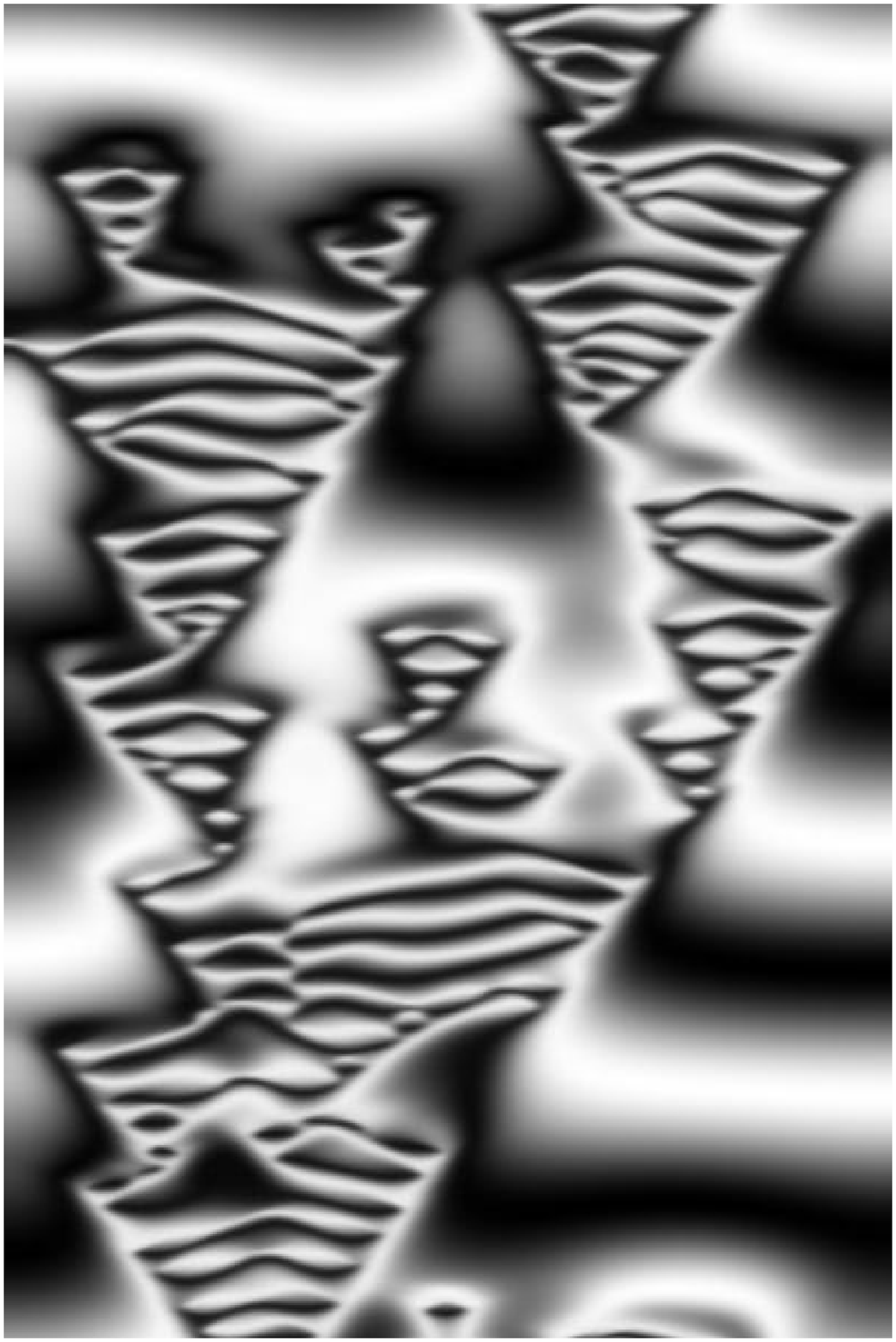}} 
& {\includegraphics*[width=2.5cm,height=4cm,angle=-90,origin=rB]
{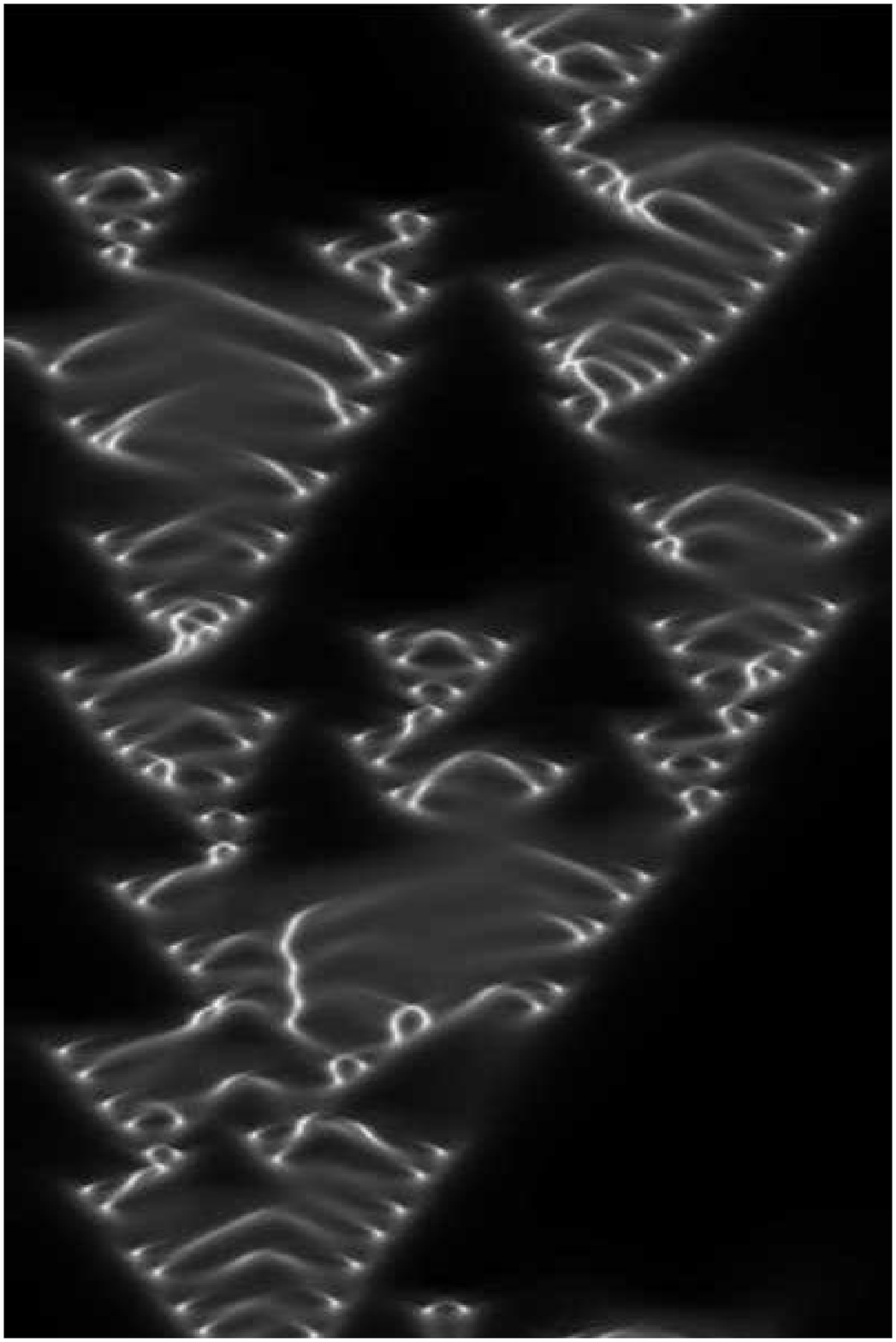}} 
& {\includegraphics*[width=2.5cm,height=4cm,angle=-90,origin=rB]
{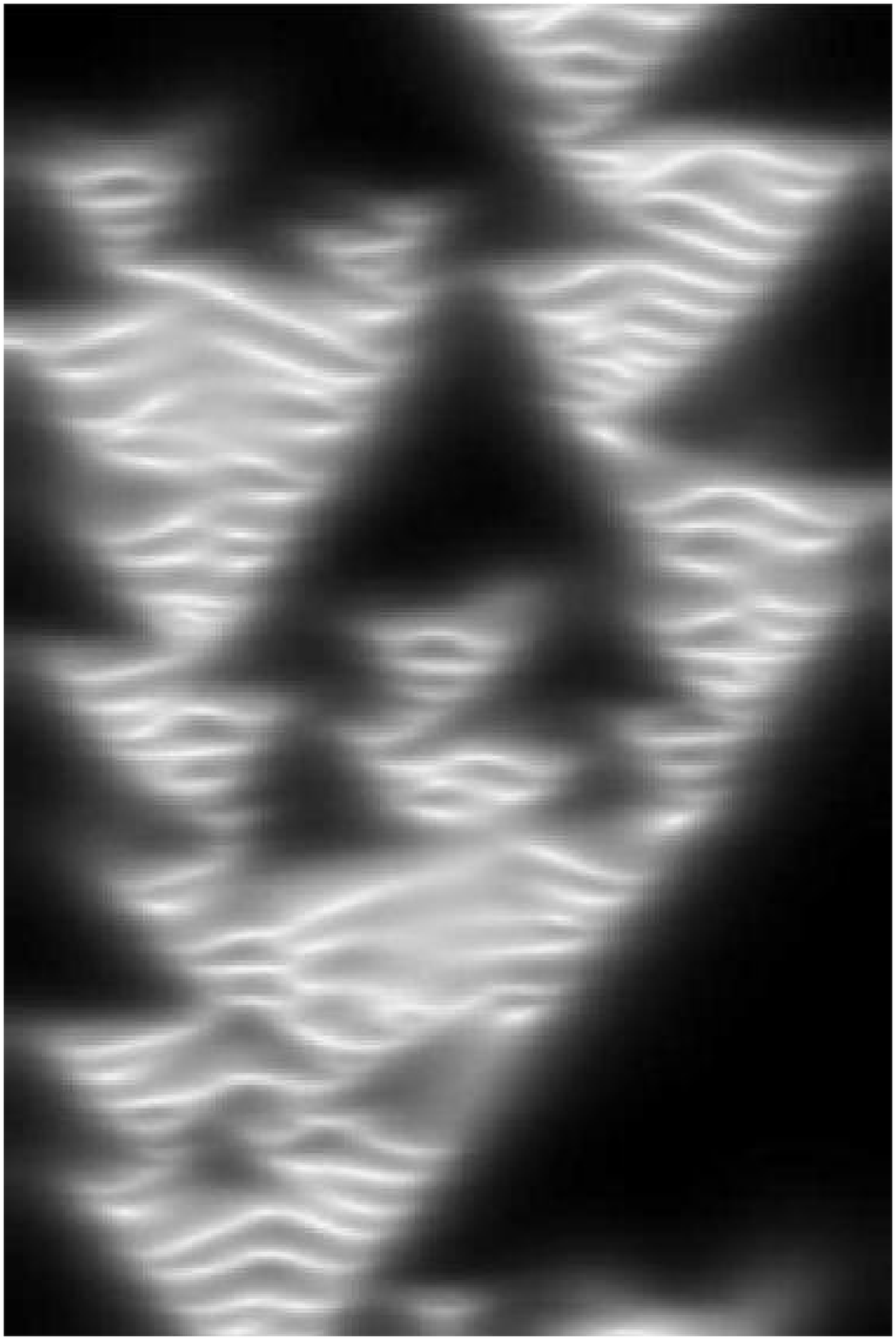}}\\
\begin{minipage}[b]{1ex}
\textbf{7}\\\mbox{ }\\\mbox{ }
\end{minipage}
& {\includegraphics*[width=2.5cm,height=4cm,angle=-90,origin=rB]
{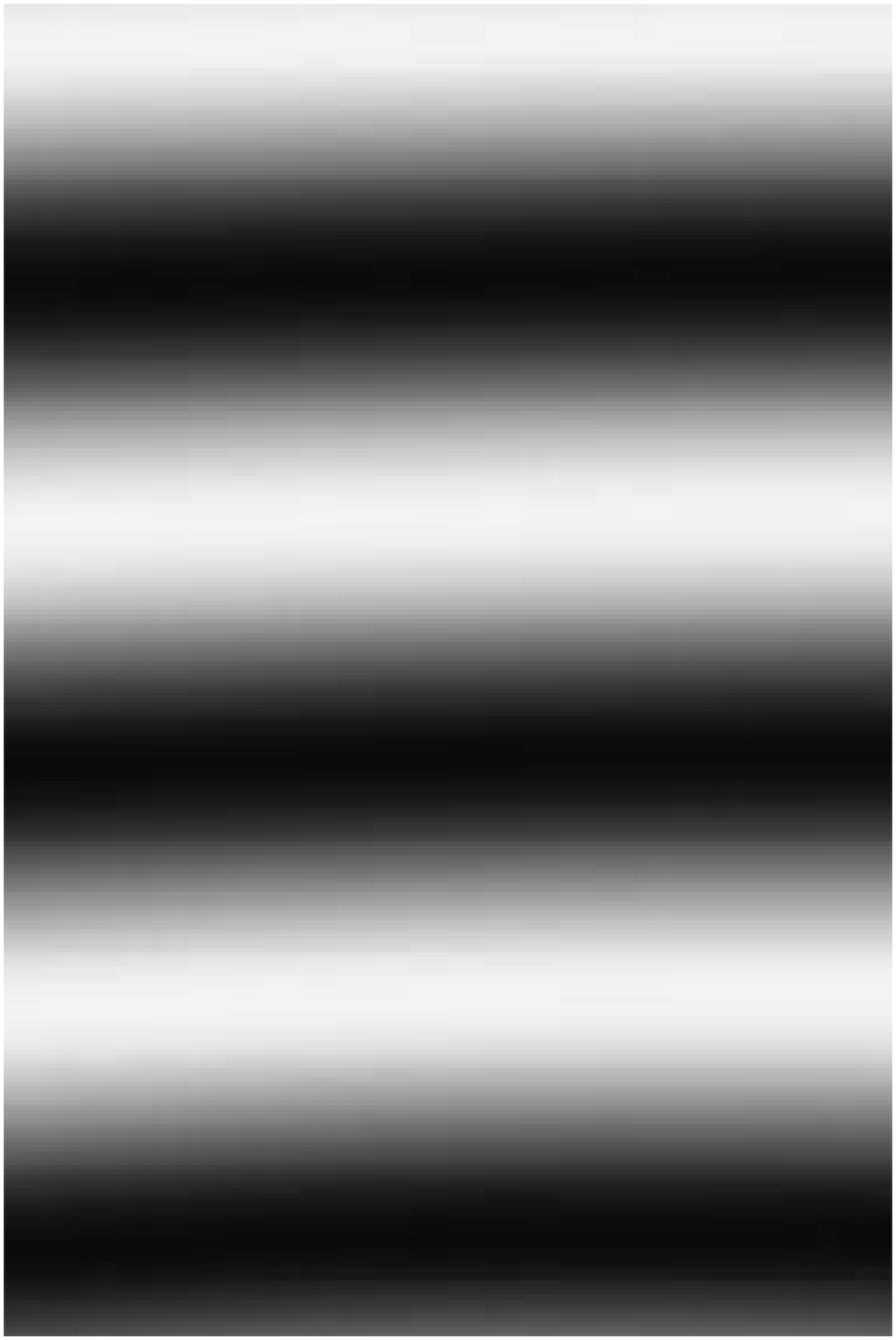}} 
& {\includegraphics*[width=2.5cm,height=4cm,angle=-90,origin=rB]
{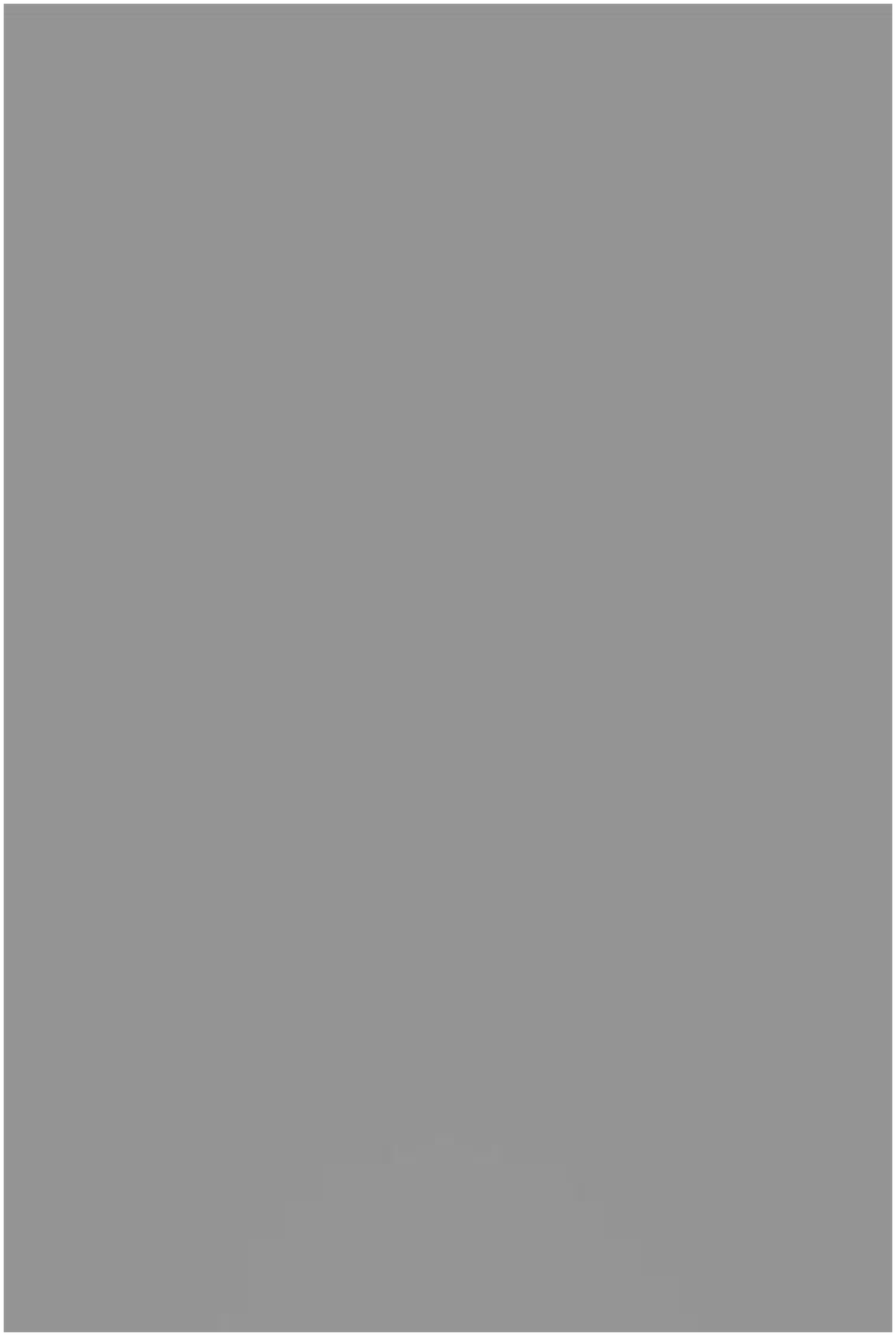}} 
& {\includegraphics*[width=2.5cm,height=4cm,angle=-90,origin=rB]
{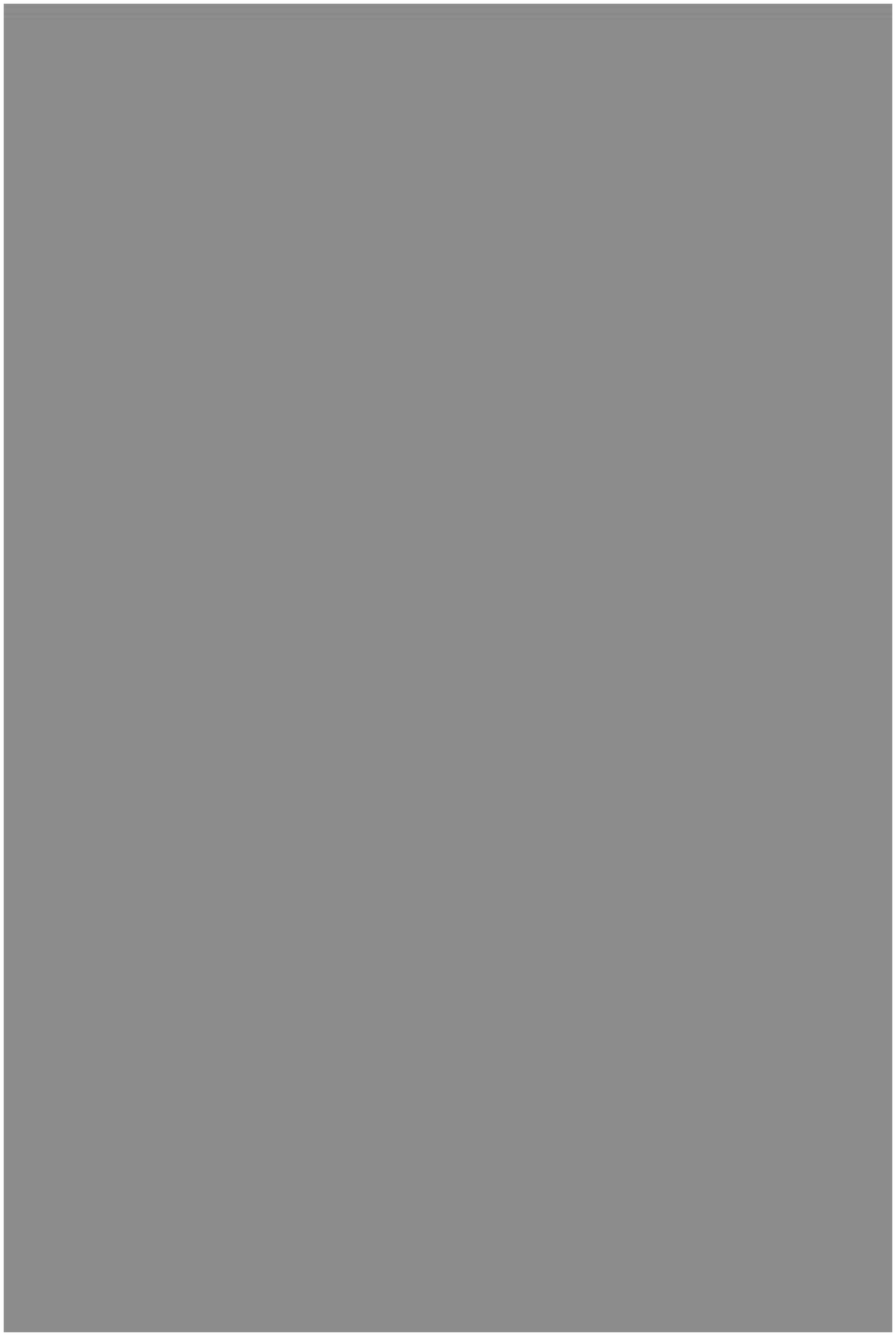}} \\
\end{tabular}
\caption{Spatiotemporal diagrams displaying evolution of 
Re($\eta$), $\rho=|\eta| $ and $r=|z| $ in typical patterns 
observed in the regions 1-6 for the one-dimensional system. 
The respective parameter values are given in Fig. \ref{phasediagram}. 
The displayed space and time intervals are (1) 
$L=100,$ $T=166,$ (2) $L=100,$ $T=166,$ (3) $L=100,$ $T=250,$ 
(4) $L=200,$ $T=500,$ (5) $L=100,$ $T=250,$ (6) $L=200,$ $T=500,$ 
and (7) $L=200,$ $T=250.$
The contrast level is adjusted individually in each plot to ensure best
visualization of the pattern.}
\label{patterns}
\end{figure}

Comparing properties of observed patterns, we can qualitatively divide 
them into several groups occupying different regions in the parameter 
plane. Each group is marked by its own symbol and the respective 
regions in Fig. \ref{phasediagram} are
numbered. Examples of typical spatiotemporal patterns, observed in
one-dimensional simulations within each region, are displayed in 
Fig. \ref{patterns}.
These simulations were all performed for a system of the total length 
$L=200$ and for a total time varying between 500 to 3000 time units. 
To clearly present the observed patterns, only parts of their entire 
evolution are however displayed.

In region 1, the slow limit cycle is Benjamin-Feir stable, while the 
rapid limit cycle is unstable. The system is found here in the regime 
of fully developed amplitude turbulence characterized by creation of 
multiple amplitude defects. Increasing parameter $\beta$, we enter a 
region 2 closer to the stability boundary where turbulence becomes 
intermittent. We observed the emergence of larger groups of 
synchronized oscillators which become able to reach the rapid limit 
cycle and perform harmonical oscillations for several periods. 
Inside the small parameter region 3 near the stability boundary,
the frequency and the amplitude of oscillations correspond almost 
everywhere to those of the rapid limit cycle. 
However, the system does not undergo
complete synchronization. Lines of amplitude defects travelling 
through the system act as wave sources here.

In regions 5 and 7, both lying above the stability boundary of rapid
oscillations, uniform oscillations are observed starting from random 
initial conditions. 
The oscillations are rapid inside region 5 and slow inside
region 7. These two domains are separated by regions 4 and 6 in the
parameter plane, where competition between two oscillation mode takes 
place leading to complex spatiotemporal regimes.

The patterns inside region 4 are characterized by a background of rapid
chaotic oscillations, with small amplitude and numerous amplitude 
defects.
On this highly desynchronized background, {\em bursts of 
synchronization\/} emerge. Such bursts consist of large groups of 
elements which suddenly start to oscillate together with a large 
amplitude and a small frequency corresponding to the stable slow 
limit cycle. 
However, these groups do not keep synchronized over a long time: after 
less than one oscillation period the turbulence overwhelms again. 
As already noticed in our discussion of
birhythmicity, the amplitude of the coupling field $z$ is very small 
in the rapid limit cycle, while in the slow limit cycle it gets a 
larger value comparable to the amplitude of $\eta $. Analyzing patterns
in region 4, it can be seen that during the synchronization bursts, 
the amplitude $|z|$ is indeed much larger than in the turbulent 
background. Note that such patterns
with synchronization bursts have also been observed for some parameter
values below the stability boundary of rapid uniform oscillations.

Region 6 lies near region 7, where slow uniform oscillations are 
winning the competition. 
Here, the patterns can be described as exhibiting {\em bursts
of desynchronization\/} on the background of slow uniform oscillations.
Inside such bursts, the coupling field $z$ is strongly decreased in 
amplitude and only short-range diffusive coupling between oscillators 
is effective.

\begin{figure}
\begin{center}
\scalebox{0.5}[0.5]{\includegraphics*{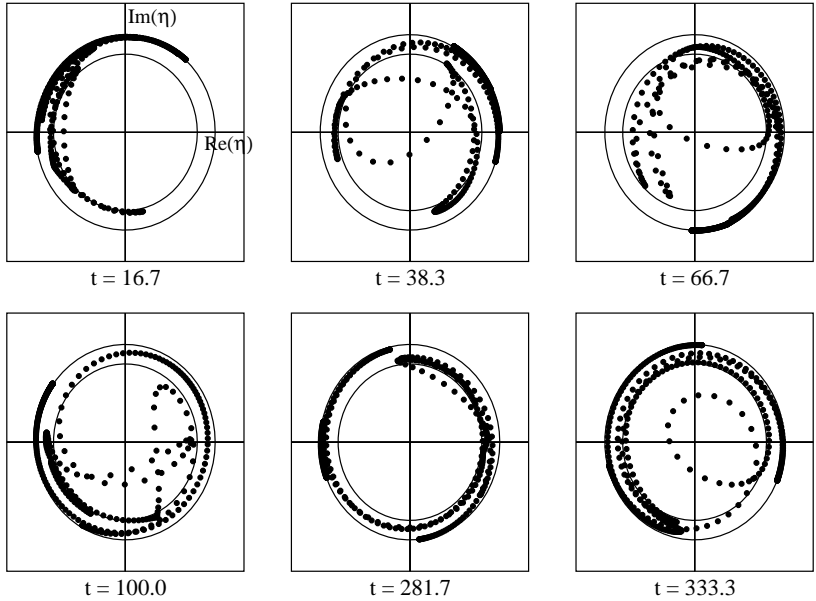}}
\caption{Selected snapshots of the phase portraits for the pattern 6 in
Fig. \ref{patterns}. 
The circles show two different limits cycles in the birhythmic system.}
\label{phaseportr_snap}
\end{center}
\end{figure}

To further illustrate this regime, phase portraits of the patterns in 
region 6 were constructed at different time moments. Fig. 
\ref{phaseportr_snap} shows several subsequent snapshots from the 
recorded video. 
Each point in a phase portrait represents a state of one of the 
oscillators. The spatial information (i.e., spatial distances between 
the displayed oscillators) is lost in this
representation, but the motions performed by the oscillator population 
are seen more clearly. The two circles indicate the two coexisting 
limit cycles.
The cycle with a smaller oscillation amplitude is rapid, whereas the 
cycle with a larger amplitude is slow.

We can see from these snapshots that the system is usually divided into
two groups of oscillators, occupying the two coexisting attractors 
(limit cycles). Some oscillators are found in the vicinity of the 
origin $\eta =0$.
They correspond to amplitude defects generated at the border between 
spatial regions with different oscillation frequencies (bright dots in 
the spatiotemporal plot of $|\eta|$ for pattern 6 in Fig. 
\ref{patterns}). Such defects are not present at all times, and the 
respective points are not, for example, seen in the phase portraits at 
$t=16.7$ and $t=281.7$. 
The oscillators sitting on the rapid limit cycle with the smaller 
amplitude belong to the desynchronization bursts.

\begin{figure}
\begin{center}
\begin{tabular}{ccc}
\begin{minipage}[b]{2em}
Re$(\eta)$\\\mbox{ }\\\mbox{ }
\end{minipage}
&{\includegraphics[width=2.5cm]{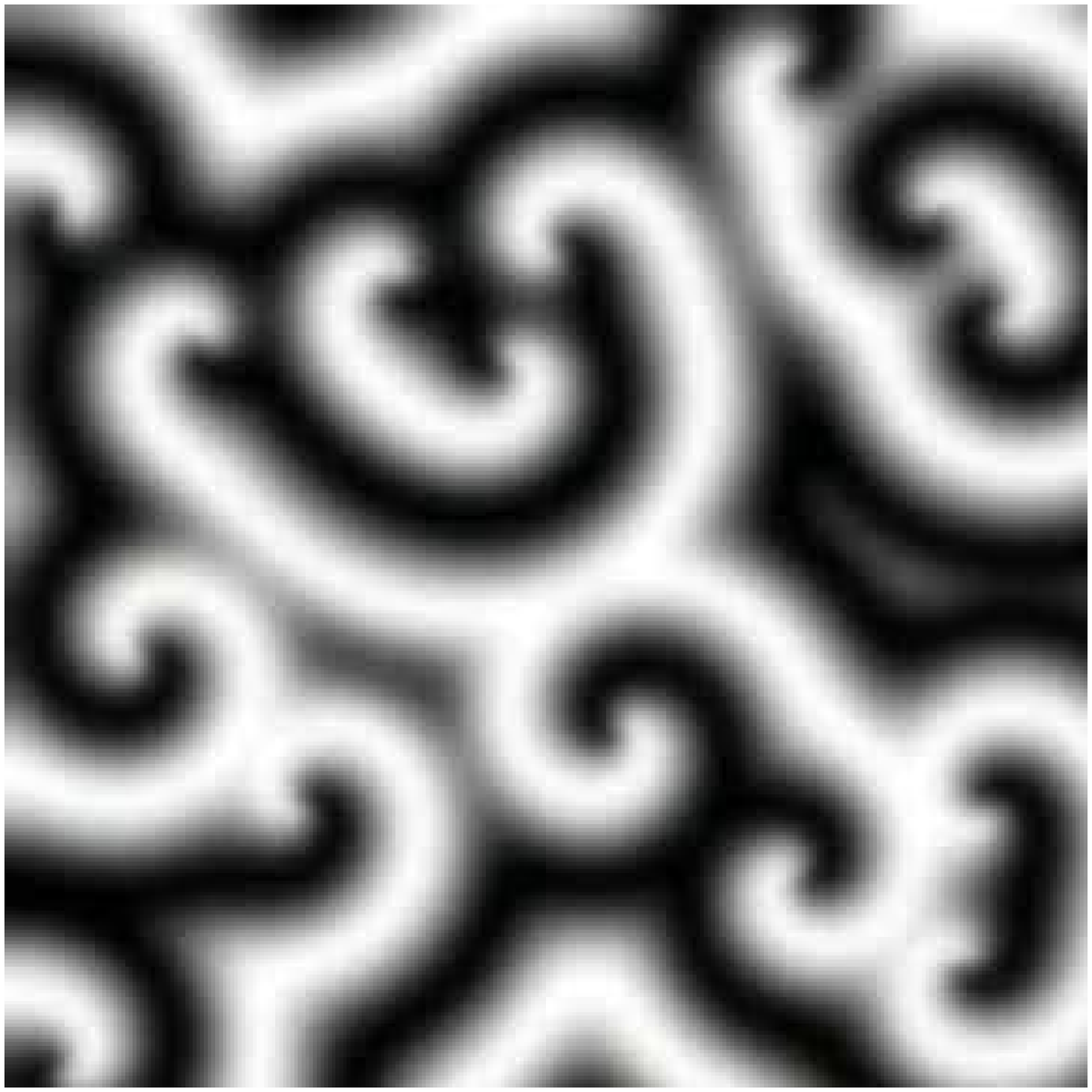}}
& {\includegraphics[width=2.5cm,height=6cm,angle=-90,origin=rB]
{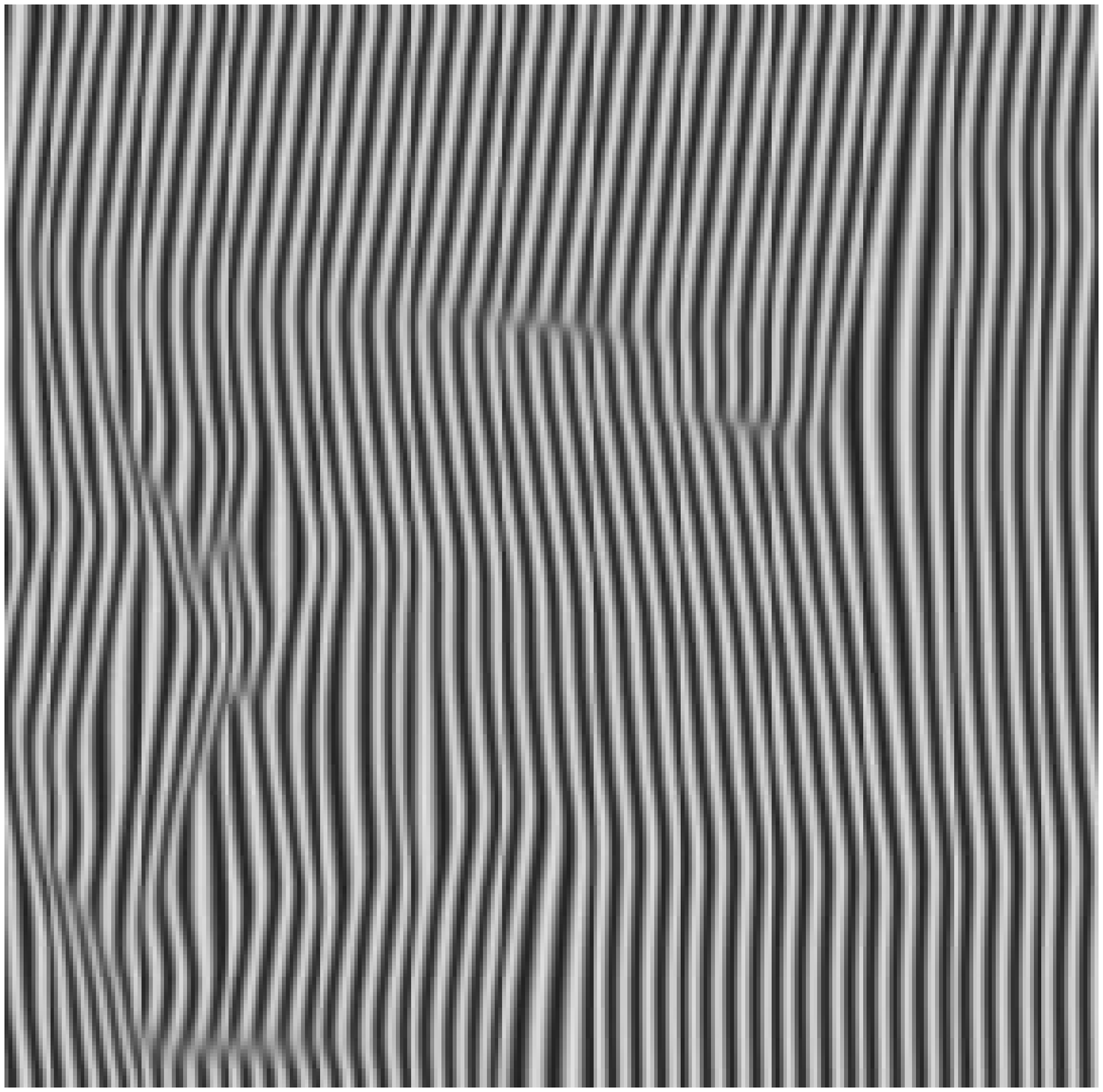}}\\
\begin{minipage}[b]{2em}
$|\eta|$\\\mbox{ }\\\mbox{ }
\end{minipage}
&{\includegraphics[width=2.5cm]{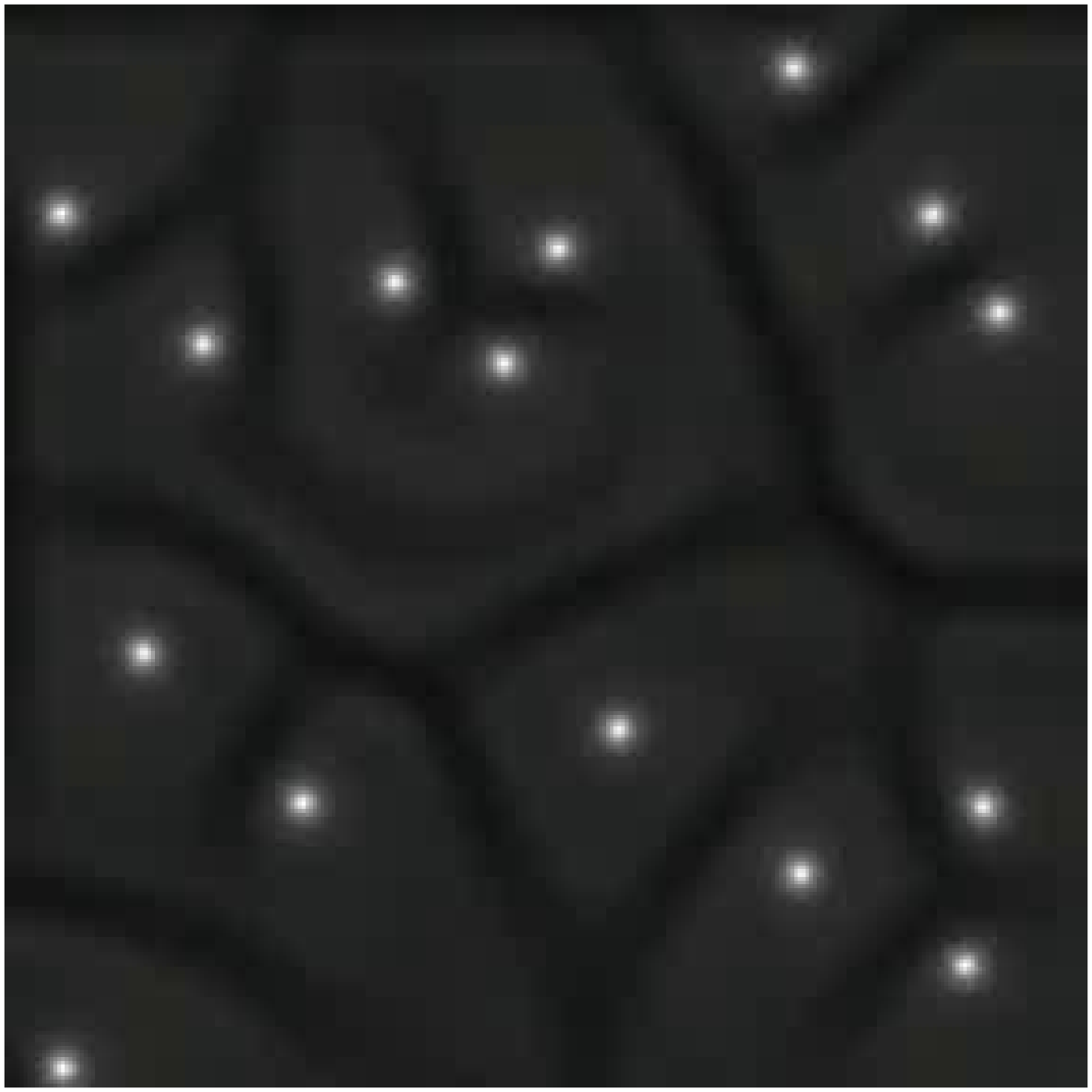}}
& {\includegraphics[width=2.5cm,height=6cm,angle=-90,origin=rB]
{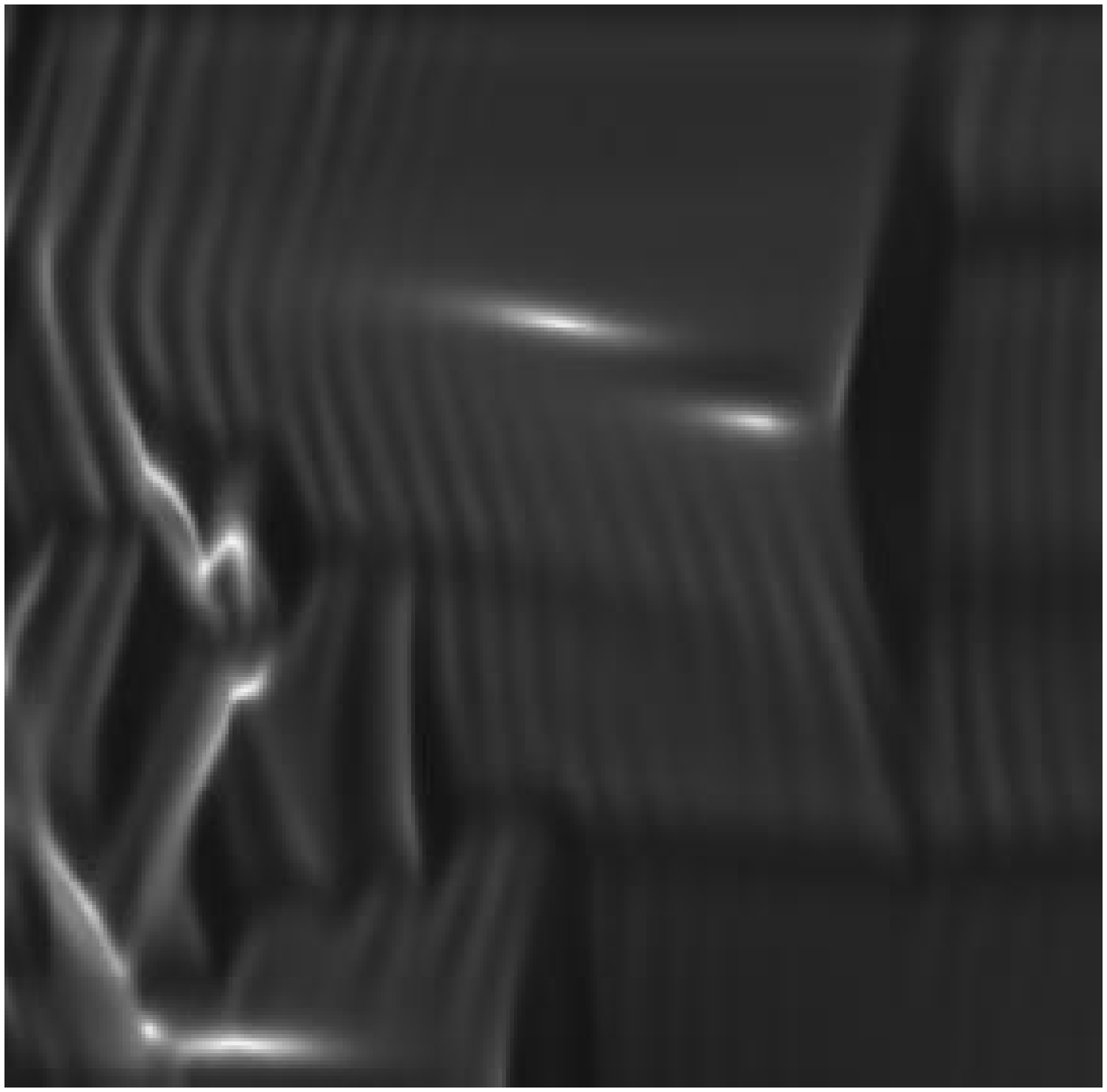}}\\
\begin{minipage}[b]{2em}
$|z|$ \\\mbox{ }\\\mbox{ }
\end{minipage}
&{\includegraphics[width=2.5cm]{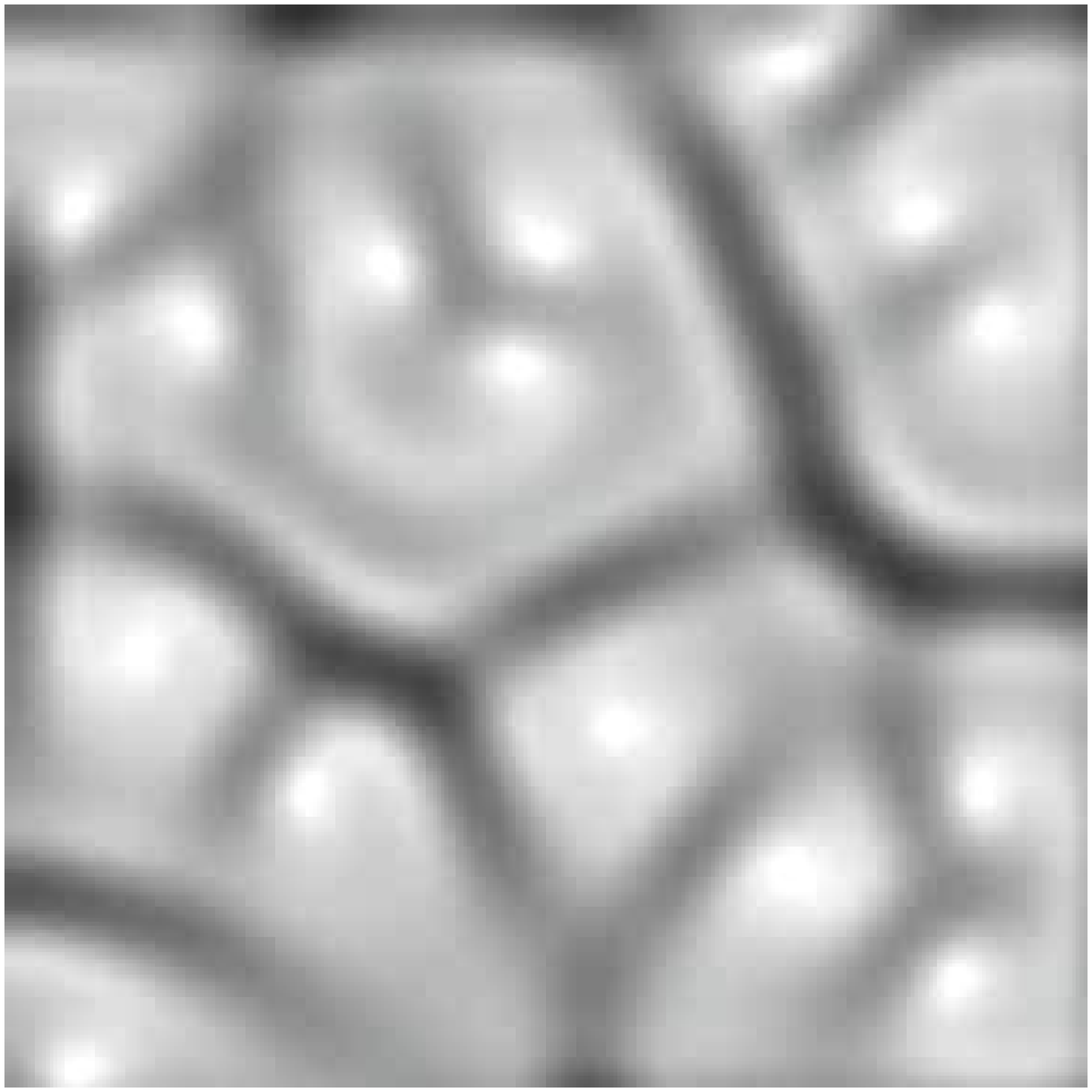}}
& {\includegraphics[width=2.5cm,height=6cm,angle=-90,origin=rB]
{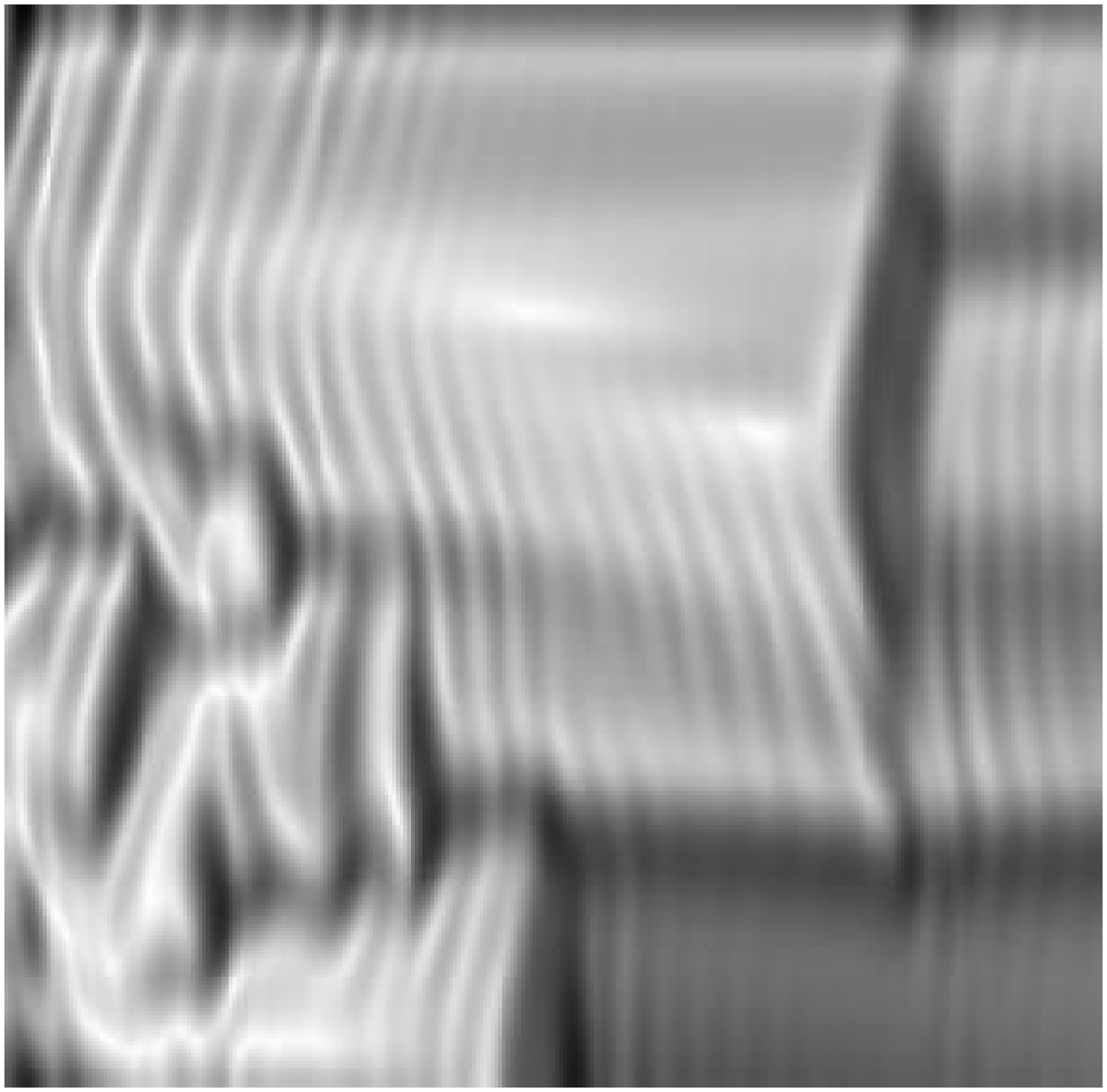}}\\
\end{tabular}
\caption{
 Multiple spiral waves in a two-dimensional system of size $120\times
120;$ the same parameter values as for the pattern 3 in
 Fig. \ref{patterns}. The right
panels are space-time plots showing evolution of the pattern along one
horizontal cross section; the displayed time interval is $T=480$}
\label{patt3_2d}
\end{center}
\end{figure}

For selected parameter values, two-dimensional simulations of the 
system have additionally been performed. 
Figure \ref{patt3_2d} displays the behavior of a
two-dimensional system at the same parameters as for the pattern 3 in 
Fig. \ref{patterns}.
The space-time diagrams in the right panels show the pattern 
development along a horizontal cross section. Starting from random
initial conditions, a population of rotating spiral waves develops. 
After a transient, a stable configuration of rotating spirals is 
reached in two dimensions.

\begin{figure}
\begin{center}
\begin{tabular}{ccc}
\begin{minipage}[b]{2em}
Re$(\eta)$\\\mbox{ }\\\mbox{ }
\end{minipage}
&{\includegraphics*[width=2.5cm]{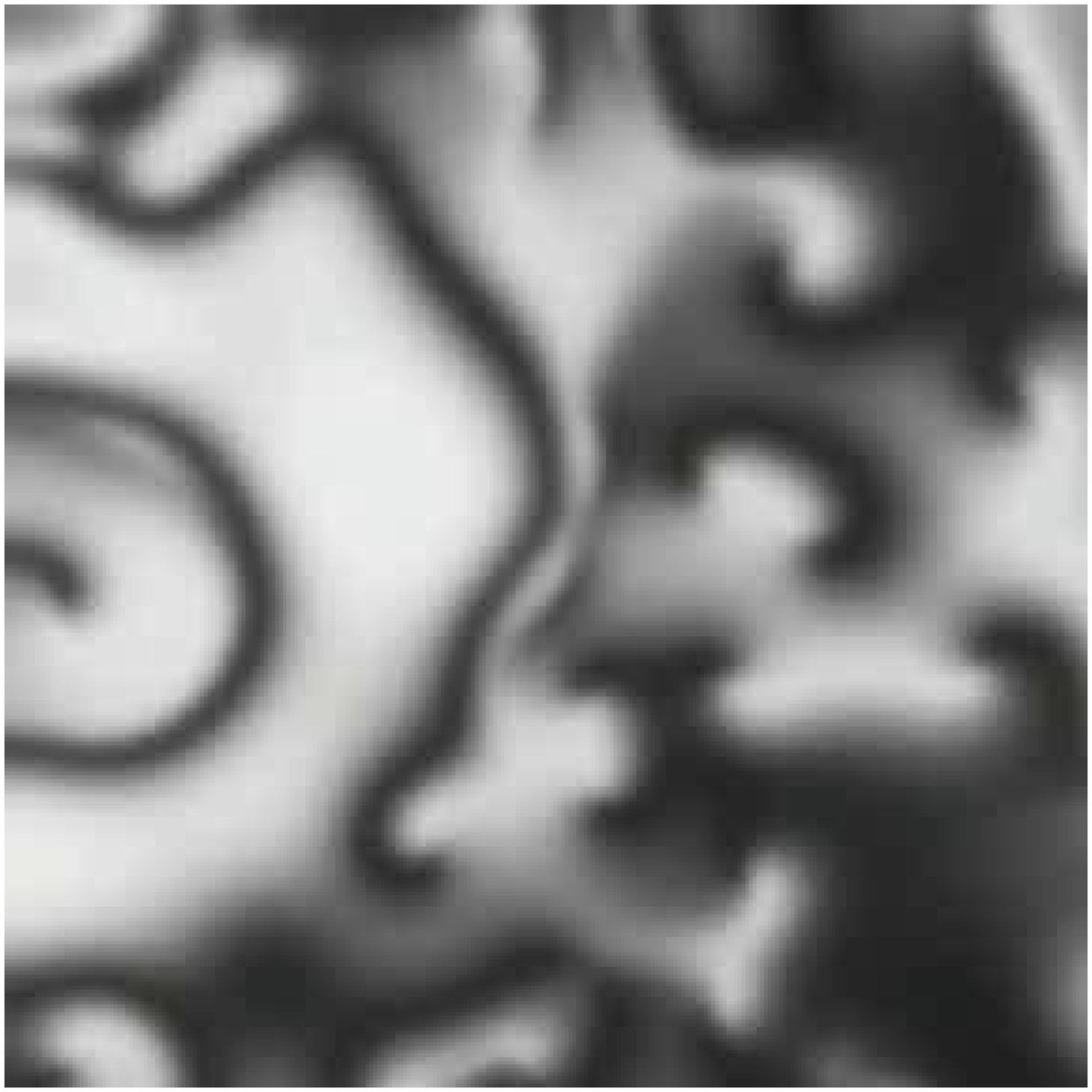}}
& {\includegraphics*[width=2.5cm,height=6cm,angle=-90,origin=rB]
{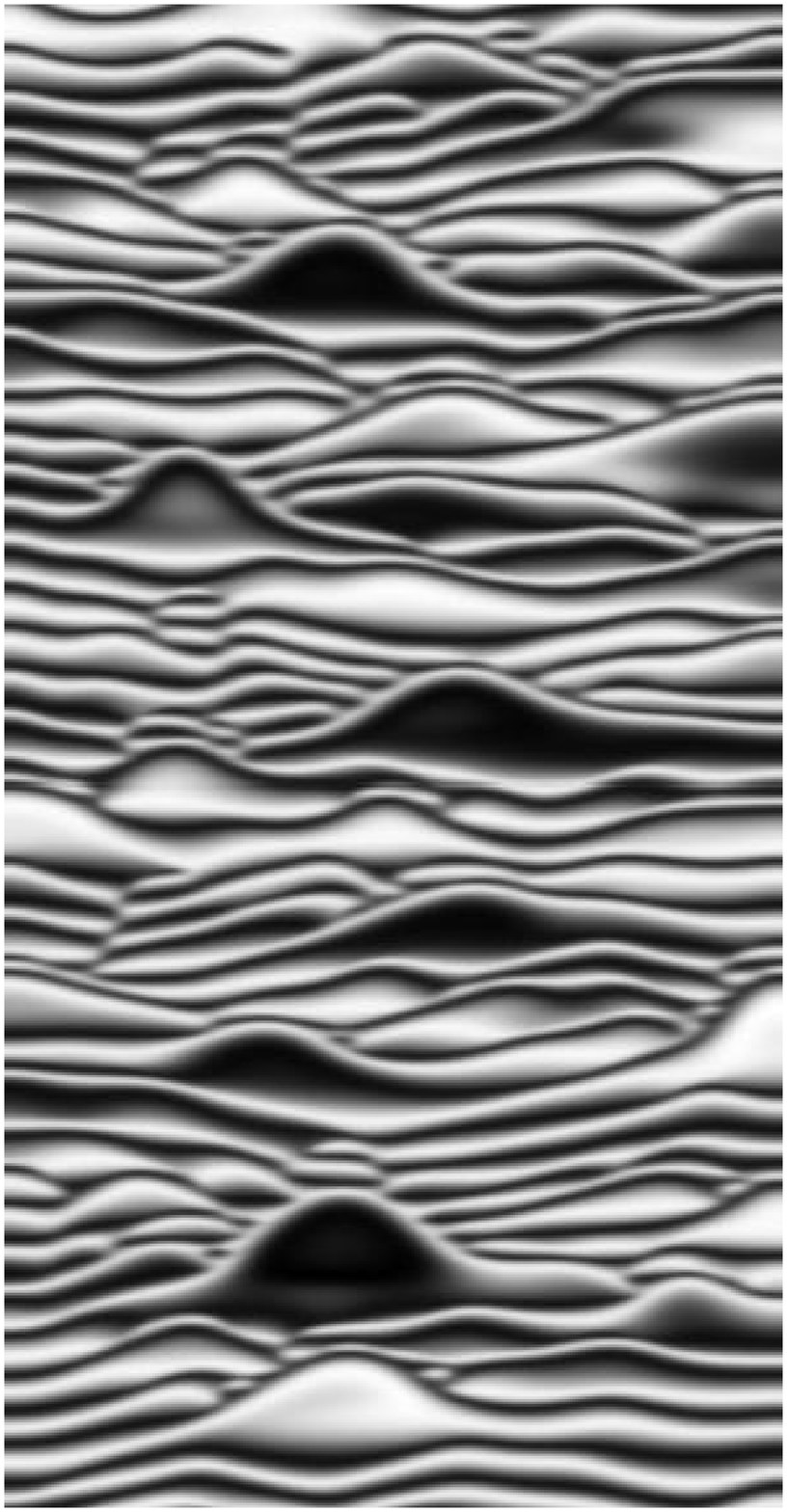}}\\
\begin{minipage}[b]{2em}
$|\eta|$\\\mbox{ }\\\mbox{ }
\end{minipage}
&{\includegraphics*[width=2.5cm]{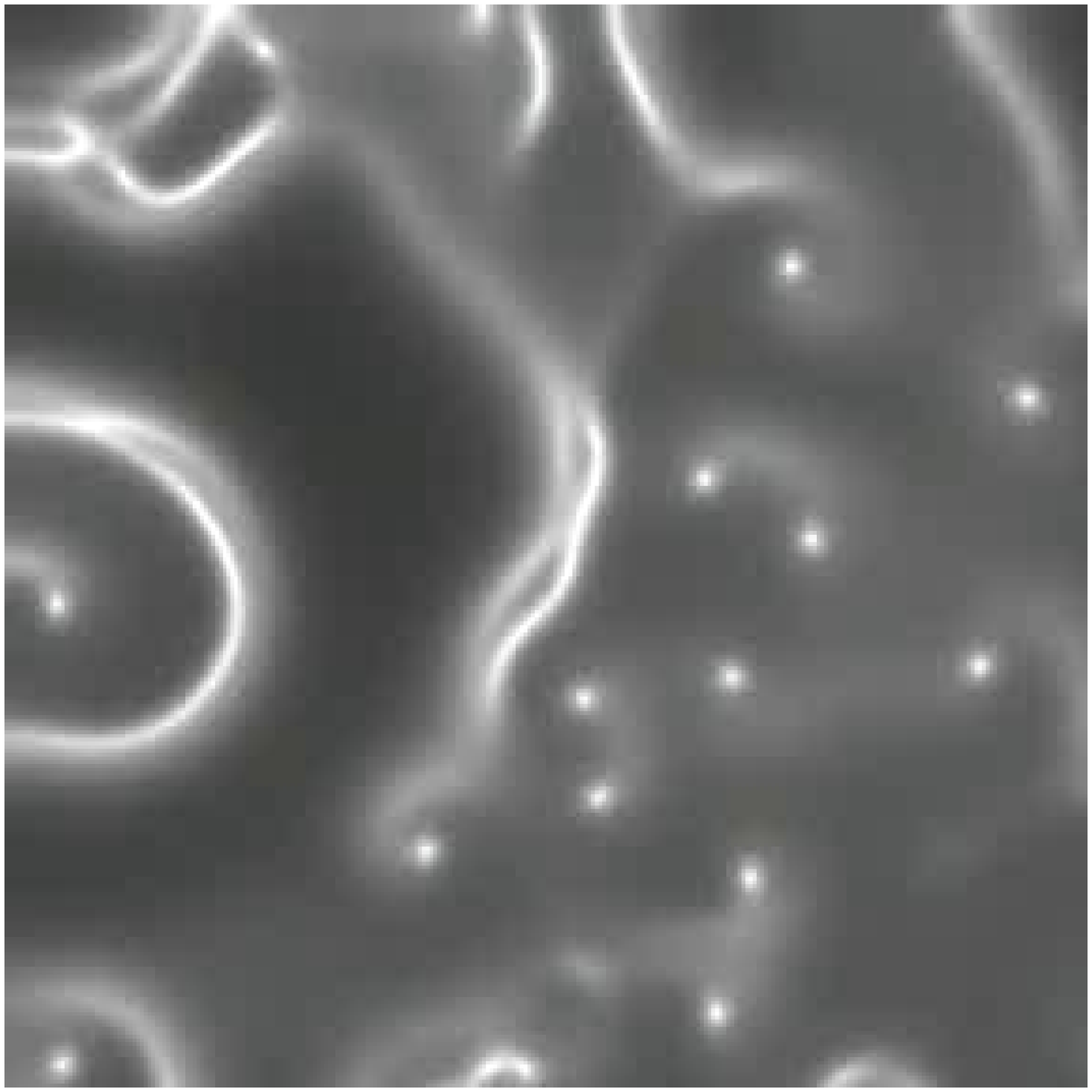}}
& {\includegraphics*[width=2.5cm,height=6cm,angle=-90,origin=rB]
{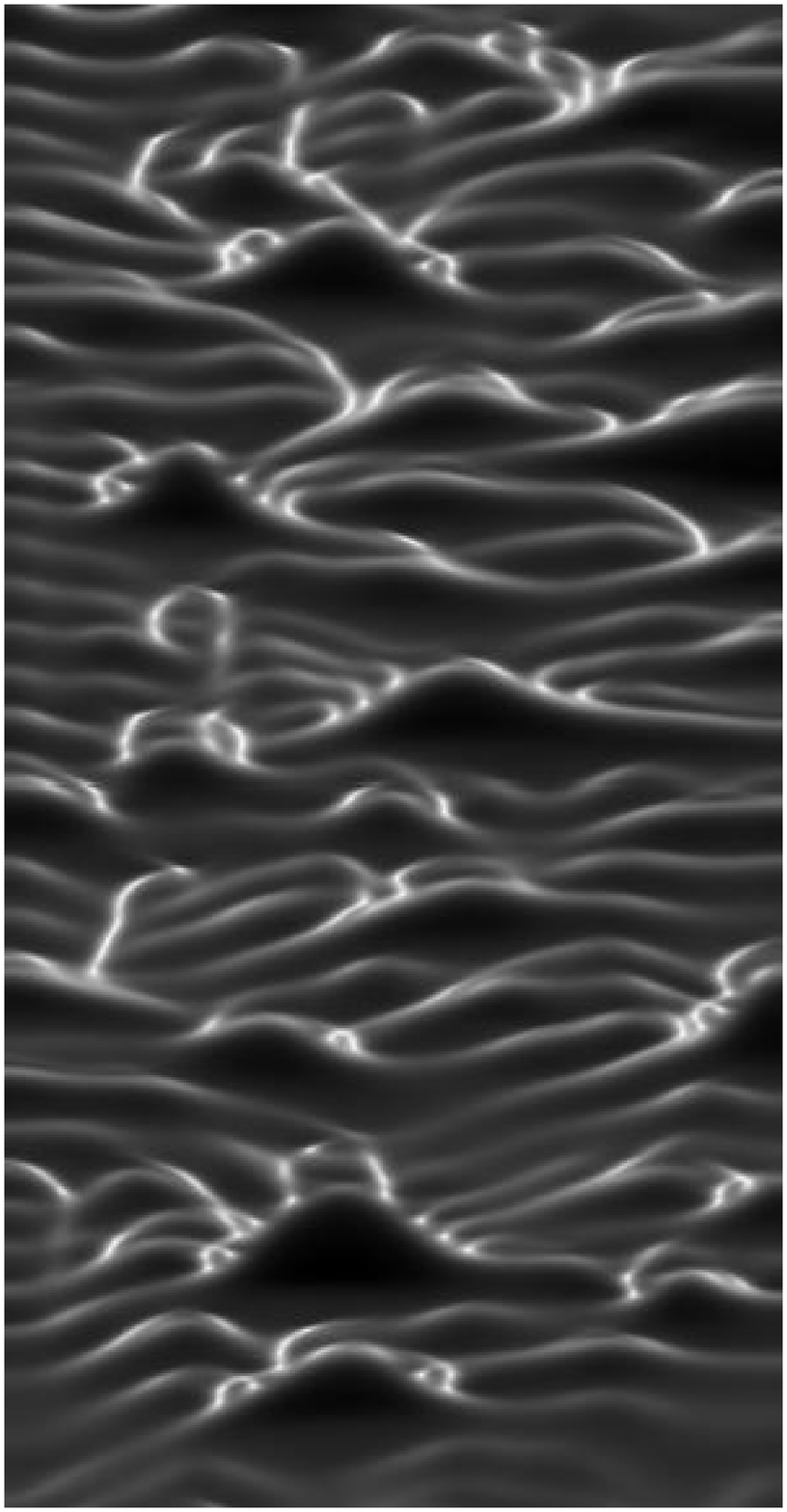}}\\
\begin{minipage}[b]{2em}
$|z|$ \\\mbox{ }\\\mbox{ }
\end{minipage}
&{\includegraphics*[width=2.5cm]{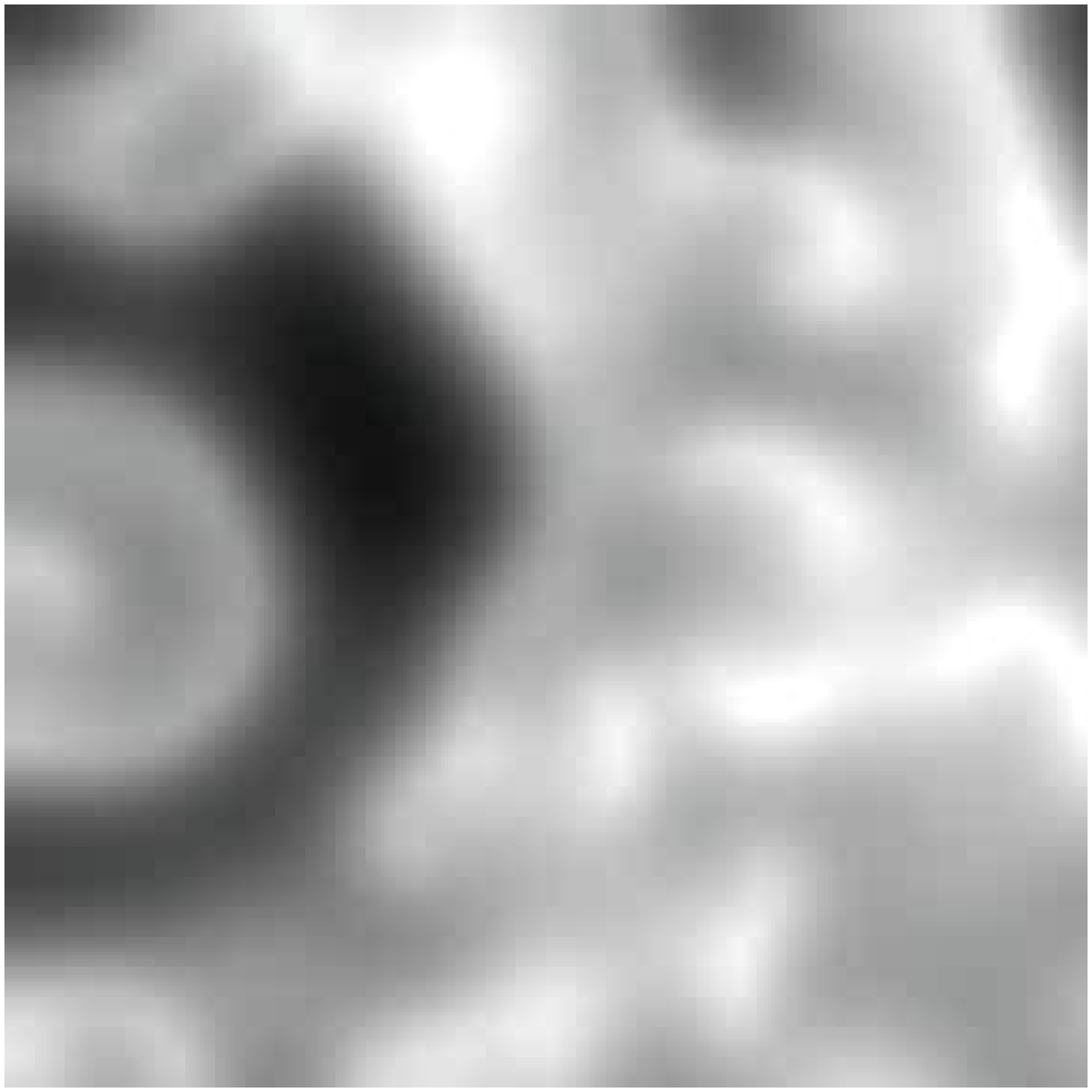}}
& {\includegraphics*[width=2.5cm,height=6cm,angle=-90,origin=rB]
{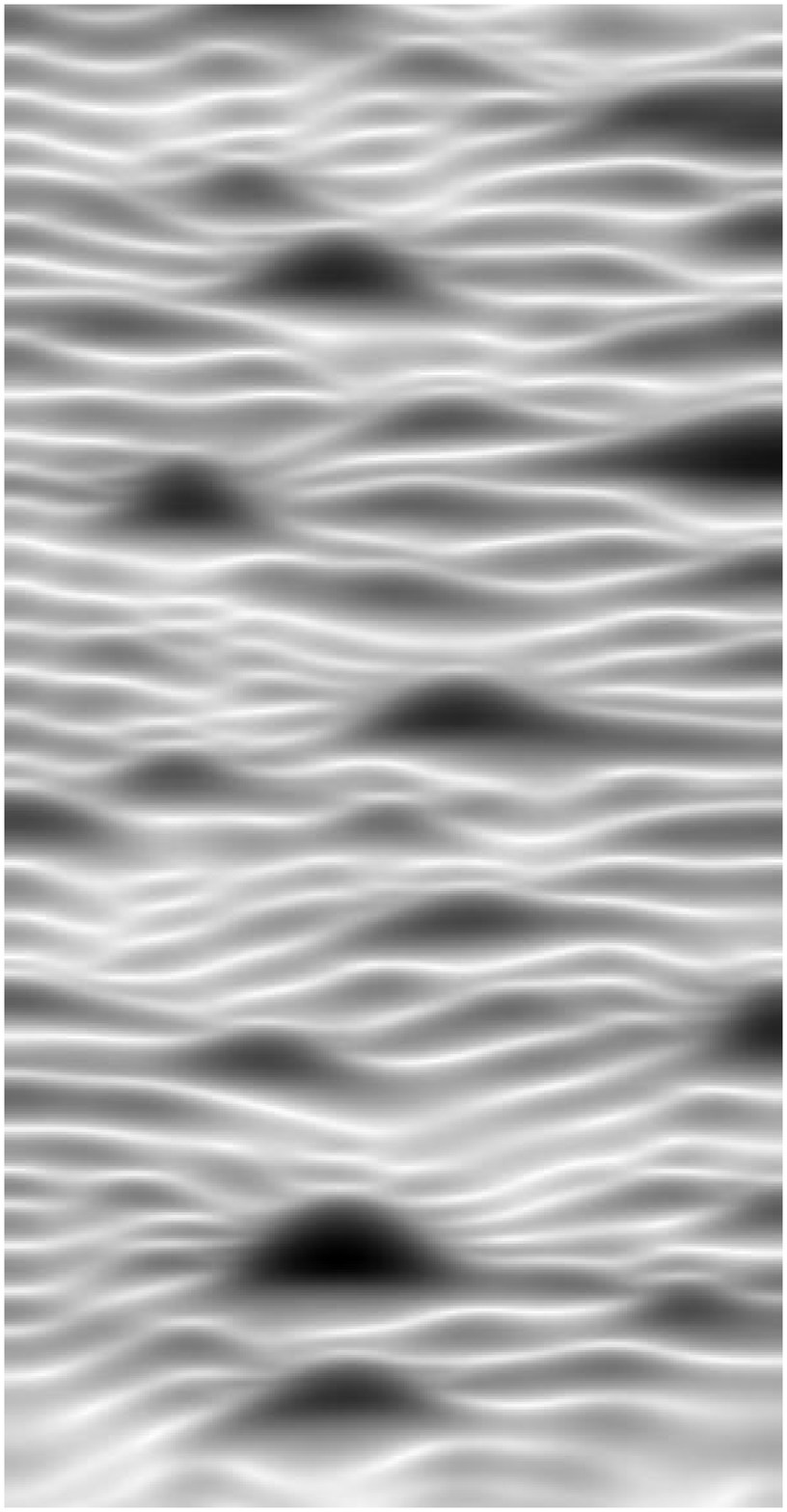}}\\
\end{tabular}
 \caption{Bursts of synchronization in a two-dimensional system of size
$120\times 120;$ the same parameter values as for the pattern 4 in 
Fig. \ref{patterns}.
The right panels are space-time plots showing evolution of the pattern 
along one horizontal cross-section; $T=480$.}
\label{patt4_2d}
\end{center}
\end{figure}

Figure \ref{patt4_2d} shows two-dimensional patterns corresponding 
to synchronization bursts (pattern 4 in Fig. \ref{patterns}). 
Inside such a burst, occupying for example the
left central region in the left panels, the coupling field $z$ is 
increased in magnitude. The space-time diagrams (right panels) reveal 
that the synchronized regions with slow oscillations have only 
relatively short lifetimes and are replaced by irregular rapid 
oscillations. However, they appear again and again in the course of 
time.

\begin{figure}
\begin{center}
\begin{tabular}{cccc}
\includegraphics*[width=2.5cm]{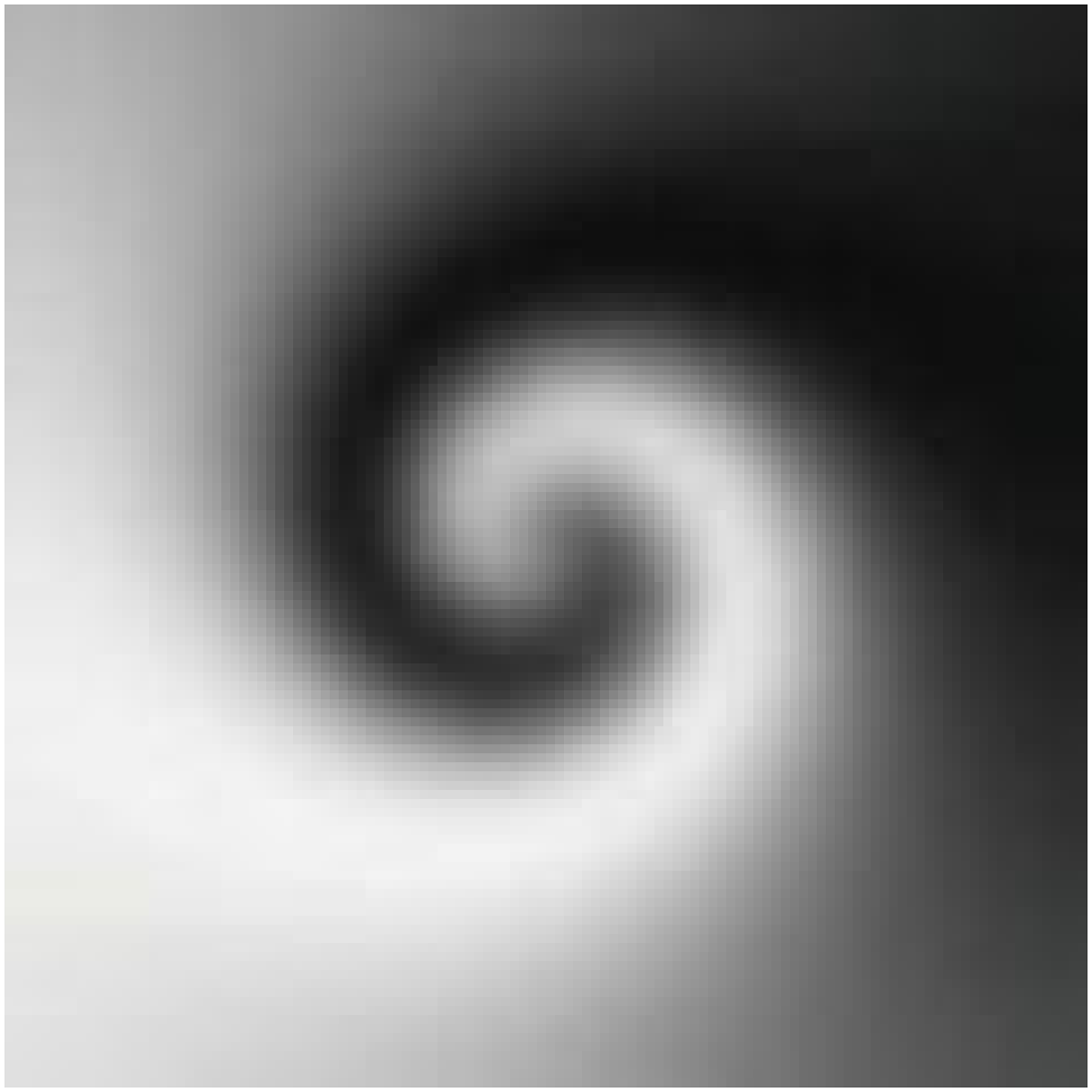}
&\includegraphics*[width=2.5cm]{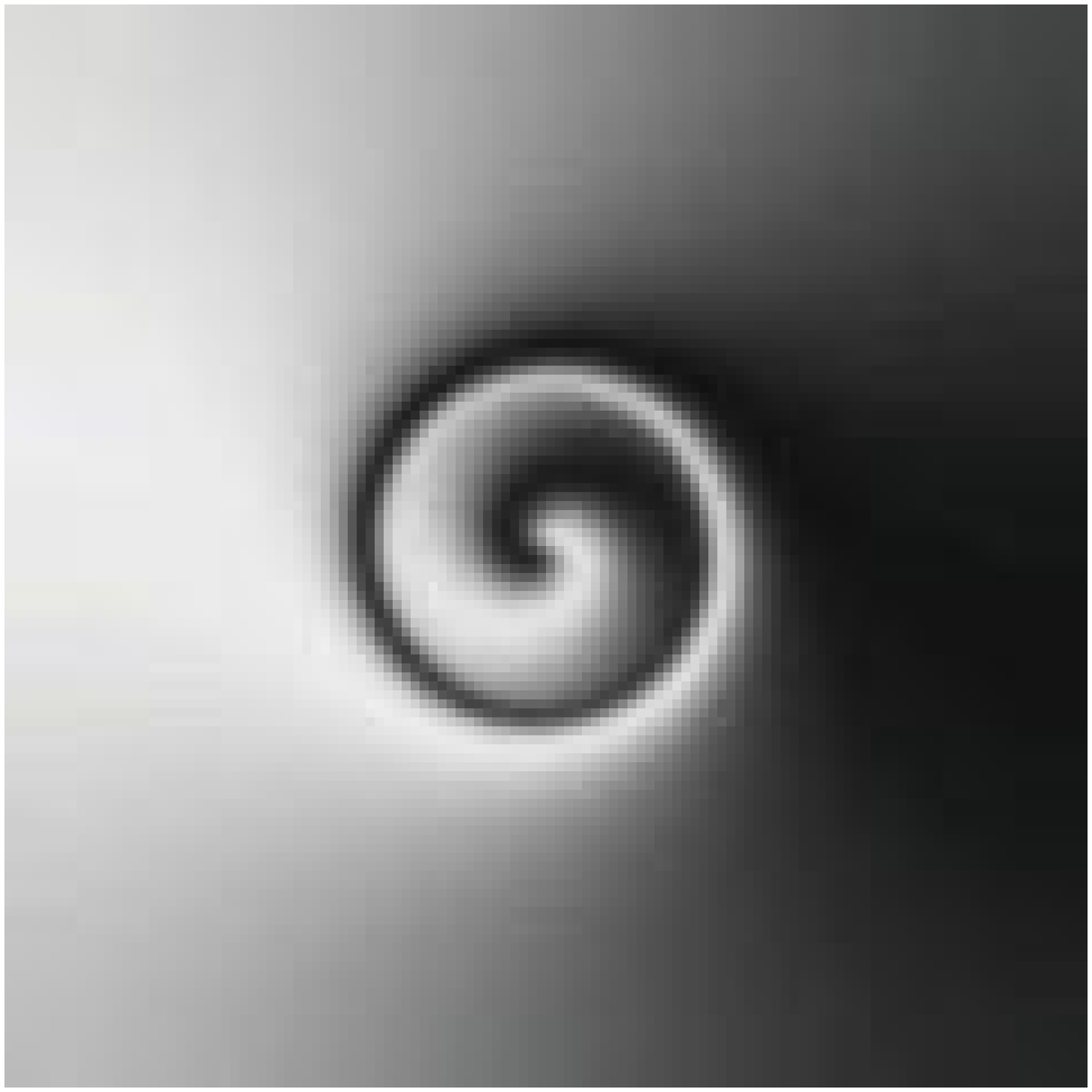}
&\includegraphics*[width=2.5cm]{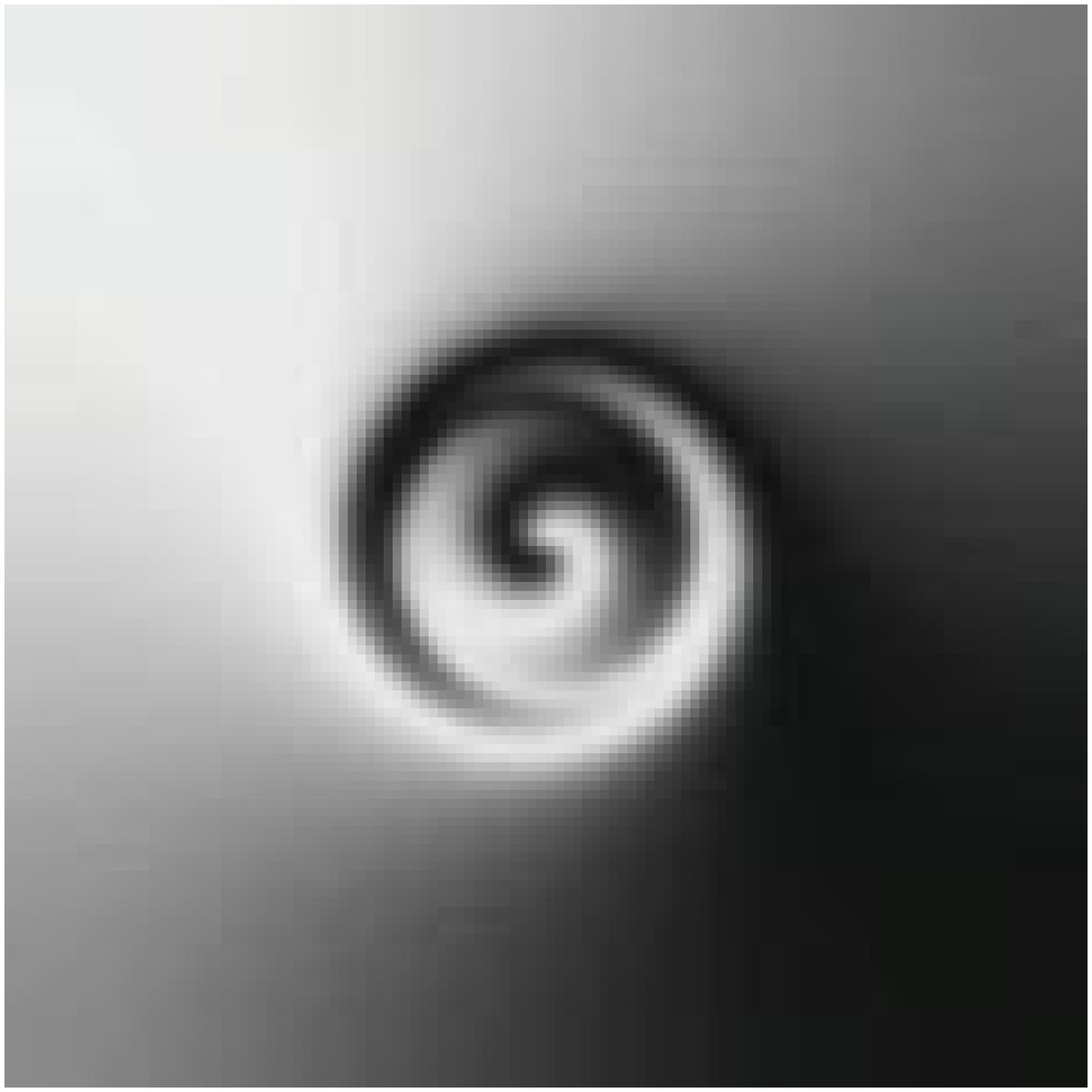}
&\includegraphics*[width=2.5cm]{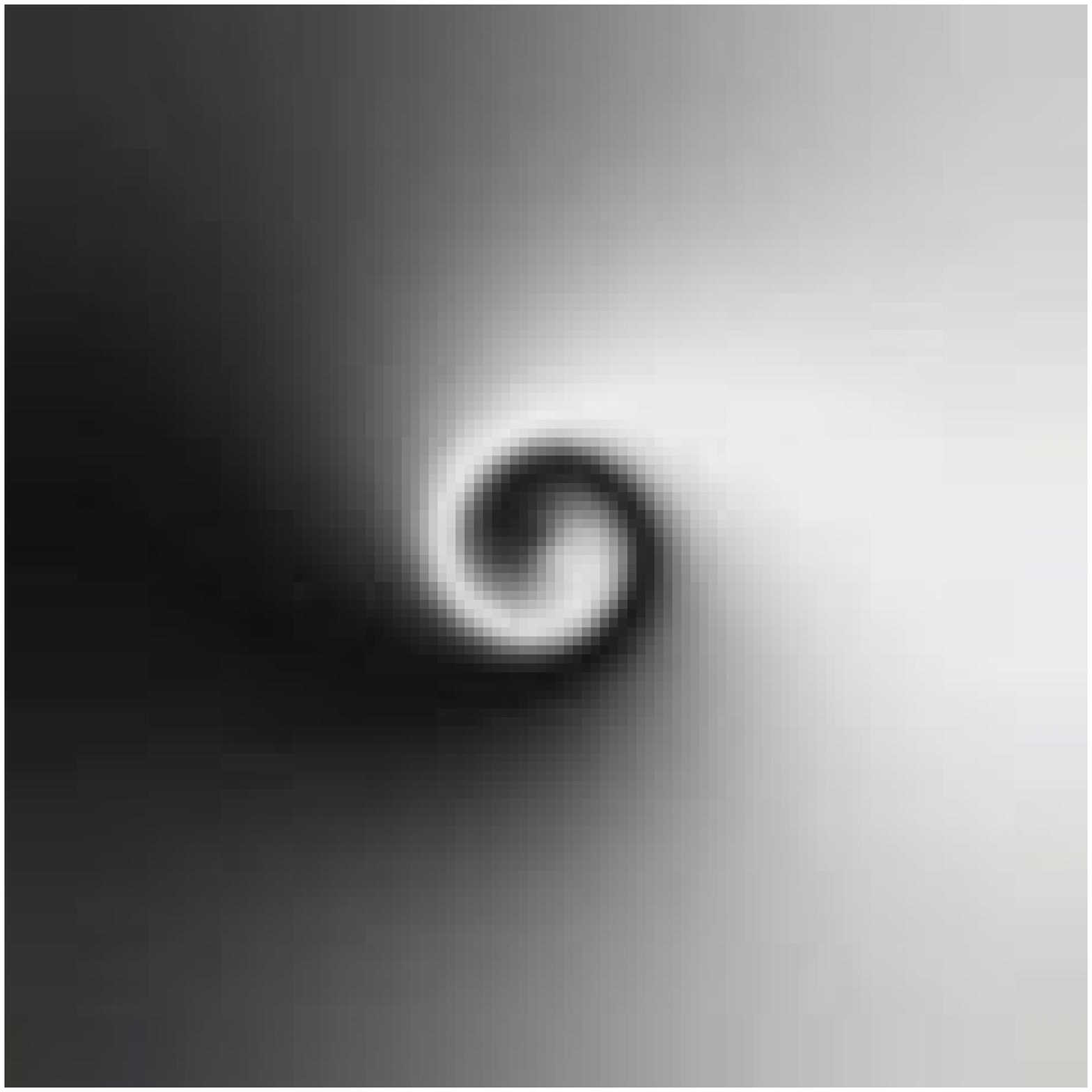}\\
\includegraphics*[width=2.5cm]{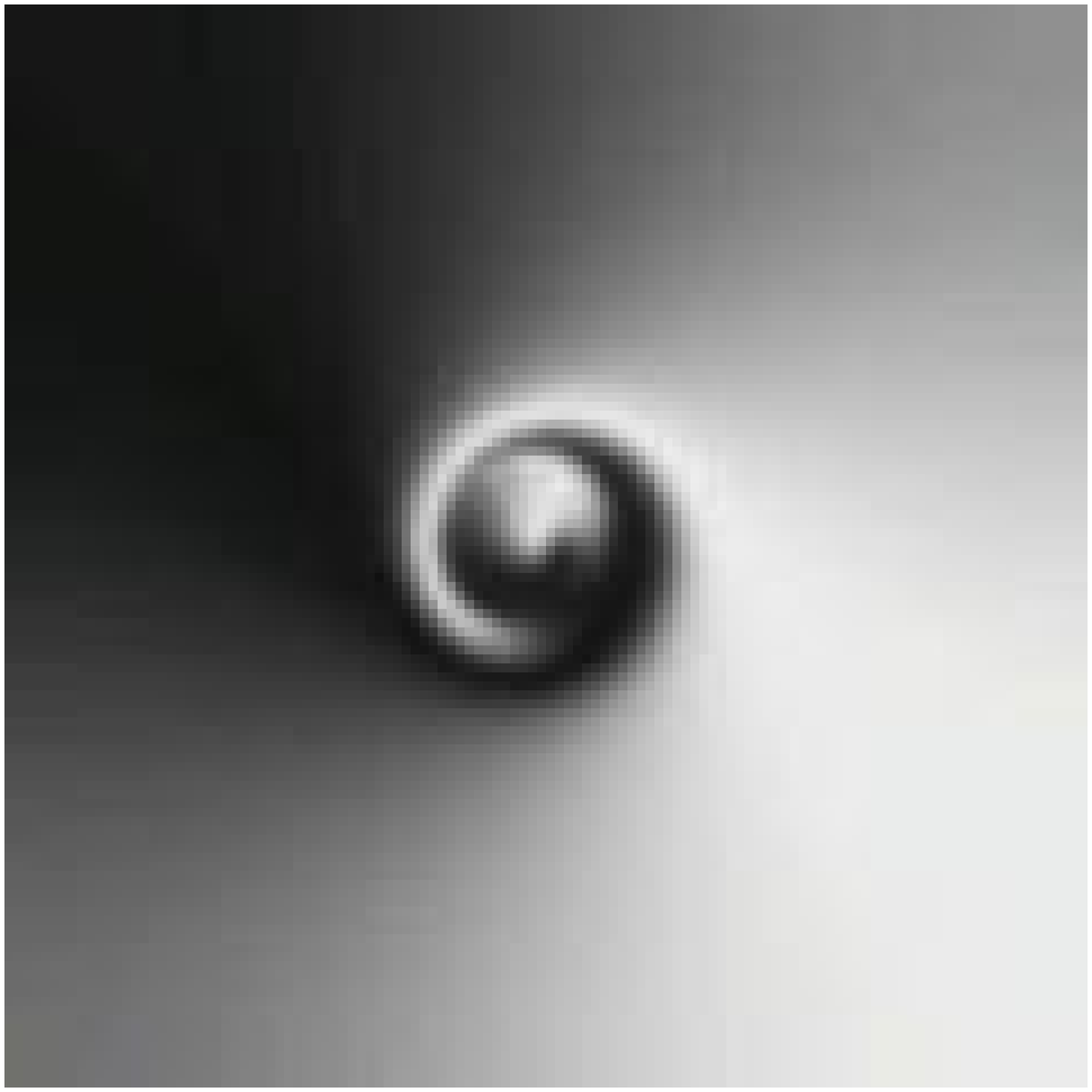}
&\includegraphics*[width=2.5cm]{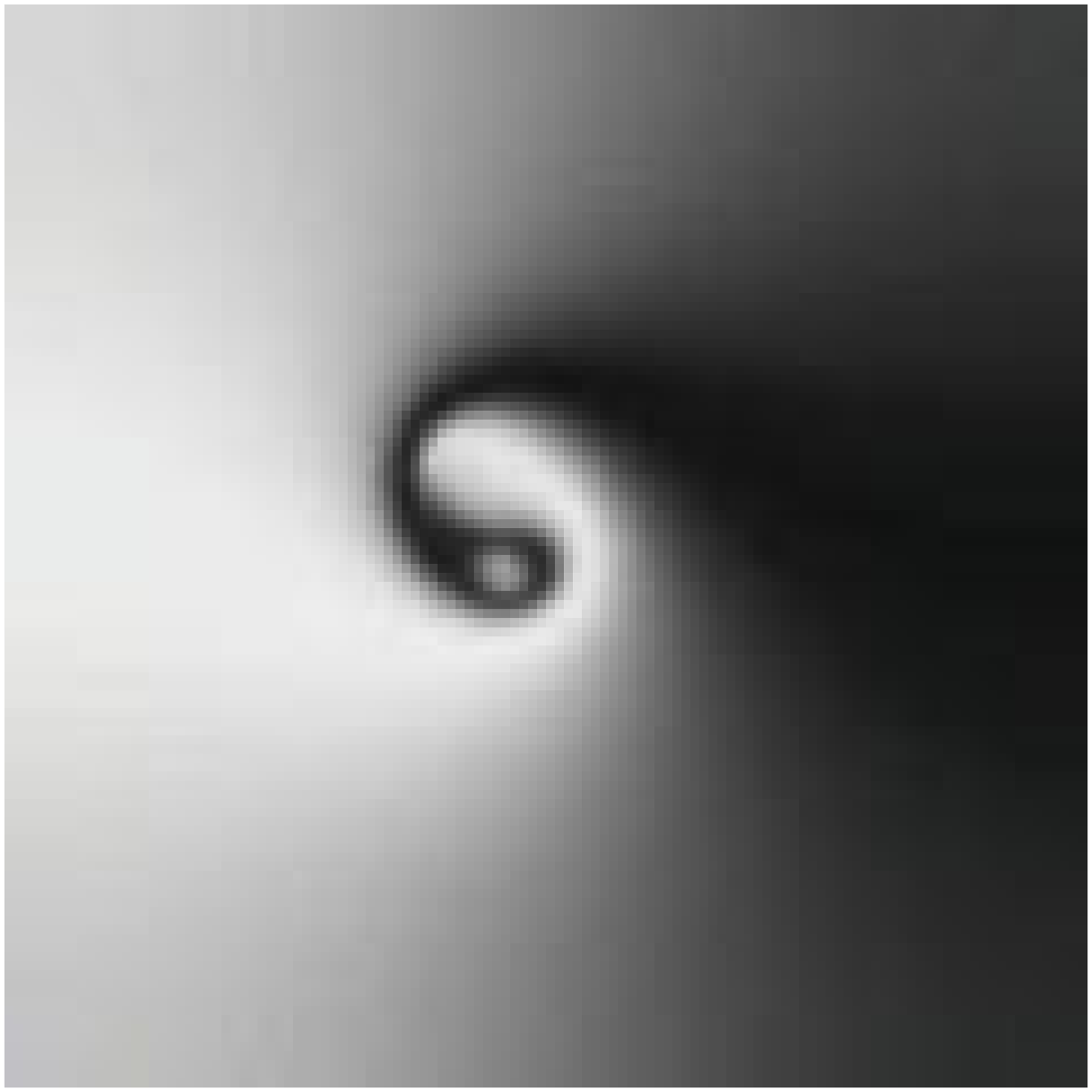}
&\includegraphics*[width=2.5cm]{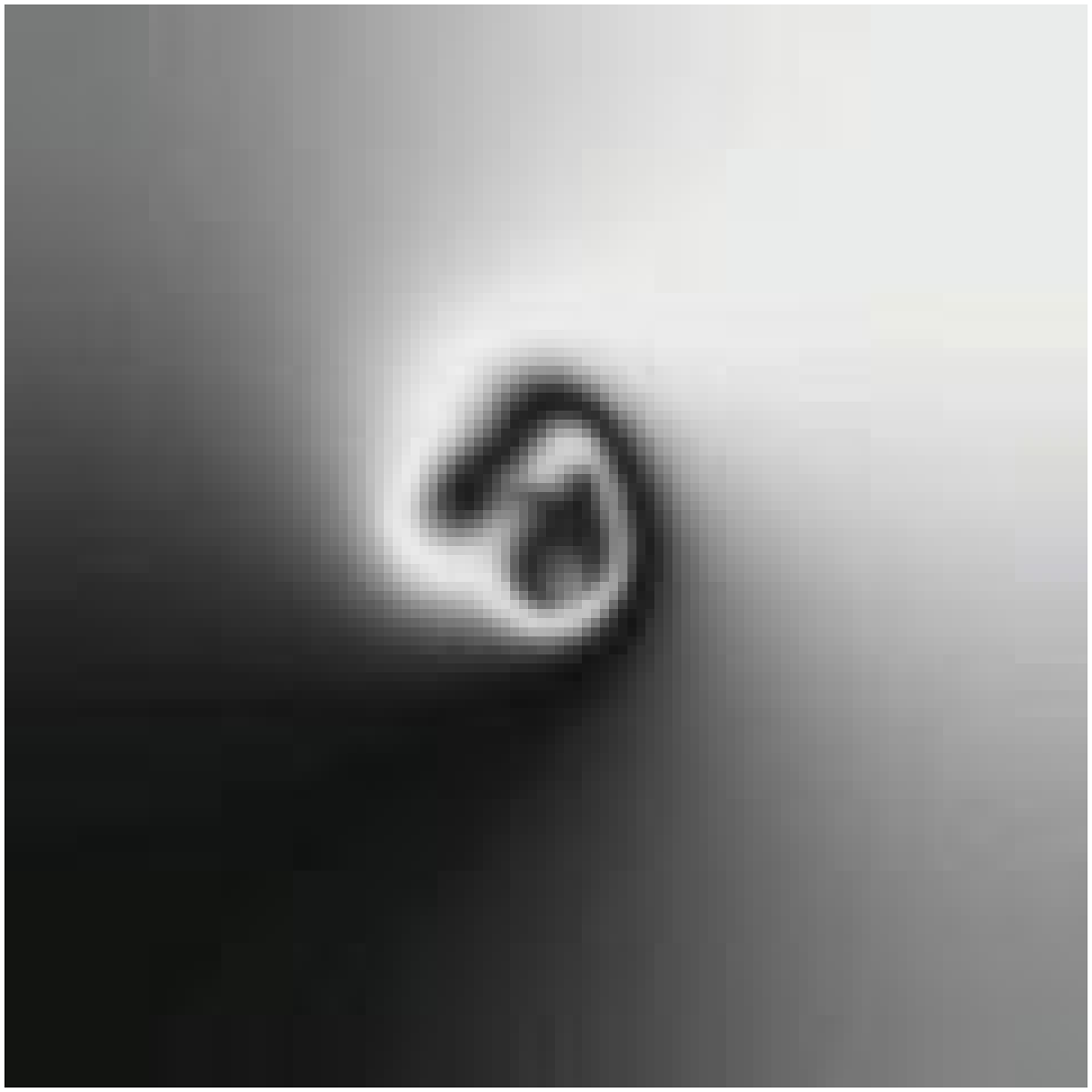}
&\includegraphics*[width=2.5cm]{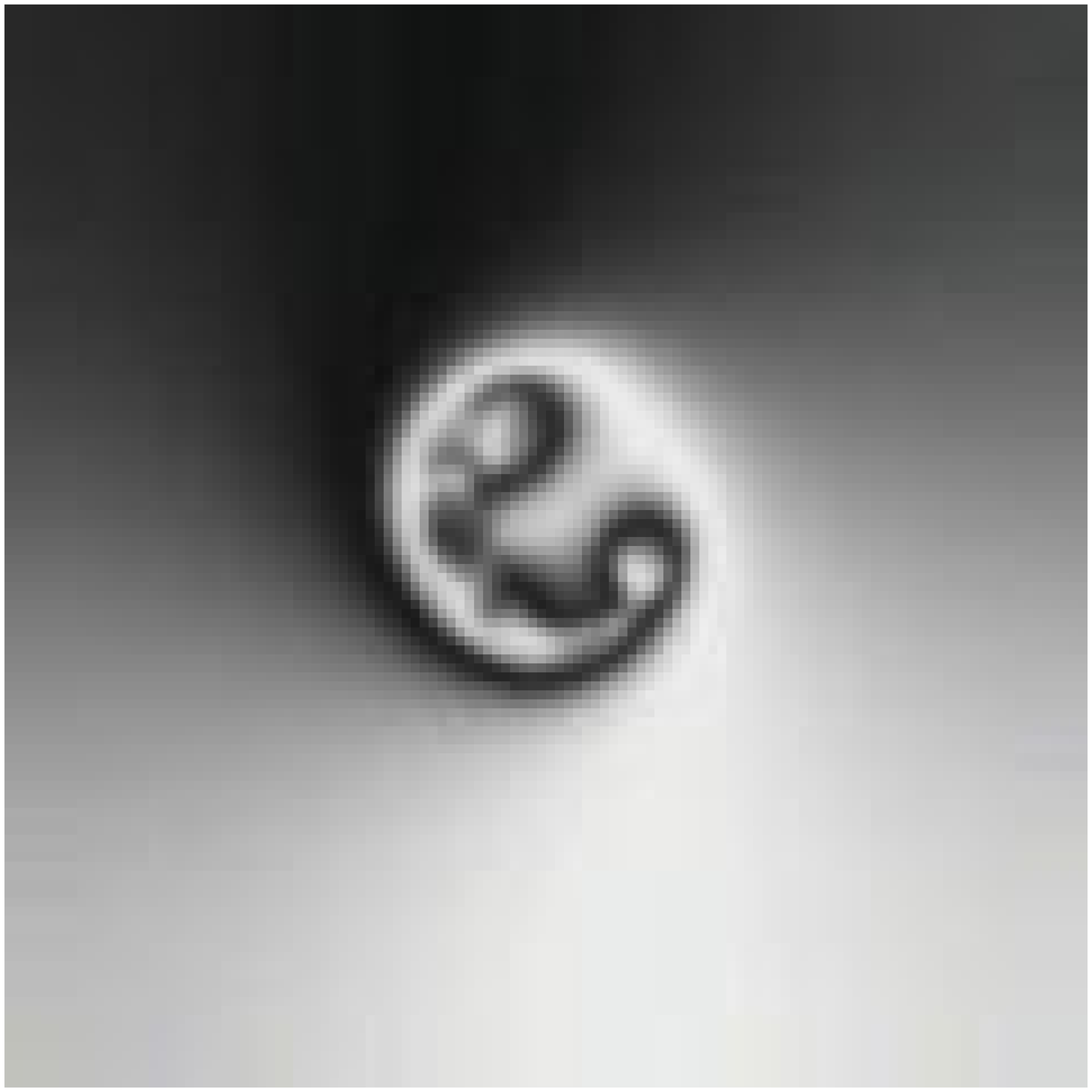}
\end{tabular}
\end{center}
\caption{Instability of a rotating spiral. Simulation for a 
two-dimensional system of size $120\times 120$ with parameters 
$\omega =2,\alpha =3,\beta =1,K=0.4,l=10,$ and $\tau =10$. 
Subsequent snapshots of the field Re($\eta$) at times 
$T$ = 2.8, 10.0, 15.2, 37.2, 96.8, 119.6, 130.0, and 139.2 are 
presented.}
\label{des_bursts_first}
\end{figure}

\begin{figure}
\begin{center}
\begin{tabular}{ccc}
\begin{minipage}[b]{2em}
Re$(\eta)$\\\mbox{ }\\\mbox{ }
\end{minipage}
&{\includegraphics*[width=2.5cm]{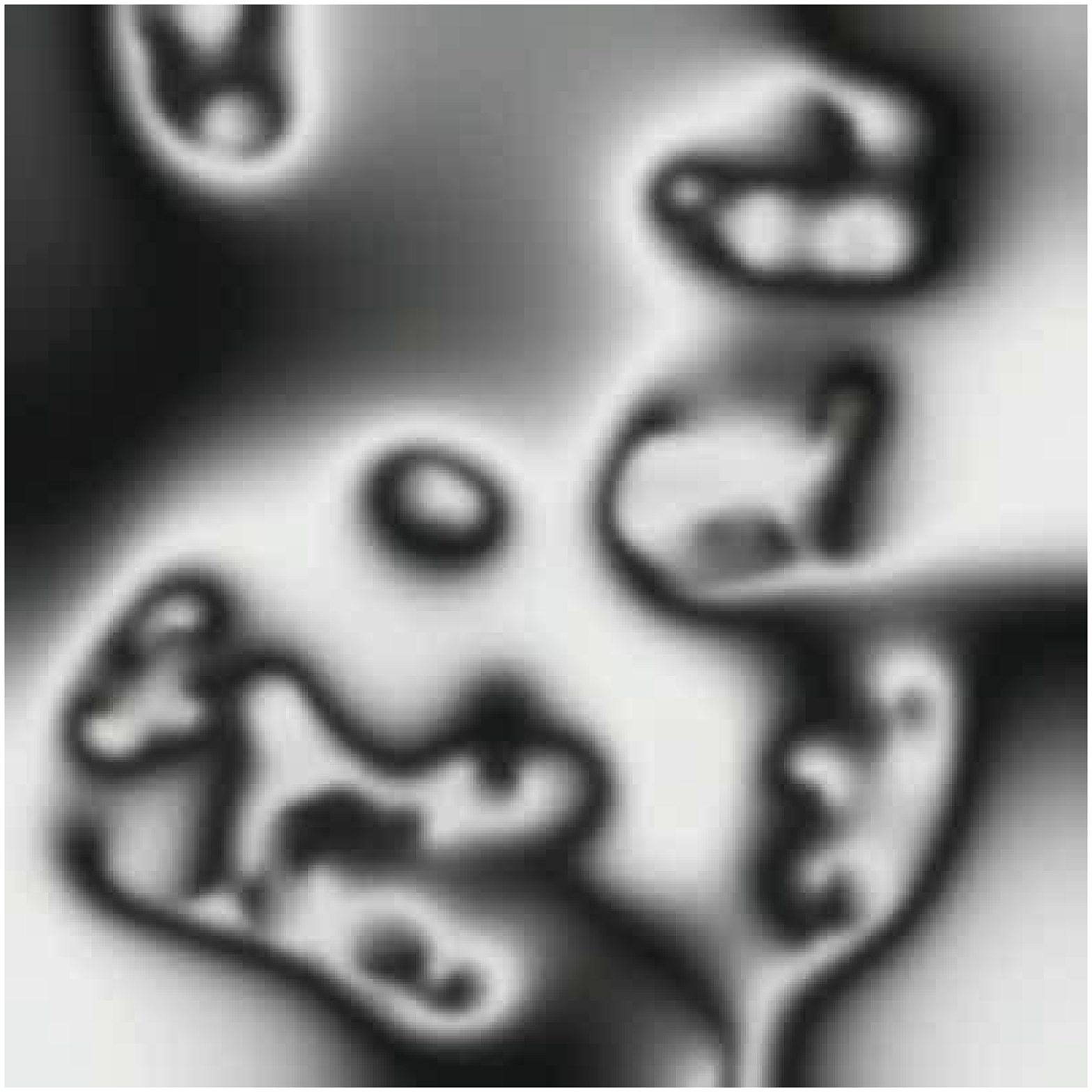}}
& {\includegraphics*[width=2.5cm,height=6cm,angle=-90,origin=rB]
{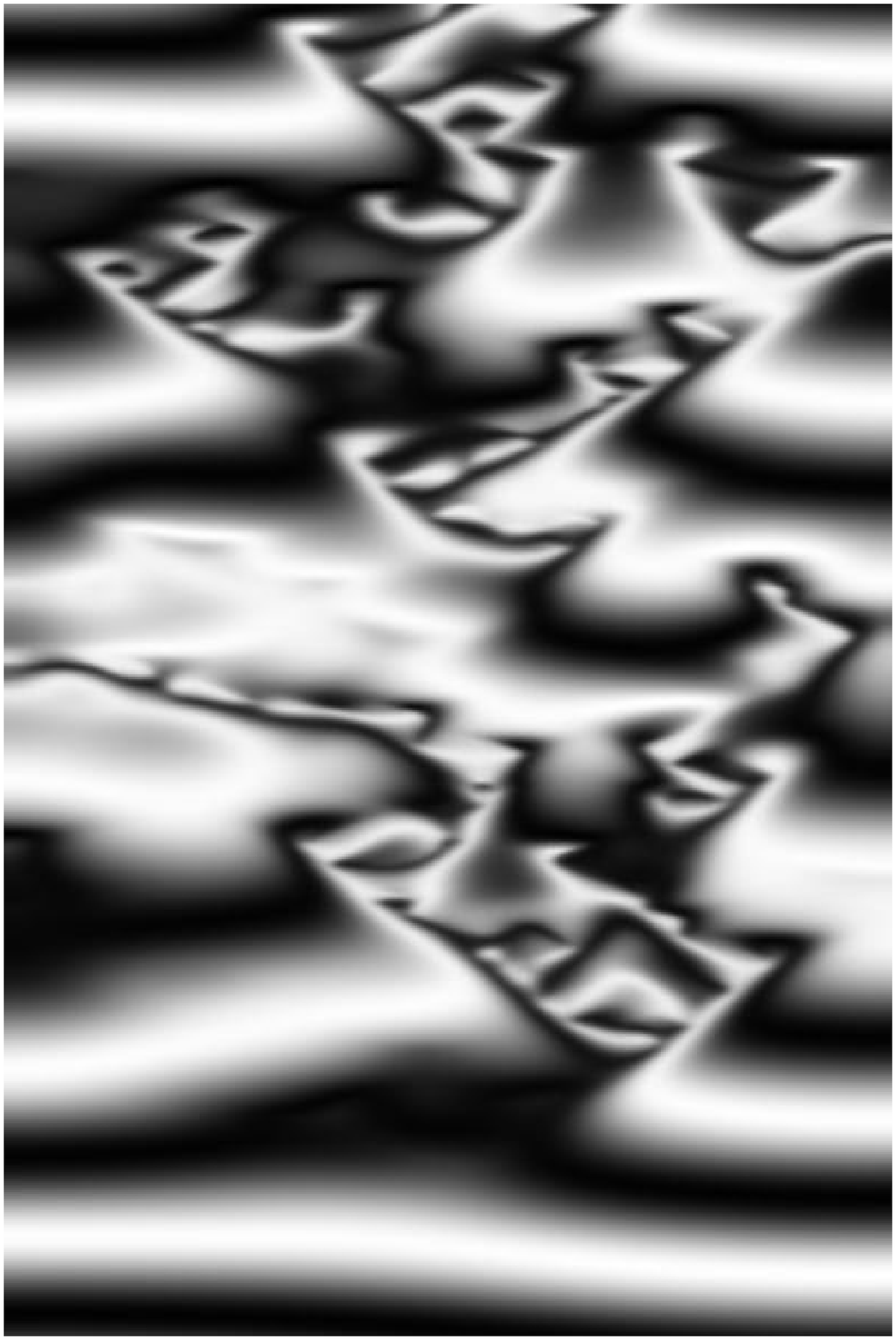}}\\
\begin{minipage}[b]{2em}
$|\eta|$\\\mbox{ }\\\mbox{ }
\end{minipage}
&{\includegraphics*[width=2.5cm]{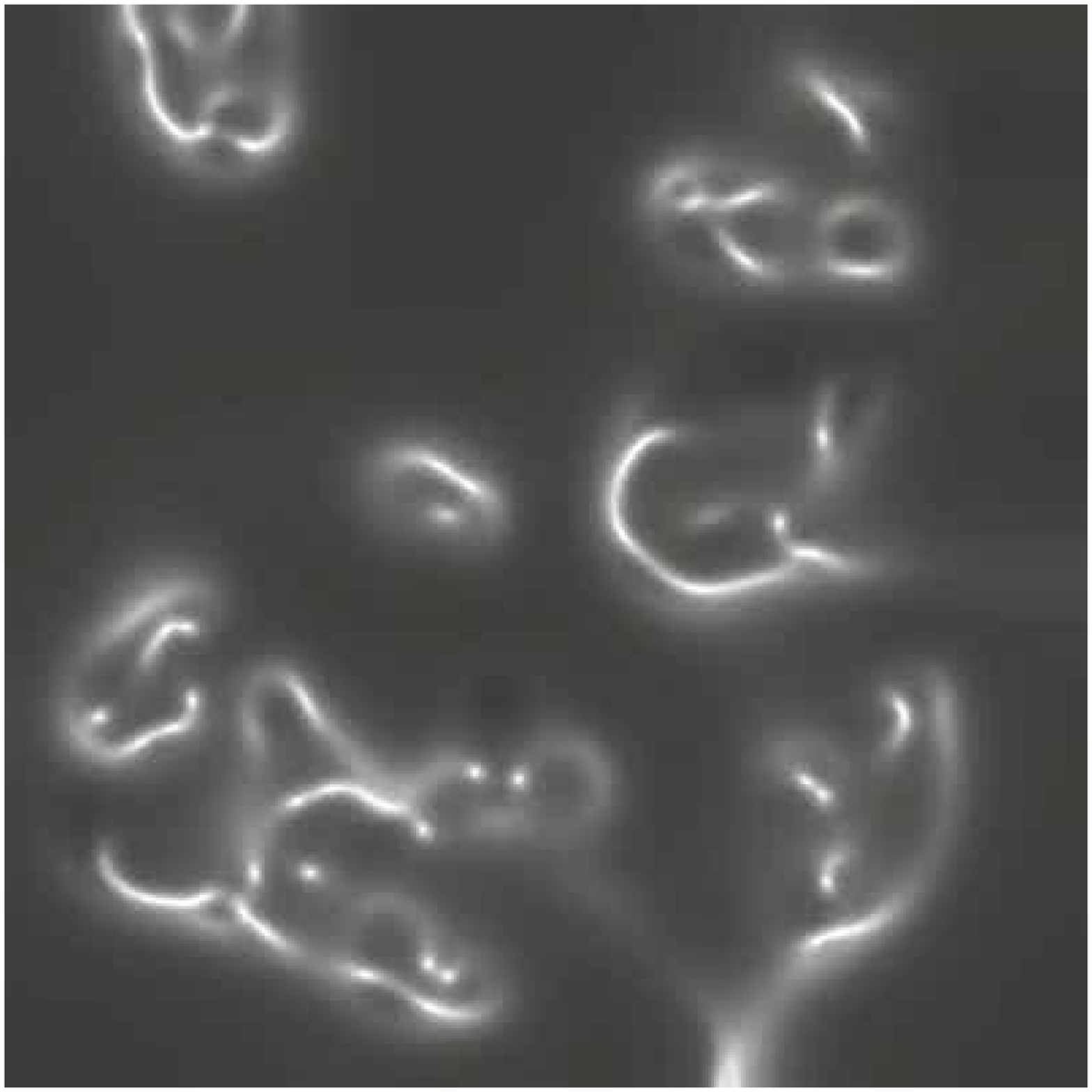}}
& {\includegraphics*[width=2.5cm,height=6cm,angle=-90,origin=rB]
{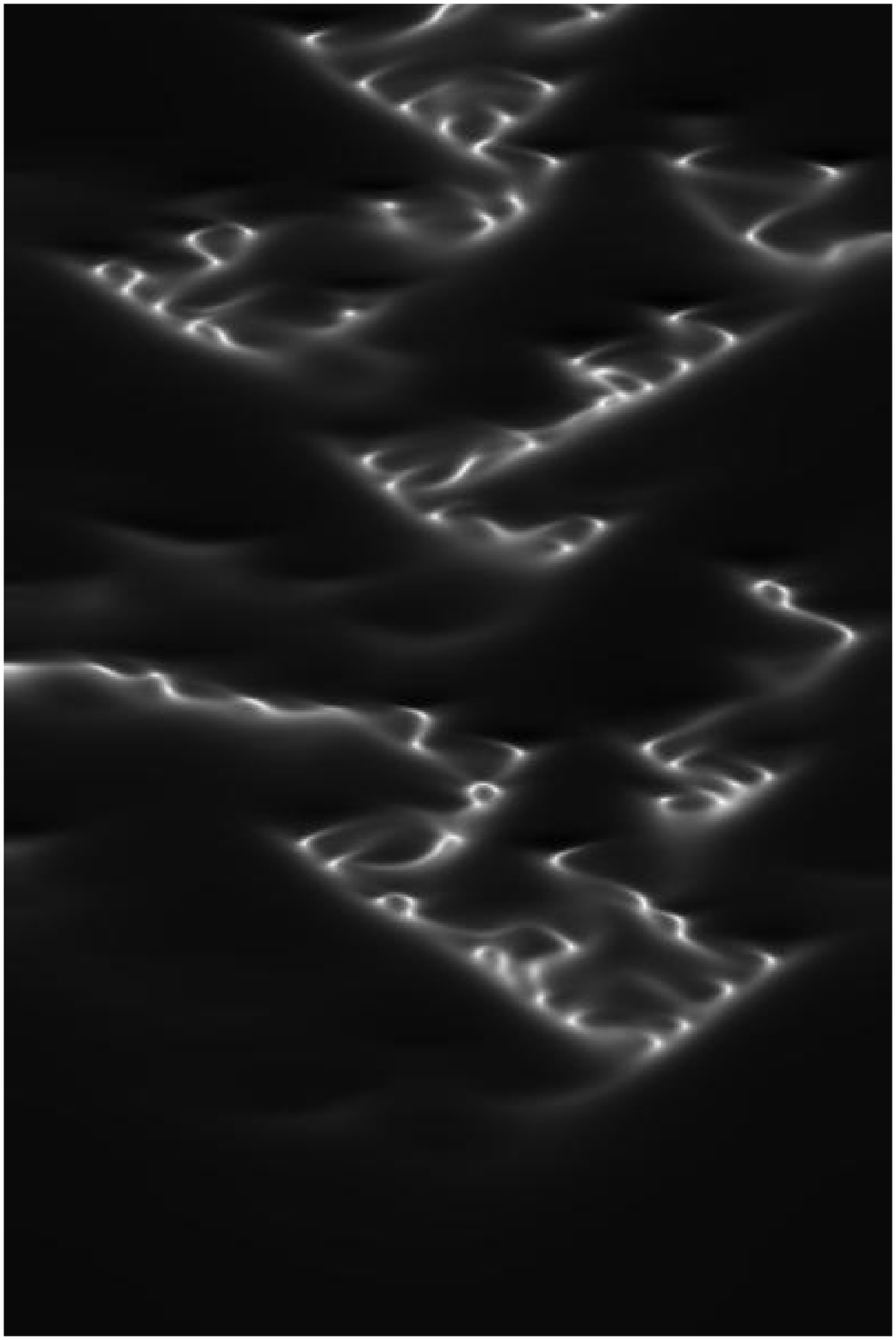}}\\
\begin{minipage}[b]{2em}
$|z|$ \\\mbox{ }\\\mbox{ }
\end{minipage}
&{\includegraphics*[width=2.5cm]{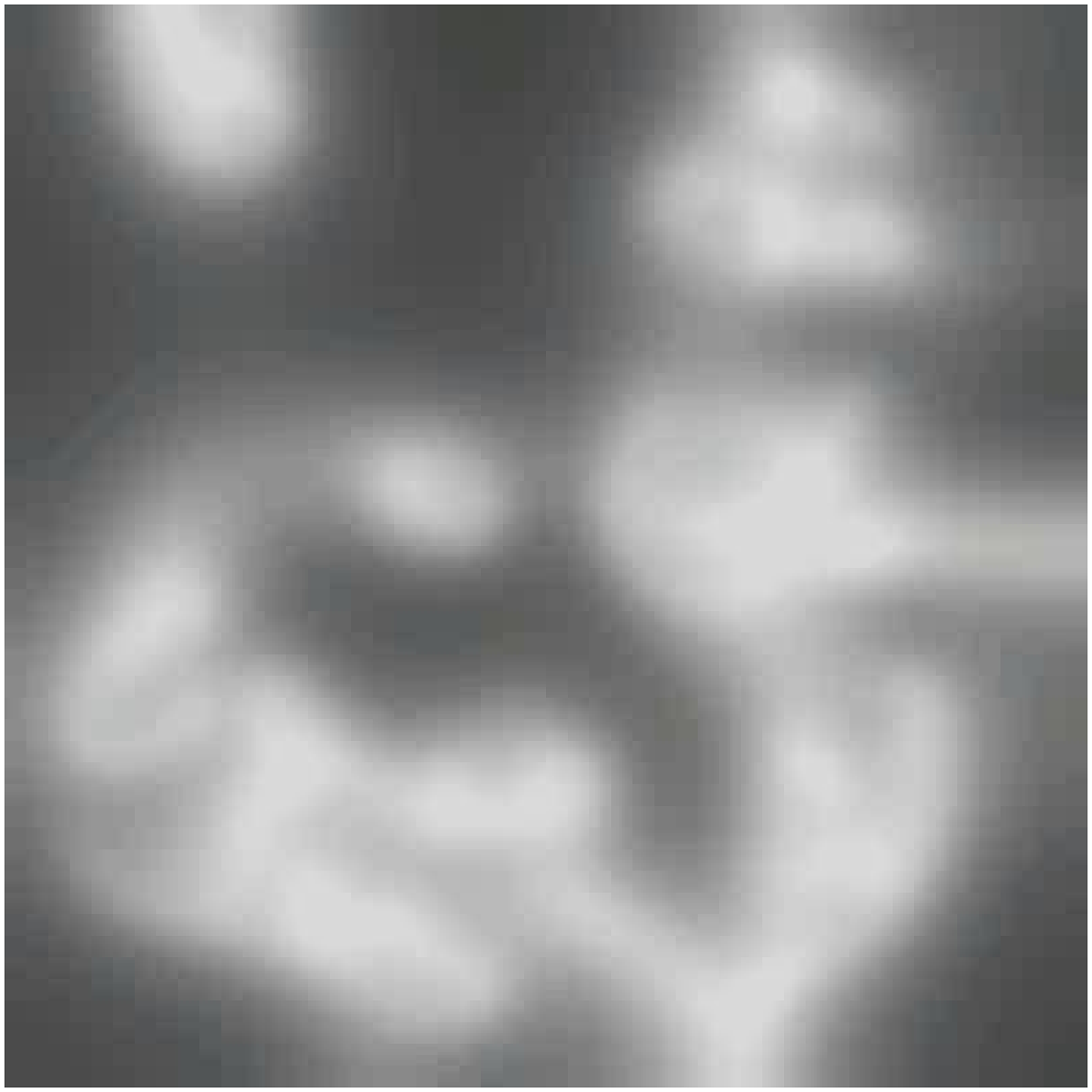}}
& {\includegraphics*[width=2.5cm,height=6cm,angle=-90,origin=rB]
{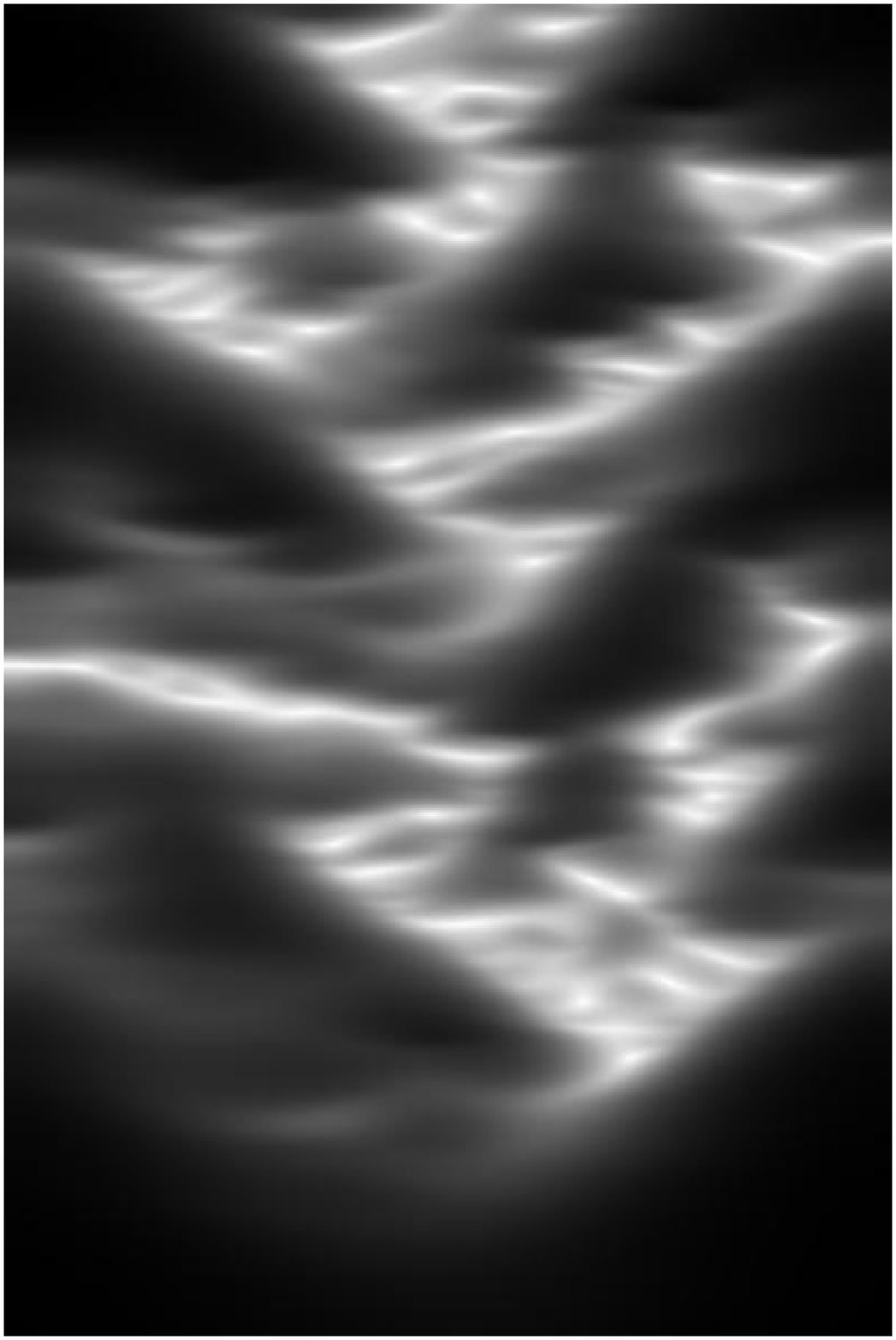}}\\
\end{tabular}
 \caption{
Bursts of desynchronization in a two-dimensional system of size 
$120\times 120;$ continuation of the simulation presented in Fig. 
\ref{des_bursts_first}. 
The parameters are $\omega =2,\alpha =3,\beta =1,K=0.4,l=10,$ and 
$\tau =10$; $T=720$}
\label{des_bursts_second}
\end{center}
\end{figure}

A behavior, which can be described as bursts of desynchronization, was 
found in our two-dimensional simulations even outside of the 
birhythmicity region, where only slow uniform oscillations are possible
(Fig. \ref{des_bursts_first} and \ref{des_bursts_second}). 
We started here
with the initial condition, which is commonly used to generate rotating
spirals. A rotating spiral was indeed first formed. However, some
instability has then developed beginning from the its central region, 
where the oscillation amplitude $z$ was decreased (see Fig. 
\ref{des_bursts_first}).
The development of this instability has resulted 
(Fig. \ref{des_bursts_second}) in complete destruction of the spiral 
and the appearance of relatively small chaotic domains on the 
background of almost uniform slow oscillations. 

Typical snapshots of spatial distributions of the variables Re$(\eta)$,
$|\eta|$ and $|z|$ in this pattern are displayed in the left panels 
in Fig. \ref{des_bursts_second}. 
Inside the domains, the coupling field $z$ is
reduced in magnitude and these small spatial regions are filled with
irregular rapid variations of the complex field $\eta$. Thus, they can
be classified as desynchronization bursts. The space-time diagrams 
through one horizontal cross section of the medium (right panels in 
Fig. \ref{des_bursts_second}) show
that such domains can spread or shrink, and travel through the medium. 
The edges of these structures are marked by the appearance of amplitude
defects where $|\eta|$ is close to zero.

Remarkably, if we keep fixed all other parameter values, but eliminate
diffusional coupling between the oscillators (the Laplacian term in 
equation (\ref{model:one})), evolution starting from the same initial 
conditions leads to
spiral waves with phase-randomized cores, similar to those which were
previously considered \cite{shima,chapter9}. Thus, the inclusion of
diffusive coupling has a strong effect on pattern formation in the 
system.
Though patterns resembling spiral waves with phase-randomized core are
indeed initially developing in the diffusively coupled case, they are
unstable and, after a transient, lead to the development of 
intermittent spatiotemporal regimes with desynchronization bursts.

\section{Discussion}

Our study can be viewed as an extension of a series of detailed
investigations of systems with nonlocal coupling between oscillators
performed by Y. Kuramoto with his coworkers. We have chosen the model 
(\ref{model}), which belongs to the previously discussed class, and
focused our analysis on the effects of coupling inertiality and 
diffusion in this model. 
The principal effect of the inertiality in nonlocal coupling
is that the system becomes birhythmic and has uniform slow and rapid
oscillations as two different coexisting attractors.

For rapid oscillations, the inertial field $z$ responsible for nonlocal
coupling is not excited and the system is essentially reduced to an 
array of only diffusively coupled oscillators with amplitudes $\eta$. 
In contrast to this, the nonlocal coupling field $z$ is involved in 
slow oscillations and nonlocal coupling between oscillators is in 
operation under these conditions.

Our stability analysis based on the phase dynamics equation has shown 
that, though slow uniform oscillations are always stable in the 
considered model, rapid uniform oscillations may become modulationally 
unstable near the birhythmicity boundary. 
Traveling fronts, separating spatial regions with two different 
oscillation modes, have been seen in our numerical simulations.
Note that the fronts separating oscillating and non-oscillating regions
have been recently observed in the vicinity of a subcritical 
Hopf bifurcation \cite{coullet04}.

Irregular intermittent spatiotemporal regimes of two different kinds 
have been numerically observed. In the patterns with synchronization 
bursts, the background is occupied by rapid chaotic oscillations. 
On this background, short-living small domains with slow and rather 
uniform oscillations spontaneously develop. 
Inside such domains, the nonlocal coupling field is
much larger in magnitude than for the rapidly oscillating background. 
The characteristic size of such domains was comparable to the 
nonlocality radius in the model. This kind of intermittent turbulence 
is to some extent similar to the behavior observed for a system near 
the Andronov homoclinic bifurcation, exhibiting bistability  between 
a stationary and an oscillatory state \cite{coullet97,coullet98}. 
However, the developing domains are filled here not by a stationary
state, but by slow oscillations.

In the spatiotemporal patterns with desynchronization bursts, the 
background is occupied by slow oscillations which are almost uniform. 
On this background, a cascade of desynchronization bursts develops. 
Each burst represents a localized turbulent spot filled with rapid 
irregular oscillations. 
Inside it, the nonlocal coupling field is reduced in
magnitude. Such patterns correspond to the regimes of intermittent 
localized turbulence previously seen for the complex Ginzburg-Landau 
equation with global coupling \cite{batt97}, in the experiments with 
global delayed feedback in
surface chemical reactions \cite{science} and in the respective
realistic simulations \cite{bertram1}. In some cases, they
develop in the situations when rotating spiral waves with 
phase-randomized cores have been seen \cite{shima,chapter9} in systems 
where only nonlocal coupling was present.
Birhythmicity as a cause of weak turbulence has recently been
investigated in a biological model of glycolitic oscillations
\cite{batt04}.

Stimulating discussions with Prof. Y. Kuramoto are gratefully 
acknowledged.
We thank M. Stich for the discussion of questions related to
birhythmicity.

\section*{Appendix A}
Starting from system (\ref{four}) we linearize the first three 
equations
around the values $\rho_0, r_0, \psi_0$ of the uniform oscillations. 
The system that we obtain can be written as
\begin{subequations}
\begin{eqnarray}
\dot{\delta\rho}&=&a_1\delta\rho+b_1\delta r+c_1\delta\psi
+d_1\nabla\Theta^2+e_1\nabla^2\Theta\\
\dot{\delta r}&=&a_2\delta\rho+b_2\delta r+c_2\delta\psi
+d_2\nabla\Theta^2\\
\dot{\delta\psi}&=&a_3\delta\rho+b_3\delta r+c_3\delta\psi
+d_3\nabla\Theta^2+e_3\nabla^2\Theta\\
\dot\Theta&=&a_{4}\delta\rho+b_{4}\delta r
+c_{4}\delta\psi+d_{4}\nabla\Theta^2+e_{4}\nabla^2\Theta+f_4
\label{linear}
\end{eqnarray}
\end{subequations}
where the coefficients are given in the following table
\begin{center}
\begin{tabular}{|l||l|}
\hline
$a_1 = 1-3 \rho_0^2-K$ 
& $a_3 = 2\alpha\rho_0+\left[\frac{K r_0}{\rho_0^2}-r_0\frac{1}{\tau}
\right]\so$\\
\hline
$b_1 = K\co$
& $b_3 = \so\left[-\frac{K}{\rho_0}+\frac{\rho_0}{\tau r_0^2}\right]$\\
\hline
$c_1 = -K r_0\so$
& $c_3 = \co\left[-K \frac{r_0}{\rho_0}+\frac{\rho_0}{\tau r_0}
\right]$\\
\hline
$d_1 = -\frac{\rho_0}{4}$
& $d_3 = -l^2\frac{r_0}{4\tau}$\\
\hline
$e_1 = \frac{\zeta}{2}\rho_0\beta$
& $e_3 = \frac{1}{2}\left(1-\frac{l^2}{\tau}\right)$\\
\hline\hline
$a_2 = \frac{1}{\tau}\co$
& $a_4 = 2\alpha\rho_0+\left[K\frac{r_0}{\rho_0^2}+\frac{1}
{\tau\rho_0}\right]\so$\\
\hline
$b_2 = -\frac{1}{\tau}$
& $b_4 = \so\left[-\frac{K}{\rho_0}-\frac{\rho_0}{\tau r_0^2}\right]$\\
\hline
$c_2 = -\frac{\rho_0}{\tau}\so$
& $c_4 = \co\left[-K\frac{r_0}{\rho_0}+\frac{\rho_0}{\tau r_0}
\right]$\\
\hline
$d_2 = -l^2\frac{r_0}{4\tau}$
& $d_{4} = \frac{\beta}{4}$\\
\hline
$e_2 = 0$
& $e_4 = \frac{1}{2}\left(1+\frac{l^2}{\tau}\right)$\\
\hline\hline
\multicolumn{2}{|c|}{$f_4 = -\omega+\alpha\rho_0^2+
\left[-K\frac{r_0}{R_0}+\frac{\rho_0}{\tau r_0}\right]\so$} \\
\hline
\end{tabular}
\end{center}

Now, since we are in the approximation where $\rho, r, \psi$ adjust 
adiabatically to $\Theta$, we can assume 
$\dot{\delta\rho}=\dot{\delta r}=\dot{\delta\psi}=0$, so that we get 
from (\ref{linear})
\begin{eqnarray}
\delta \rho &=& \frac{b_3 c_2 d_1-b_2 c_3 d_1-b_3 c_1 d_2+b_1 c_3 d_2
+b_2 c_1 d_3-b_1 c_2 d_3}
{-a_3 b_2 c_1 + a_2 b_3 c_1 + a_3 b_1 c_2-a_1 b_3 c_2 + a_2 b_1 c_3 
- a_1 b_2 c_3}\nabla\Theta^2\nonumber\\
& &+\frac{b_3 c_2 e_1-b_2c_3e_1+b_2c_1e_3-b_1c_2e_3}
{-a_3 b_2 c_1 + a_2 b_3 c_1 + a_3 b_1 c_2-a_1 b_3 c_2 + a_2 b_1 c_3 
- a_1 b_2 c_3}\nabla^2\Theta\\
\delta r &=& \frac{a_3c_2d_1-a_2c_3d_1-a_3c_1d_2+a_1c_3d_2+a_2c_1d_3
-a_1c_2d_3}
{a_3b_2c_1-a_2b_3c_1-a_3b_1c_2+a_1b_3c_2+a_2b_1c_3-a_1b_2c_3}
\nabla\Theta^2\nonumber\\
& &+\frac{a_3c_2e_1-a_2c_3e_1+a_2c_1e_3-a_1c_2e_3}
{a_3b_2c_1-a_2b_3c_1-a_3b_1c_2+a_1b_3c_2+a_2b_1c_3-a_1b_2c_3}
\nabla^2\Theta\\
\delta\psi &=& \frac{a_3 b_2 d_1-a_2b_3d_1-a_3b_1d_2+a_1b_3d_2
+a_2b_1d_3-a_1b_2d_3}
{-a_3 b_2 c_1+a_2b_3c_1+a_3b_1c_2-a_1 b_3 c_2 -a_2 b_1 c_3 
+ a_1 b_2 c_3}\nabla\Theta^2\nonumber\\
& &+\frac{a_3 b_2 e_1-a_2 b_3 e_1+a_2 b_1 e_3-a_1 b_2 e_3}
{-a_3 b_2 c_1+a_2b_3c_1+a_3b_1c_2-a_1 b_3 c_2 -a_2 b_1 c_3 
+ a_1 b_2 c_3}\nabla^2\Theta.
\end{eqnarray}
These expressions can be put into the equation for $\Theta$ to get
\begin{eqnarray}
\dot\Theta=C_0+C_1(\nabla\Theta)^2+C_2\nabla^2\Theta
\end{eqnarray}
where
\begin{eqnarray}
C_0&=&f_4\\
C_1&=&d_{4}+\left[c_4(a_3b_2d_1-a_2b_3d_1-a_3b_1d_2+a_1b_3d_2
+a_2b_1d_3-a_1b_2d_3)\right.\nonumber\\
& &-b_{4}(a_3c_2d_1-a_2c_3d_1-a_3c_1d_2+a_1c_3d_2+a_2c_1d_3
-a_1c_2d_3)\nonumber\\
& &\left.+a_{4}(b_3c_2d_1-b_2c_3d_1-b_3c_1d_2+b_1c_3d_2+b_2c_1d_3
-b_1c_2d_3)\right]/\nonumber\\
& &(-a_3b_2c_1+a_2b_3c_1+a_3b_1c_2-a_1b_3c_2-a_2b_1c_3+a_1b_2c_3)\\
C_2&=&e_{4}+\left[c_{4}(a_3b_2e_1-a_2b_3e_1+a_2b_1e_3-a_1b_2e_3)
\right.\nonumber\\
& &-b_{4}(a_3c_2e_1-a_2c_3e_1+a_2c_1e_3-a_1c_2e_3)\nonumber\\
& &\left.+a_{4}(b_3c_2e_1-b_2c_3e_1+b_2c_1e_3-b_1c_2e_3)\right]/
\nonumber\\
& &(-a_3b_2c_1+a_2b_3c_1+a_3b_1c_2-a_1b_3c_2-a_2b_1c_3+a_1b_2c_3)
\end{eqnarray}

% REFERENCES


\begin{thebibliography}{9}

\bibitem{kurbook}
Y. Kuramoto,
{\em Chemical Oscillations, Waves and Turbulence\/}, 
(Springer, New York, 1984).

\bibitem{winfree}
A. T. Winfree,
{\em J. Theor. Biol. \/} \textbf{16} (1967) 15.

\bibitem{ertl}
R. Imbihl and G. Ertl,
{\em Chem. Rev.\/} \textbf{95} (1995) 697.

\bibitem{global}
G. Veser, F. Mertens, A. S. Mikhailov, and R. Imbihl, 
{\em Phys. Rev. Lett.\/} \textbf{71} (1993) 935-939. 

\bibitem{enzymes}
A. S. Mikhailov and B. Hess, 
{\em J. Phys. Chem.\/} \textbf{100} (1996) 19059--19065.

\bibitem{kur95}
Y. Kuramoto,
{\em Prog. of Theor. Phys.\/} \textbf{94} (1995) 321--330.

\bibitem{kur96}
Y. Kuramoto and H. Nakao,
{\em Phys. Rev. Lett.\/} \textbf{76} (1996) 4352--4355.

\bibitem{kur97}
Y. Kuramoto and H. Nakao,
{\em Physica D\/} \textbf{103} (1997) 294--313.

\bibitem{kur98}
Y. Kuramoto, D. Battogtokh and H. Nakao,
{\em Phys. Rev. Lett.\/}\textbf{81} (1998) 3543--3546

\bibitem{batt99}
D. Battogtokh,
{\em Prog. of Theor. Phys.\/} \textbf{102} (1999) 947--952.

\bibitem{kur00}
Y. Kuramoto, H. Nakao and D. Battogtokh,
{\em Physica A\/} \textbf{288} (2000) 244--264.

\bibitem{batt00}
D. Battogtokh and Y. Kuramoto,
{\em Phys. Rev. E\/} \textbf{61} (2000) 3227--3230.

\bibitem{batt02}
D. Battogtokh,
{\em Phys. Lett. A\/} \textbf{299} (2002) 558--564.

\bibitem{chapter9}
Y. Kuramoto,
in {\em Nonlinear Dynamics and Chaos: Where do we go from here?\/}, 
edited by S.J. Hogan, A.R. Champneys, B. Krauskopf, M.d. Bernardo, 
R.E. Wilson, H.M. Osinga, and M.E. Homer (Institute of Physics, 
Bristol, 2002) Chapter: Reduction methods applied to nonlocally 
coupled oscillator systems 800--819.

\bibitem{shima}
S. Shima and Y. Kuramoto,
{\em Phys. Rev. E} \textbf{69} (2004) 036213.

\bibitem{tanaka}
D. Tanaka and Y. Kuramoto,
{\em Phys. Rev. E\/} \textbf{68} (2003) 026219.

\bibitem{batt96}
D. Battogtokh and A.S. Mikhailov,
{\em Physica D\/} \textbf{90} (1996) 84.

\bibitem{batt97}
D. Battogtokh, A.S. Mikhailov, and A. Preusser
{\em Physica D\/} \textbf{106} (1997) 327.

\bibitem{kawamura}
Y. Kawamura and Y. Kuramoto, 
{\em Phys. Rev. E\/} \textbf{69} (2004) 016202. 

\bibitem{michael01}
M. Stich, M. Ipsen and A. Mikhailov,
{\em Phys. Rev. Lett.} \textbf{86} (2001) 4406.

\bibitem{michael02}
M. Stich, M. Ipsen and A. Mikhailov,
{\em Physica D} \textbf{171} (2002) 19.

\bibitem{coullet97}
M. Argentina and P. Coullet,
{\em Phys. Rev. E\/} \textbf{56} (1997) R2359-R2362.

\bibitem{coullet98}
M. Argentina and P. Coullet, 
{\em Physica A\/} \textbf{257} (1998) 45--60.

\bibitem{coullet04}
P. Coullet and L. Kramer,
{\em Chaos} \textbf{14} (2004) 244.

\bibitem{batt04}
D. Battogtokh and J.J. Tyson,
{\em Phys. Rev. E\/} \textbf{70} (2004) 026212.

\bibitem{science}
M. Kim, M. Bertram, M. Pollmann, A. von Oertzen, A.S. Mikhailov, 
H.H. Rotermund, and G. Ertl,
{\em Science\/} \textbf{292} (2001) 1357.

\bibitem{bertram1}
M. Bertram and A. S. Mikhailov, 
{\em Phys. Rev. E\/} \textbf{67} (2003) 036207.

\bibitem{bertram2}
M. Bertram, C. Beta, M. Pollmann, A. S. Mikhailov, H. H. Rotermund, 
and G. Ertl, 
{\em Phys. Rev. E\/} \textbf{67} (2003) 036208.

\end{thebibliography}
\end{document}